\documentclass[preprint,
aps,amsmath,nofootinbib,tightenlines]{revtex4}
\usepackage{graphicx}
\usepackage{bm}
\usepackage{epsfig}
\usepackage{graphics,color}
\usepackage{amssymb}
\usepackage{afterpage, ifthen}

\def\bmag#1{{|{\mathbf #1}|}}

\def\Dsl{\hbox{/\kern-.6000em D}} 

\def\dsl{\,\raise.15ex\hbox{/}\mkern-13.5mu D}
\def\bsigma{\mbox{\boldmath $\sigma$}}

\def\psip#1{\psi_{\mathbf{#1}}}
\def\chip#1{\chi_{\mathbf{#1}}}
\def\bsigma{\mbox{\boldmath $\sigma$}}

\def\ltap{\ \raise.3ex\hbox{$<$\kern-.75em\lower1ex\hbox{$\sim$}}\ }
\def\gtap{\ \raise.3ex\hbox{$>$\kern-.75em\lower1ex\hbox{$\sim$}}\ }
\def\OMIT#1{}

\def\lsim{\mathrel{\raise.3ex\hbox{$<$\kern-.75em\lower1ex\hbox{$\sim$}}}}
\def\gsim{\mathrel{\raise.3ex\hbox{$>$\kern-.75em\lower1ex\hbox{$\sim$}}}}

\def\msb{{\overline{\rm MS}}}

\def\CL{{\cal L}}

\newcommand{\nn}{\nonumber}

\newcommand{\bmk}{\mathbf k}
\newcommand{\bmp}{\mathbf p}
\newcommand{\bmpp}{{\mathbf p^\prime}}

\newcommand{\bmq}{\mathbf q}
\newcommand{\bmr}{\mathbf r}

\newcommand{\bmD}{\mathbf D}

\newcommand{\bmsigma}{\mathbf \bsigma}

\newcommand{\bmx}{\mathbf x}

\newcommand{\mG}{m_t\Gamma_t}

\def\CA{{\cal A}}
\def\CO{{\cal O}}
\def\CV{{\cal V}}
\newcommand{\bmkp}{{\mathbf k^\prime}}
\newcommand{\Ord}[1]{{\cal O}\!\left(#1\right)}
\newcommand{\arsinh}{\sinh^{-1}}
\newcommand{\im}{{\mbox{Im}}}
\newcommand{\re}{{\mbox{Re}}}
\newcommand{\En}{{\hat E}}
\makeatletter
\newcommand\figcaption{\def\@captype{figure}\caption}
\newcommand\tabcaption{\def\@captype{table}\caption}
\makeatother

\catcode`\@=11
\def\slash{\mathpalette\make@slash}
\def\make@slash#1#2{\setbox\z@\hbox{$#1#2$}%
  \hbox to 0pt{\hss$#1/$\hss\kern-\wd0}\box0}
\catcode`\@=12 

\begin{document}


\preprint{ \vbox{ \hbox{MPP-2010-22} \hbox{TKK-10-14} \hbox{TPP10-15}
\hbox{SFB/CPP-10-18}
}}

\title{\phantom{x}\vspace{0.5cm} 
Phase Space Matching and Finite Lifetime Effects for 
Top-Pair Production Close to Threshold
\vspace{1.0cm} }

\author{Andr\'e H. Hoang}
  \affiliation{Max-Planck-Institut f\"ur Physik (Werner-Heisenberg-Institut) 
  F\"ohringer Ring 6, D-80805 M\"unchen, Germany 
\footnote{Electronic address: ahoang@mppmu.mpg.de}}

\author{Christoph~J.~Rei\ss er}
\affiliation{Institut f\"ur Theoretische Teilchenphysik\\
Karlsruhe Institute of Technology (KIT), \\
D-76128 Karlsruhe, Germany
\footnote{Electronic address: reisser@kit.edu}}

\author{Pedro Ruiz-Femen\'\i a}
\affiliation{Institut f\"ur Theoretische Teilchenphysik und Kosmologie\\
RWTH Aachen University\\
D-52056 Aachen, Germany \vspace{1cm}
\footnote{Electronic address: ruiz@physik.rwth-aachen.de}}


\begin{abstract}
\vspace{0.5cm}
\setlength\baselineskip{18pt}

The top-pair $t\bar t$ production cross section close to threshold in $e^+e^-$
collisions is strongly affected by the small lifetime of the top quark. 
Since the cross section is defined through final states containing the top
decay products, a consistent definition of the cross section depends on
prescriptions how these final states are accounted for the cross section. 
Experimentally, these prescriptions are implemented for example through cuts on 
kinematic quantities such as the reconstructed top quark invariant masses. As long as these 
cuts do not reject final states that can arise from the decay of a top and an
anti-top quark with a small off-shellness compatible with the nonrelativistic
power-counting, they can be implemented through
imaginary phase space matching conditions in NRQCD. The prescription-dependent
cross section can then be determined from the optical theorem using the $e^+e^-$
forward scattering amplitude. We compute the phase space matching conditions
associated to cuts on the top and anti-top 
invariant masses at next-to-next-to-leading logarithmic (NNLL) order and
partially at next-to-next-to-next-to-leading logarithmic (N${}^3$LL) order
in the nonrelativistic expansion and, together with finite lifetime and
electroweak effects known 
from  previous work, analyze their numerical impact on the $t\bar t$ cross
section. We show that the phase space matching contributions are essential to
make reliable NRQCD predictions, particularly for energies below the peak
region, where the cross section is small.
We find that irreducible background contributions associated to
final states that do not come from top decays are strongly suppressed and can
be neglected for the theoretical predictions.

\end{abstract}

\maketitle


\newpage

%
%
%
\section{Introduction}
\label{sectionintroduction}

The measurement of the line shape of the total cross section
$\sigma(e^+e^-\to t\bar{t})$ for top quark
pair production at energies around the top-antitop threshold ($\sqrt{q^2}\sim
350\,$GeV) constitutes a major part of the top physics program at a future
linear collider (LC). The rise and the form of the cross section allow for
precise measurements of the top quark mass $m_t$ in a threshold mass
scheme~\cite{Hoang:2000yr},  
the top quark width $\Gamma_t$, the top-Yukawa coupling $y_t$ and the strong 
coupling $\alpha_s$.
In view of the expected experimental precision at the
LC~\cite{Martinez:2002st}, theoretical uncertainties for the 
predictions at the level of ${\rm d}\sigma/\sigma\sim 2-3\%$ in the peak and the
continuum region are desired~\cite{Hoang:2006pc}. In the energy region below the
peak where the cross section is becoming tiny the theory error should not
exceed the level of around $5$~fb~\cite{Martinez:2002st}. 

From the theoretical perspective the QCD dynamics of the top quark pair for
threshold energies is quite nontrivial since the small relative
velocity of the top-antitop pair $v\ll 1$ leads
to a proliferation of physical scales that need to be accounted for: the top
mass $m_t$, the relative three-momentum $\bmp\sim m_t v$ and the nonrelativistic
kinetic energy $E\sim m_t v^2$ of the top quark pair.
In the standard QCD perturbative expansion singular terms $\propto
(\alpha_s/v)^n$ and $\propto(\alpha_s\ln v)^n$ arise from the ratios of these
scales. Since the Coulomb-like dynamics enforces the power-counting
$v\sim\alpha_s(m_t v)$ the proper treatment of these terms requires
resummations within a power-counting framework with a simultaneous expansion in
$v\sim\alpha_s\ll 1$.  This can be achieved with nonrelativistic QCD (NRQCD), a
low-energy effective field theory (EFT) of QCD that separates the different 
quantum fluctuations that are relevant for the kinematic situation of 
heavy nonrelativistic quark pairs.
Within the fixed-order approach, which achieves a systematic summation of terms 
$\propto\alpha_s^n v^m$, complete NNLO  predictions (i.e. $n+m\le 3$) have been
made~\cite{Hoang:2000yr}. 
For available NNNLO results we refer to
Refs.~\cite{Hoang:2003ns,Beneke:2005hg,Eiras:2006xm,Beneke:2008ec,Beneke:2008cr,Kiyo:2008mh,Anzai:2009tm,Smirnov:2009fh,Smirnov:2010zc}.
Using a proper low-scale
short-distance threshold mass scheme with cut-off scale $R\sim
m_t\alpha_s$~\cite{Hoang:1998nz,Beneke:1998rk,Hoang:2008yj,Hoang:2009yr} 
(which avoids the pole mass renormalon) the energy where the cross section rises
is stable in perturbation theory. However, the fixed-order predictions suffer
from large normalization uncertainties at the level of 10-20~\% which indicate
potentially large 
logarithmic terms. These normalization uncertainties are particularly
problematic for measurements of the top width and the Yukawa coupling $y_t$. 
This problem of potentially large logarithmic terms is addressed in 
renormalization group improved NRQCD calculations of the cross section, which
account for a systematic summation of terms $\propto\alpha_s^n v^m \ln^k v$.
Renormalization group improved QCD results are fully known at NLL order 
(i.e.~$n+m-k\le 2$). At NNLL order (i.e.~$n+m-k\le 3$) all ingredients are known
except for the NNLL renormalization group evolution of the Wilson coefficient of
the leading order top pair current~\cite{Hoang:2000ib,
Hoang:2001mm,Pineda:2006ri}. 
Partial results for the NNLL order anomalous dimension of the current have been
computed in Ref.~\cite{Hoang:2003ns,Hoang:2006ht}.  
At the present stage
renormalization group improved results have a normalization uncertainty of 
6-10~\%~\cite{Hoang:2003xg,Pineda:2006ri}.

Electroweak effects and in particular the top quark decay play an equally
important role. Already at leading order it is important to account for the top
decay width since it widens the top-antitop bound state resonances and turns the
threshold cross section into a smooth lineshape. Despite this fact the
determination of subleading electroweak and finite lifetime effects have
received somewhat less attention in the literature in the past.
Theoretically, for predictions of an inclusive cross section the top width acts 
as
an infrared cutoff for the top energy and thus allows for a
perturbative computation for all threshold energies.  
We have $\Gamma_t\approx (G_F/8\sqrt{2}\pi)m_t^3\approx 1.5$~GeV, which scales
like $g_2^2 m_t \sim g_1^2 m_t$ where $g_1$ and $g_2$ are the U(1) and SU(2)
couplings, and we also find numerically that 
$\Gamma_t\sim m_t \alpha_s^2$. With  $g_1^2\sim g_2^2\sim\alpha_s$ it is
therefore natural to adopt the power-counting  
\begin{align}
\label{powercounting}
v\sim \alpha_s\sim \alpha_{\rm qed}^{1/2} \ll 1
\end{align}
when making perturbative theoretical predictions.

Electroweak interactions are responsible for various effects that
can be categorized into four classes: (a) ``Hard" electroweak effects, 
which includes hard, point-like corrections related to the
$t\bar{t}$ production mechanism by virtual photons and Z exchange,
as well as electroweak corrections to the hard matching conditions of the NRQCD 
operators and potentials, (b) electromagnetic effects for the
luminosity spectrum of the $e^+e^-$ initial state, (c) electromagnetic
corrections to the low-energy nonrelativistic dynamics of the $t\bar{t}$ pair
and its decay products and (d) effects related to the finite top quark
lifetime. 
The corrections from class (a) can be determined by standard methods through
matching for top quarks in the on-shell limit and are real numbers. Up to NNLL
order they have been discussed and implemented in
Ref.~\cite{Hoang:2006pd}. Earlier work can be found in
Refs.~\cite{Guth:1991ab,Grzadkowski:1986pm}. The QED beam effects from class (b) are taken into account by a
convolution of the ``partonic'' cross section with an on-shell $e^+e^-$ pair in
the initial state with the collider's luminosity spectrum. The luminosity
spectrum accounts for initial state radiation, the 
accelerator-dependent beam energy spread and the beam-strahlung, and effects
coming from beam-beam interactions. Since the luminosity spectrum is for the
most part determined either experimentally or from experimental simulations we
do not consider it any further in this work.  
The photon interactions of the $t\bar{t}$ pair belonging to class (c) are
quite similar to the gluonic corrections and can be incorporated in the same way
into NRQCD through potentials and ultrasoft interactions. The most important
of these effects are the QED corrections to the Coulomb potential contributing 
at
NLL order.   

The finite lifetime effects of class (d) are the main purpose of this
work. 
A proper treatment of instability effects entails that for the definition
of the cross section, final states compatible with the top decay chains are
accounted for. Since this necessarily also includes final states that do not
arise from top decays, but are experimentally indistinguishable, the cross
section - and potentially also the theoretical methods to compute it - are
dependent on what selection prescriptions are employed. In general, as long as
selection cuts do not reject final states that can arise from the decay of a top
and an anti-top quark with a small off-shellness compatible with the power
counting of a nonrelativistic (anti-)top quark propagator\footnote{We call such
selection prescriptions ``inclusive'' throughout our work.}, they can be
implemented through imaginary contributions to the Wilson coefficients of the
NRQCD operators. This means that the top decay products are integrated out
together with the information on the selection cuts. The resulting NRQCD
Lagrangian is formally non-Hermitian.  
The cross section can then be obtained via the optical theorem, i.e.~from the
imaginary part of the $e^+e^-\to e^+e^-$ forward scattering
amplitude~\cite{Fadin:1987wz,Fadin:1988fn,Hoang:2004tg}. In this work we adopt
this inclusive approach.  

\begin{figure}[t]
  \begin{center}
  \includegraphics[width=0.6\textwidth]{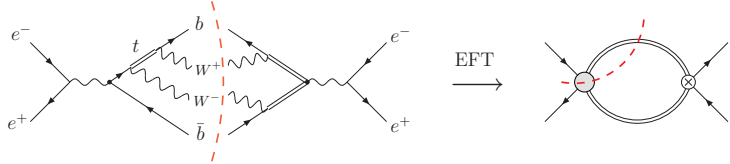}
  \caption{Interference of the double-resonant diagram with a single-resonant
    one, as described in the full and effective theories. The (grey) blob in the
    $t\bar{t}$ production vertex in the EFT diagram represents the NNLL matching
    condition which results from integrating out the lower $\bar bW^-$ loop. The cut
    through this vertex represents the contribution of the absorptive part of
    the matching condition in the optical theorem. The (red) dashed
  line means we extract the imaginary part of the forward scattering amplitude
  or, equivalently, that we perform the phase space integration over the
  particles in the cut. Note that we use double lines for representing the top quarks.} 
  \label{fig2}
  \end{center}
\end{figure}

One can distinguish two
different types of imaginary contributions to the NRQCD Wilson coefficients:
(1) contributions arising from cuts of full theory diagrams through top
decay final states and (2) contributions describing the selection
prescriptions, which we will also frequently refer to as cuts in the
following.~\footnote{The different meanings of the word ``cut'' used frequently 
in this work should be clear from the context.} The type-1 contributions
describe decays of (anti)top quark modes that are propagating in NRQCD. In the
matching procedure they arise from cuts of full theory diagrams through (anti)top
decay final states. The leading type-1 NRQCD term is the 
well known on-shell width contribution,
\begin{equation}
\delta{\cal L} = \sum_{\bmp} \psi_\bmp^\dagger \, \frac{i}{2}\Gamma_t \, 
\psi_\bmp
+ \sum_{\bmp} \chi_{\bmp}^\dagger \, \frac{i}{2} \Gamma_t \, \chi_\bmp 
\,,
\label{eq:bilinear}
\end{equation}
which in the matching procedure comes from the $bW$ cut of the full theory top
quark on-shell self-energy, and which contributes at leading-log (LL)
order in NRQCD. Here $\Gamma_t$ 
is the top quark on-shell width, and $\psi_{\bmp}$ and
$\chi_{\bmp}$ represent Pauli spinor field operators that destroy top and
antitop quarks, respectively. This term leads to the NRQCD (anti)top propagator
of the form 
\begin{align}
\label{eq:prop}
\frac{i}{p_0-\frac{\bmp^2}{2m_t}+i\frac{\Gamma_t}{2}}
\end{align}
and can be readily implemented
into computations for stable top quarks supplemented by the replacement rule 
$E\to E+i\Gamma_t$, where $E=\sqrt{s}-2m_t$ is the c.m.~energy with respect to
the two-particle threshold~\cite{Fadin:1987wz,Fadin:1988fn}.
The type-1 contributions to the Wilson coefficients up to NNLL order and
neglecting the width of the W-boson were
determined in Ref.~\cite{Hoang:2004tg}. As a new higher order feature these
include interference contributions 
from double-resonant ($e^+e^-\to t\bar t\to b\bar b W^+W^-$) 
and single-resonant ($e^+e^-\to t\bar b W^-\,, \bar t b W^+\to b\bar b W^+W^-$)
amplitudes as illustrated in Fig.~\ref{fig2}. Electroweak gauge-invariance is
maintained during the matching procedure by the inclusion of the imaginary
contribution of the top quark wave function renormalization Z-factor arising
from the $bW$ intermediate state in the top self-energy of the full theory. 

An important new theoretical aspect is the emergence of
imaginary anomalous dimensions caused by UV-divergences in the NRQCD $t\bar t$
phase space integrations~\cite{Hoang:2004tg}. These UV-divergences originate
from the Breit-Wigner behavior of the top and antitop propagators of
Eq.~(\ref{eq:prop}) which, upon being cut, lift the on-shell dispersion relation
$p_0=\bmp^2/2m_t$. In the forward scattering 
amplitude these propagators allow (anti)top intermediate states contributing
to the NRQCD cross sections which have arbitrarily large invariant mass. Once
$v^2$-suppressed NNLL order operators are inserted, this behavior causes
UV-divergences which are compensated by imaginary counter-terms associated
with $(e^+e^-)(e^+e^-)$ forward-scattering operators. The resulting NLL order
imaginary anomalous dimensions of the Wilson coefficients of the
$(e^+e^-)(e^+e^-)$ forward-scattering operators sum $\ln v$ terms from the top
decay phase space. It was shown in Ref.~\cite{Hoang:2004tg} that the
type-1 contributions to the imaginary parts of the NRQCD Wilson coefficients are
numerically important for the normalization of the cross section and for the
correct prediction of the c.m.~energy of the $t\bar t$ quasi-resonance peak. 
The existence of phase space divergences in NRQCD inclusive cross section
computations for top pair production at threshold indicates that - due to the
finite lifetime - predictions beyond the leading order approximation need
additional short-distance information to be defined unambiguously. This
short-distance information is provided by the experimental
selection criteria for the final states that are accounted for the cross
section determination, and also incorporate background contributions related to diagrams 
without an intermediate $t\bar{t}$ pair.\footnote{Also outside the framework of effective
theories such selection criteria are in general necessary, and prescription-free
cross section definitions for unstable particle production do no exist. The
numerical impact of the selection prescriptions can, however, be frequently
neglected if the width of the involved particles is much smaller than other
relevant kinematic scales.} 
In NRQCD this phase-space short-distance information is incorporated in the
imaginary contributions to the Wilson coefficients of type-2 we have already
mentioned above. We call the computation of these contributions the ``phase
space matching'' procedure. Many different feasible inclusive cross section
definitions can 
be devised.

In this work we determine and analyze the type-2 imaginary contributions to the
NRQCD Wilson coefficients at NNLL and N${}^3$LL order for cuts on the invariant
masses of the 
reconstructed top and antitop quarks, $M_{t}$ and $M_{\bar t}$, respectively. 
For simplicity we neglect the width of the $W$ bosons and combinatorial
background in the reconstruction. The combinatorial background can be estimated
from Monte-Carlo simulations. We demonstrate that for moderate top invariant
mass cuts  
\begin{align}
\label{eq:invarmasscuts}
|M_{t,\bar t}-m_{t}|\,\le\, \Delta M_t
\end{align}
with $\Delta M_t \sim 15 - 35$~GeV, the phase space matching conditions are
dominated by the NRQCD phase space contributions, i.e.\ they can be computed
from the difference 
between the (potentially) divergent NRQCD phase space integrations without any
cuts and the ones with the cuts in Eq.~(\ref{eq:invarmasscuts}) being
imposed. 
This is because using the $\overline{\mbox{MS}}$ scheme in NRQCD diagrams 
involving the unstable top propagator of Eq.~(\ref{eq:prop}) largely
overestimates the contributions from unphysical phase space 
regions that are parametrically away from the potential, soft and ultrasoft 
regions that can be described by NRQCD.\footnote{
A similar feature arises also for the computation matching conditions
within NRQCD for a stable quark, see Ref.~\cite{Hoang:1999zc} for a comparison of
a cutoff scheme with $\overline{\mbox{MS}}$.} 
Thus the main numerical effect of the phase space matching procedure is to
remove these unphysical contributions
and the phase space matching procedure can be carried out within
NRQCD itself. The remaining hard contributions, which require the evaluation of 
multi-leg full theory diagrams, are smaller than $5$~fb and can be neglected
in view of the expected experimental precision. This simplifies the computations
substantially and makes the determination of higher order QCD corrections
feasible. Since one can assume that the situation is similar for the threshold
production of other heavy unstable colored particles in new physics models, it is
straightforward to generalize our results for such processes, also for the
hadron collider environment. 

An important conceptual aspect of the invariant mass cuts defined in
Eq.~(\ref{eq:invarmasscuts}) is that already for moderate cuts 
$\Delta M_{t}\sim 15-35$~GeV the cut $\Delta \bmp$ on the nonrelativistic
(anti)top three-momentum $\bmp$ is $\Delta \bmp \sim \sqrt{2m_t \Delta
  M_{t}}\sim 100$~GeV, and thus represents a hard scale of the order $m_t$. 
This justifies the implementation of the phase space effects into the matching
conditions of the Wilson coefficients. From a technical point of view, the
method of computating the phase 
space matching conditions within NRQCD is quite similar to using a cutoff
scheme to regularize ultraviolet (UV) divergences in loop diagrams for the
renormalization. The procedure therefore leads to power-counting breaking effects which
could spoil the nonrelativistic expansion. We find that this is not the case
partly because the momentum cutoff $\Delta \bmp$ is still sufficiently smaller
than the top mass $m_t$. In analogy to the usual renormalization and matching
computations beyond the one-loop level, the phase space matching procedure also
requires the determination of phase space matching coefficients of subdiagrams
at higher orders in the loop expansion. Many technical details of the
calculations carried out in this work are given in Ref.~\cite{ReisserPhd}.

The outline of this paper is as follows: 
In Section~\ref{sectionnotation} we review the NRQCD framework which we use 
to describe the top-antitop resonance region, and collect previous results on
electroweak 
and finite lifetime effects in the cross section. In Section~\ref{sec:basic}
we discuss the main concepts of the NRQCD phase space matching procedure with 
special emphasis on the definition of the inclusive cross section. In 
Section~\ref{subsectionphasespacematchingNRQCD} we compute the 
imaginary type-2 matching conditions for the Wilson coefficients of the $(e^+e^-)(e^+e^-)$
forward scattering operators with $\alpha_s=0$ and a cut $\Delta M_t$ on the
reconstructed invariant masses of the top and antitop quarks, including also
${\cal O}(v^2)$ relativistic corrections. The structure of the nonrelativistic
expansion of the phase space matching contributions for the inclusive NRQCD
cross section is examined in Sec.~\ref{subsectionfulltheory} by a comparison
with the full Standard Model tree-level predictions for the processes  
$e^+e^-\to t\bar{t} \to b\bar{b}\,W^+W^-$ and $e^+e^-\to b\bar{b}\,W^+W^-$
obtained from Madgraph~\cite{Alwall:2007st}. The ${\cal O}(\alpha_s)$ QCD
corrections to the phase space matching conditions arising from ultrasoft gluon
exchange and potential interactions are computed in Sec.~\ref{sectionQCD}. The
calculations allow to determine the complete set of NNLL 
phase space matching contributions and of a part of the N${}^3$LL
corrections. The results are discussed in detail and analyzed specifically
with respect to the convergence of the $\alpha_s$-expansion, and we also discuss
non-perturbative effects. 
Details on the computation  of the ${\cal O}(\alpha_s)$ ultrasoft corrections to
the  phase space matching coefficients are relegated to
Appendix~\ref{app:QCDinterference}. Finally, in Sec.~\ref{sectionanalysis}, we
analyze the numerical impact of the phase space matching contributions to the
inclusive NRQCD cross section and compare their size to the other types of
electroweak and finite lifetime corrections. Readers only interested in the main
concepts of the phase space matching and the numerics of the final results might
jump from here directly to Sec.~\ref{sec:basic} and then to
Sec.~\ref{sectionanalysis}.

\section{NRQCD Formalism and Previous Results} 
\label{sectionnotation}

For the work in this paper we use the vNRQCD effective field theory
formalism~\cite{Luke:1999kz,Manohar:2000hj,Hoang:2002yy} to describe the nonrelativistic top-antitop dynamics
relevant for the inclusive threshold cross section. In this section we briefly
review the basic vNRQCD notation and ingredients with special emphasis on
electroweak and finite lifetime effects in the cross section. We also outline
the previous results concerning the type-1 imaginary contributions to the vNRQCD
Wilson coefficients determined in Ref.~\cite{Hoang:2004tg}. We note that the results are
general and can in principle also be implemented  with minimal notational
modifications within other formalisms such as pNRQCD~\cite{Brambilla:1999xf}. 

\vskip 5mm
\noindent
{\bf \underline {Basics:}}
\nopagebreak
\\[2mm]
The vNRQCD Lagrangian contains heavy quark bilinear, potential, soft and
ultrasoft operator terms. Up to the NNLL order the bilinear terms read
\begin{eqnarray} 
\label{Lke}
 {\mathcal L}_{\rm bilinear}(x) &=& \sum_{\bmp}
   \psip{\bmp}^\dagger(x)   \biggl\{ i D^0 - {(\bmp-i\bmD)^2 \over 2 m_t} 
   +\frac{{\bmp}^4}{8m_t^3}  
   + \frac{i}{2} \Gamma_t \bigg( 1 - \frac{\bmp^2}{2 m_t^2} \bigg) 
   - \delta m_t \biggr\} \psip{\bmp}(x) 
\nonumber\\ & & \qquad
+ \, (\psip{\bmp} \to\chip{\bmp})\,,
\end{eqnarray}
where the fields 
$\psip{\bmp}$ and $\chip{\bmp}$ destroy top and antitop quarks with
label momentum ${\bmp}$, and $\Gamma_t$ is the top quark
width defined at the top quark pole. The velocity counting of the (anti)top
quark fields for $d=4$ is $\psip{\bmp}\sim \chip{\bmp}\sim v^{3/2}$.
The term $\propto
\Gamma_t\frac{\bmp^2}{2m_t}$ is the top lifetime dilation correction. 
We also included the ultrasoft
gauge covariant derivative $D^\mu=(D^0,-\bmD)=\partial^\mu + i g A^\mu$ where
$A^\mu$ is the ultrasoft gauge field. The dependence on $D^\mu$ is directly  tied
to the (anti)top three-momentum label $\bmp$ to all orders of perturbation theory
through reparametrization invariance~\cite{Luke:1992cs}.    
We use the $v$-counting $D^0\sim m_t v^2\sim\Gamma_t$, and the ultrasoft
interactions start contributing in matrix elements at N$^3$LL order.
The term $\delta m_t$ is a residual mass term which is of order $v^2$ in a
short-distance threshold mass scheme~\cite{Hoang:2000yr}. The LL order terms in
Eq.\,(\ref{Lke}) lead to the top/antitop propagator in Eq.\,(\ref{eq:prop}).

The leading order potential term contains the well known 
Coulomb interaction,
\begin{eqnarray}
{\cal L}_{\rm pot} & = &
-\sum_{\bmp,\bmp^\prime} \frac{{\cal V}_c^{(s)}(\nu)}{(\bmp-\bmp^\prime)^2}\,
\psip{\bmp^\prime}^\dagger \psip{\bmp} \chip{-\bmp^\prime}^\dagger\chip{-\bmp}
\,,
\label{Lpot}
\end{eqnarray} 
where 
\begin{eqnarray}
{\cal V}_c^{(s)}(\nu)=-4\pi C_F \alpha_s(m_t\nu)
\label{Vccoeff}
\end{eqnarray} 
is the Coulomb Wilson coefficient for a color singlet heavy quark pair and $\nu$
is the vNRQCD velocity renormalization scale. The velocity renormalization scale
$\nu$ is introduced to conveniently describe the correlated renormalization
group evolution of soft ($\sim m_t v$) and ultrasoft 
($\sim m_t v^2$) effects, and its natural scaling to sum large logarithmic terms
is $\nu\sim v\sim\alpha_s$.  The evolution of the Coulomb  Wilson coefficient
differs from the running of the strong coupling starting at NNLL order due to
ultrasoft gluon corrections. There are also radiative corrections to the Coulomb
interaction arising from soft gluon loops. In vNRQCD they contribute through
matrix elements involving T-products of soft gluon operators. These QCD
corrections to the Coulomb potential are known at 
${\cal O}(\alpha_s)$~\cite{Fischler:1977yf} and 
${\cal O}(\alpha_s^2)$~\cite{Schroder:1998vy,Peter:1996ig,Peter:1997me}, and 
have recently been determined even at 
${\cal O}(\alpha_s^3)$~\cite{Anzai:2009tm,Smirnov:2009fh,Smirnov:2010zc}. 
A discussion on the $v^2$-suppressed vNRQCD potentials is given in 
Refs.~\cite{Manohar:2000hj,Hoang:2002yy}. We note
that there are also potentials generated by the electroweak interactions. The
dominant one is the QED contribution to the Coulomb potential which can be
easily implemented by writing 
${\cal V}_c^{(s)}(\nu) = -4\pi C_F \alpha_s(m_t\nu) - 4\pi\alpha(m_t\nu)$.
According to Eq.~(\ref{powercounting}) this QED effect contributes at NLL
order. There are no ${\cal O}(\alpha\alpha_s)$ QED NNLL corrections that
contribute to the Coulomb potential. 
There are also higher order electroweak potentials initiated e.g.\ by the
exchange of the Higgs~\cite{Jezabek:1993tj,Harlander:1995dp,Eiras:2006xm} or the
Z-boson. 

Eqs.~(\ref{Lke}) and~(\ref{Lpot}) yield the LL nonrelativistic
top-antitop dynamics. For predictions of the inclusive cross section 
one important ingredient is the zero-distance Green function. At LL order and
using dimensional regularization with $d=4-2\epsilon$, the unrenormalized
zero-distance Green function has the form ($a=C_F\alpha_s$)
\begin{eqnarray}
 G^0(a,v,m_t,\nu) & = &
 \frac{m_t^2}{4\pi}\left\{\,
 i\,v - a\left[\,\ln\left(\frac{-i\,v}{\nu}\right)
 -\frac{1}{2}+\ln 2+\gamma_E+\psi\left(1\!-\!\frac{i\,a}{2\,v}\right)\,\right]
 \,\right\}
 \nonumber \\ & &
 +\,\frac{m_t^2\,a}{4 \pi}\,\,\frac{1}{4\,\epsilon}
\,,
\label{deltaGCoul}
\end{eqnarray}  
where
\begin{eqnarray}
v & = &
\sqrt{\frac{\sqrt{s}-2(m_t+\delta m_t)+i\Gamma_t}{m_t}}
\,,
\end{eqnarray}
$\psi(x)\equiv d/dx \ln \Gamma(x)$ is the digamma function,
and $\sqrt{s}$ is the $e^+e^-$ c.m.\ energy.  
The divergent $1/\epsilon$ term in Eq.~(\ref{deltaGCoul}) is a UV-divergence in
the real part of the LL order Green function and does not play any role for the
production of stable particles. 

\vskip 5mm
\noindent
{\bf \underline {Top pair production currents:}}\\[2mm]
Top-antitop pair production and annihilation is described by currents involving
the top and antitop fields. For cross section predictions up to NNLL order one
needs the leading and subleading $^3S_1$ currents ${\cal O}_{\bmp,1}^j$ and 
${\cal O}_{\bmp,2}^j$, respectively, and the $^3P_1$
current ${\cal O}_{\bmp,3}^j $, which is
$\bmp/m_t$-suppressed compared to the leading $S$-wave current. 
These currents have the form~\cite{Hoang:2000ib,Hoang:2001mm}
\begin{eqnarray}{\cal O}_{\bmp,1}^j & = &
\psi_{\bmp}^\dagger\, \sigma^j (i\sigma^2)
  \chi_{-\bmp}^*\,,
\qquad
{\cal O}_{\bmp,2}^j  = 
\frac{1}{m_t^2}\,\psi_{\bmp}^\dagger\,\bmp^2\, \sigma^j (i\sigma^2)
  \chi_{-\bmp}^*
\,,\nonumber\\
  {\cal O}_{\bmp,3}^j & = &
\frac{-i}{2m_t}\,\psi_{\bmp}^\dagger\,[\sigma^j,\bmsigma\cdot\bmp]\, (i\sigma^2)
  \chi_{-\bmp}^*
\,.
\label{currents}
\end{eqnarray}
To ensure electroweak gauge invariance at subleading order it is
necessary to include the initial electron and positron
fields, which leads to the $t\bar t$ production operators
\begin{eqnarray}
{\cal O}_{V,\bmp,\sigma} & = & 
\big[\,\bar e_{+}\,\gamma_j\,e_{-}\,\big]\,{\cal O}_{\bmp,\sigma}^j
\,,\qquad
{\cal O}_{A,\bmp,\sigma} \, = \,
\big[\,\bar e_{+}\,\gamma_j\,\gamma_5\,e_{-}\,\big]\,{\cal O}_{\bmp,\sigma}^j
\,,
\label{eettop}
\end{eqnarray}
where the index $j = 1,2,3$ is summed and the index $\sigma = 1,2,3$
distinguishes between the different currents. 
The current ${\cal O}_{V,\bmp,1}$ contributes at LL order for the inclusive
cross section, and ${\cal O}_{V,\bmp,2}$ and  ${\cal O}_{V,\bmp,3}$ contribute
at NNLL order. Because the effective theory is
constructed such that it describes $t \bar t$ production only in the
threshold region and in the c.\,m.\ frame, only those initial
$e^+e^-$ states%
\footnote{
Here, $a_{\tau}(\bmk)$ and
$a_{\tau'}^c(\bmkp)$ are operators for the annihilation of an electron
and a positron with spin $\tau$, $\tau'$ and
3-momentum $\bmk$, $\bmkp$, respectively, and $|0\rangle$ is the
vacuum state.}
 $a_{\tau'}^{c\dagger}(\bmkp)a_{\tau}^s\dagger(\bmk) |0\rangle$
are allowed that fulfill $s\equiv(k+k')^2 \approx
4m_t^2$ and $\bmk = -\bmkp$. For simplicity we assume electron and
positron to travel along the $z$-direction, therefore the explicit
form of their 4-momenta is
\begin{eqnarray}
k^\mu &=& \left(\frac{\sqrt{s}}{2}, \frac{\sqrt{s}}{2}\, {\bf \hat e}_z
\right)\,,
\qquad
k^{\prime\mu} \,=\, \left(\frac{\sqrt{s}}{2},- \frac{\sqrt{s}}{2}\, {\bf \hat 
e}_z
\right)\,,
\label{electronmomenta}
\end{eqnarray}
${\bf \hat e}_z$ being the unit vector in $z$-direction. 
The fields $e_{-}$ and $e_{+}$ in Eqs.~(\ref{eettop}) are now defined as
\begin{eqnarray}
e_{-}(x) & = & \sum\limits_{\tau,\,\sqrt{s}} 
a_{\tau}(\bmk)\, u_{\tau}(\bmk)\, e^{-i \hat k\cdot x} \,,
\qquad
e_{+}(x) \, = \, \sum\limits_{\tau,\,\sqrt{s}} a_{\tau}^{c\dagger}(\bmkp)\,
v_{\tau}(\bmkp)\, e^{i \hat k'\cdot x} 
\,,
\label{electronfields}
\end{eqnarray}
where $u_{\tau}(\bmk)$ and $v_{\tau}(\bmkp)$ denote Dirac spinors for electron
and positron, respectively, and the momenta $\bmk$ and $\bmkp$ 
refer to Eqs.~(\ref{electronmomenta}). The sum over the
c.\,m.\ energy $\sqrt{s}$ 
is restricted to the threshold region. For simplicity we do not
sum over the angles of the electron and positron
momenta. The phase factors in Eqs.~(\ref{electronfields}) are defined such that
they describe only the $t\bar t$ low-energy fluctuations,
\begin{eqnarray}
\label{kdef}
\hat k^\mu  & = & \left(\frac{\sqrt{s}}{2}-m_t, \frac{\sqrt{s}}{2}\,
     {\bf \hat e}_z 
\right)\,,
\qquad
{\hat k}^{\prime\mu} \, = \, \left(\frac{\sqrt{s}}{2}-m_t,- \frac{\sqrt{s}}{2}\,
{\bf \hat e}_z 
\right)\,.
\end{eqnarray}
The dependence on 3-momenta in Eq.~(\ref{kdef}) vanishes after the operators
$\CO_{V/A,\bmp,\sigma}$ have been applied to the initial $e^+e^-$ state.
The operators for $t\bar t$
annihilation are
obtained from $\CO_{V/A,\bmp,\sigma}$ by Hermitian conjugation.

Due to the dependence of the intermediate photon
and $Z$ boson propagator on the c.m.\ energy in the process $e^+e^- \to
\gamma^*, Z^* \to t\bar t$,
we also introduce the $t\bar t$ production operators
\begin{eqnarray}
{\cal O}_{V,\bmp,1}^{(1)} & = & 
\big[\,\bar e_{+}\,\gamma_j\,(\En/m_t)\,e_{-}\,\big]\,{\cal O}_{\bmp,1}^j
\,,
\qquad
{\cal O}_{A,\bmp,1}^{(1)} \, = \,
\big[\,\bar
  e_{+}\,\gamma_j\,\gamma_5\,(\En/m_t)\,e_{-}\,\big]\,{\cal
  O}_{\bmp,1}^j 
\,.
\label{eettop1}
\end{eqnarray}
Here, $\En$ denotes the operator 
$\En=i\partial_0$ acting on the fields to the right and to the
left and thus picks up the kinetic energy $E \equiv
\sqrt{s} - 2m_t\sim m_t v^2$ from the initial $e^+e^-$ state. So these 
operators contribute a NNLL order. Concerning QCD effects, the operators
${\cal O}_{V/A,\bmp,1}^{(1)}$ have the same
matching conditions and renormalization group evolution as 
${\cal O}_{V/A,\bmp,1}$ of Eqs.~(\ref{eettop}) but their Wilson coefficients 
differ in the electroweak contributions.
Similar additional operators related to ${\cal O}_{V/A,\bmp,2}$ and 
${\cal O}_{V/A,\bmp,3}$ do not need to be introduced since they would give
contributions beyond N${}^3$LL order.

The contribution of the currents to the Lagrangian reads
\begin{align}
\CL_{\rm cur}  & = \,  \sum\limits_{\bmp}  \bigg[ C_{V,1}\,{\cal
  O}_{V,\bmp,1} + C_{A,1}\,{\cal
    O}_{A,\bmp,1} 
+  C_{V,1}^{(1)}\,{\cal
  O}_{V,\bmp,1}^{(1)} + C_{A,1}^{(1)}\,{\cal
    O}_{A,\bmp,1}^{(1)} \nonumber\\
&\,\qquad{} + C_{V,2}\,{\cal
  O}_{V,\bmp,2} + C_{A,2}\,{\cal
    O}_{A,\bmp,2} 
+ C_{V,3}\,{\cal
  O}_{V,\bmp,3} + C_{A,3}\,{\cal
    O}_{A,\bmp,3}  \,\Big] + \mbox{H.\,c.}\,.
\label{currentlagrangian}
\end{align}
If QED radiative corrections are neglected, the electron and positron fields
in the current operators act like classic fields and do not contribute
to the nonrelativistic $t\bar t$ dynamics. 
The Hermitian conjugation (H.\,c.) acts on the operators in the usual way
and on possible CP-violating phases in the Wilson coefficients as complex
conjugation. It does, however, {\it not} act on imaginary contributions in the  
Wilson coefficients that are related to finite-lifetime contributions.
Since we do not consider CP-phases in this work, the Wilson coefficients of
operators and their conjugated counterparts are equal. 
At NNLL order the currents run only due to QCD effects. Their matching
conditions, which we will generally parameterize at the hard matching velocity
scale $\nu=1$, contain electroweak contributions already at LL order. 
Including QCD, electroweak and finite lifetime effects up to NNLL order the
Wilson coefficients can be written in the form
\begin{align}
C_{V/A,1}(\Lambda,\nu) & = \, C_{V/A,1}^{\rm Born}\, c_1(\nu)\, (1 + 
i\,\delta\tilde c_1(\Lambda)) + i\,C_{V/A,1}^{\rm int }(1 + \delta
\tilde c_1^{\rm int }(\Lambda) ) + C_{V/A,1}^{\rm 1 loop}\,, \nonumber\\[2mm]
C_{V/A,1}^{(1)}(\nu) & = \, C_{V/A,1}^{(1),\rm Born}\,c_1(\nu)\,,
\nonumber\\[2mm]
C_{V/A,2}(\nu) & = \, C_{V/A,1}^{\rm Born}\,
c_2(\nu)\,, \nonumber\\[2mm]
C_{V/A,3}(\nu) & = \, C_{V/A,3}^{\rm Born}\,
c_3(\nu)\,.
\label{wilsoncoeff}
\\[-4mm]\nonumber
\end{align}
In Eqs.~(\ref{wilsoncoeff}) all imaginary contributions are indicated by
an explicit factor of the imaginary~$i$. The terms $c_i(\nu)$ parameterize the
QCD evolution and the hard matching conditions.  
Their explicit form at NLL order can be found in 
Refs.~\cite{Manohar:2000kr,Hoang:2002yy,Pineda:2001et}. Note that for
the NNLL running of $c_1(\nu)$ currently only the non-mixing
contributions~\cite{Hoang:2003ns} are fully known. Partial results for the NNLL
mixing contributions have 
been determined in Refs.~\cite{Hoang:2006ht,StahlhofenPhd}. 
For our numerical analysis of the phase space matching contributions to the
inclusive NRQCD cross section we need the NNLL order corrections to the 
matching coefficient $c_1(\nu=1)$,
\begin{align}
c_1(\nu=1) & =  1 + h_1^{(1)} + h_1^{(2)} + \ldots\,,
\label{c1harddef}
\end{align}
where
\begin{align}
h_1^{(1)} & =  - \frac{2\,C_F}{\pi}\: {\alpha_s(m_t)} 
\,,
\nonumber\\[2mm]
h_1^{(2)} & =   
  \alpha_s^2(m_t) \bigg[C_F^2\bigg(\frac{\ln 2}{3}-\frac{31}{24}
  -\frac{2}{\pi^2}\bigg) + C_A C_F\bigg(\frac{\ln 2}{2}-\frac{5}{8}\bigg) 
  + \frac{\kappa}{2} \bigg] -\frac{ 2 \,Q_t^2}{\pi} \alpha_{\rm qed}(m_t)
\,,
\label{c1hard}
\end{align}
and the constant $\kappa$ was given in Refs.~\cite{Hoang:1998xf,Hoang:1998uv}. 
The last term in Eq.~(\ref{c1hard}) is the one-loop QED matching correction to
$c_1(\nu=1)$, which contributes at NNLL order according to the power-counting
$\alpha_{\rm qed}\sim \alpha_s^2$. 
In Eqs.~(\ref{wilsoncoeff}) the terms $C_{V/A,i}^{\rm Born}$ denote electroweak matching
contributions from the tree level amplitude for the $e^+e^-\to \gamma,Z\to t\bar
t$ production process and $C_{V/A,1}^{\rm 1 loop}$ refers to the hard
one-loop electroweak corrections. The results for  $C_{V/A,1}^{\rm 1 loop}$ are
elaborate and can be found in Ref.~\cite{Hoang:2006pd} (see also 
Refs.~\cite{Guth:1991ab,Grzadkowski:1986pm}).   
The imaginary terms with $C_{V/A,1}^{\rm int }$ contain finite lifetime effects
arising from the interference of the dominant double resonant process $e^+e^-\to
t\bar t\to W^+W^-b\bar b$ with single resonant processes leading to the same
final state, but having only one top or one antitop at the
intermediate stage, see Fig.~\ref{fig2}. 
The terms $C_{V/A,1}^{\rm 1 loop}$ and  $i C_{V/A,1}^{\rm int }$ are obtained
from one-loop full theory diagrams and contribute at
NNLL order in the counting scheme of Eq.~(\ref{powercounting}) due to an
additional factor of $\alpha\sim v^2$. Thus at the
order we are working they are only accounted for in the leading order current 
${\cal O}_{A/V,\bmp,1}$. The terms $i\delta\tilde c_1(\Lambda)$ and $i\delta
c_1^{\rm int}(\Lambda)$ indicate imaginary matching contributions related to the
experimental selection cuts generically denoted by the argument $\Lambda$. As we
show in Sec.~\ref{sec:potentials} they are required to account for a phase space
matching correction of $t\bar t$ vertex subdiagrams that arise at ${\cal
  O}(\alpha_s)$. For inclusive
selection cuts such as the invariant mass prescription of
Eq.~(\ref{eq:invarmasscuts}) this term contributes at N${}^3$LL order.
To shorten the notation we frequently drop the matching scale ($\nu=1$)
dependence of the 
electroweak matching conditions displayed in Eqs.~(\ref{wilsoncoeff}). It is
implied that the electroweak couplings are evaluated at the same hard matching
scale as the QCD matching conditions. 

The electroweak matching contributions from the tree level amplitude of the 
$e^+e^-\to \gamma,Z\to t\bar t$ production process read~\cite{Hoang:2004tg} 
($\alpha\equiv\alpha_{\rm qed}(m_t)$)
\begin{align}
C_{V}^{\rm Born} & = \,  - 4\pi\alpha \left[\frac{Q_t}{4m_t^2} - \frac{v_e
    v_t}{4m_t^2 - M_Z^2}\right]\,,&
C_{A,1}^{\rm Born} & =  \,   - 4\pi\alpha \frac{a_e
  v_t}{4m_t^2-M_Z^2}\,,\nonumber\\[2mm]
C_{V,1}^{(1),\rm Born} & =  \,  \pi\alpha \left[\frac{Q_t}{m_t^2} - 
\frac{16v_e
    v_tm_t^2}{(4m_t^2 - M_Z^2)^2}\right]\,,&
C_{A,1}^{(1), \rm Born} &  = \, 16\pi\alpha \frac{a_e v_t
  m_t^2}{(4m_t^2-M_Z^2)^2}
\,,\nonumber\\[2mm]
C_{V,2}^{\rm Born} & =\, -1/6\, C_{V,1}^{\rm Born}\,,&
C_{A,2}^{\rm Born} & =\, -1/6\, C_{A,1}^{\rm Born}\,
\,,\nonumber\\[2mm]
C_{V,3}^{\rm Born} & = \, 4\pi\alpha \frac{v_e a_t}{4m_t^2 -
  M_Z^2}\,,&
C_{A,3}^{\rm Born} & =\, - 4\pi\alpha \frac{a_e a_t}{4m_t^2 -   M_Z^2}
\,,
\label{treematching}
\end{align}
where
\begin{eqnarray*}
v_f = \frac{t_3^f - 2Q_f s_w^2}{2s_w c_w}\,,\qquad a_f = \frac{t_3^f}{2s_wc_w}\,,
\end{eqnarray*}
the symbol $Q_f$ is the electric charge, and $t_3^f$ is the third component of the
weak isospin of the fermion $f$. The abbreviations $s_w$ and $c_w$ denote
the sine and cosine of the weak mixing angle, respectively.
The coefficients $C_{V/A,1}^{(1), \rm Born}$ arise from the ${\cal O}(v^2)$
terms in the expansion of the photon and Z boson propagators near threshold
using $s=4m_t^2(1+E/m_t+\ldots)$, where the dots represent terms of order $E^2$
and higher. An alternative approach is to keep the exact relativistic form for
the photon and Z propagators. This leads to the expressions  
\begin{align}
C_{V,1}^{\rm Born} & = \,  - 4\pi\alpha \left[\frac{Q_t}{s} - \frac{v_e
    v_t}{s - M_Z^2}\right]\,,&
C_{A,1}^{\rm Born} & =  \,   - 4\pi\alpha \frac{a_e
  v_t}{s-M_Z^2}\,,\nonumber\\[2mm]
C_{V,1}^{(1),\rm Born} & =  \, 0 
\,,&
C_{A,1}^{(1), \rm Born}&  = \, 0
\nonumber\\[2mm]
C_{V,2}^{\rm Born} & =\, -1/6\, C_{V,1}^{\rm Born}\,,&
C_{A,2}^{\rm Born} & =\, -1/6\, C_{A,1}^{\rm Born}\,
\,,\nonumber\\[2mm]
C_{V,3}^{\rm Born} & = \, 4\pi\alpha \frac{v_e a_t}{s -
  M_Z^2}\,,&
C_{A,3}^{\rm Born} & =\, - 4\pi\alpha \frac{a_e a_t}{s -   M_Z^2}
\,.
\label{treematchingv2}
\end{align}
For our numerical examinations we use these alternative definitions for
$C_{V/A,i}^{\rm Born}$ and $C_{V/A,i}^{(1),\rm Born}$ ($i = 1,2,3$) unless
noted otherwise.

The imaginary interference coefficients $C_{V/A,1}^{\rm int }$ are determined
from the $bW$ cuts in the one-loop electroweak corrections to the $e^+e^-\to
t\bar t$ amplitude with on-shell stable external (anti)top quarks. They also
contain the imaginary contribution of the 
(anti)top wave function renormalization Z-factor, and this term is essential to
maintain gauge-invariance. They have the form~\cite{Hoang:2004tg}
\begin{align}
i C_{V,1}^{\rm int } & = \, 
-i\,\frac{\alpha^2 \pi |V_{tb}|^2}{12 m_t^2 s_w^2 x(4c_w^2-x)(1+x)}\,
\bigg[\,
\frac{3x(1+x)}{(1-x)}\bigg(1+\frac{x-4}{4s_w^2}\bigg)\,\ln\Big(\frac{2-x}{x}\Big)
\nonumber\\[2mm] & \,
+\,Q_e Q_t (1-x)(4-x)(1+2x)(1+x+x^2)
\nonumber\\[3mm] & \,
+\, Q_e(x-1)(1+4x+2x^2+2x^3)
\, +\,  Q_t(1-x)(1+2x)(1+x+x^2)
\nonumber\\[2mm] & \,
-\,\frac{1}{2}(1+12x+9x^2+2x^3)
\,+\,\frac{1}{8s_w^2}(2+41x+28x^2-x^3+2x^4)
\,\bigg]
\,,
\label{CintV}
\end{align}
\begin{align}
i C_{A,1}^{\rm int } & = \,
i\,\frac{\alpha^2 \pi |V_{tb}|^2}{12 m_t^2 s_w^2 x(4c_w^2-x)(1+x)}\,
\bigg[\,
\frac{3x(1+x)}{(1-x)}\bigg(1+\frac{x-4}{4s_w^2}\bigg)\,\ln\Big(\frac{2-x}{x}\Big)
\nonumber\\[3mm] & \,
+\, Q_t(1-x)(1+2x)(1+x+x^2)
\nonumber\\[2mm] & \,
-\, \frac{1}{2}(1+12x+9x^2+2x^3)
\,+\,\frac{1}{8s_w^2}(2+41x+28x^2-x^3+2x^4)
\,\bigg]
\,,
\label{CintA}
\end{align}
where $x\equiv M_W^2/m_t^2$.

\vskip 5mm
\noindent
{\bf \underline {Forward scattering operators:}}\\[2mm]
The $(e^+e^-)(e^+e^-)$ forward scattering operators are required to renormalize
the phase space divergences. Their evolution accounts for the summation of $\ln
v$ terms that arise specifically from finite-lifetime effects. The matching
conditions for their Wilson coefficients account for selection prescriptions
applied on the observed final states to define the inclusive cross section.
These phase space matching conditions depend on the c.m.\ energy, and we
therefore need to define a set of operators capable to reproduce this
energy-dependence within NRQCD. The forward scattering operators are
defined as
\begin{align}
\label{eeeeop}
\tilde{\cal O}_V^{(n)} & = \,
-\big[\,\bar e_{-}\,\gamma^\mu\,e_{+}\,\big]\,
\big[\,\bar e_{+}\,\gamma_\mu\,(\En/m_t)^n\, e_{-}\,\big]
\,,
\nonumber\\[2mm]
\tilde{\cal O}_A^{(n)} & = \,
-\big[\,\bar e_{-}\,\gamma^\mu\,\gamma_5 \, e_{+}\,\big]\,
\big[\,\bar e_{+}\,\gamma_\mu\,\gamma_5\,(\En/m_t)^n\,e_{-}\,\big]
\,,
\end{align}
and thus the $\tilde{\cal O}_{V/A}^{(n)}$ 
pick up $n$ powers of the $t\bar t$ kinetic energy $E=\sqrt{s}-2m_t$ 
from the initial $e^+e^-$ state. 
The normalization of the electron and positron fields ensures that we have
\begin{eqnarray*}
\frac14 \sum \limits_{\tau,\tau^\prime}\, \Big\langle\, 0\, \Big|\,
a_{\tau}(\bmk)\, a^{c}_{\tau^\prime}(\bmkp) \, \tilde
\CO_{V/A}^{(n)} \,a^{c\dagger}_{\tau^\prime}(\bmkp) \,
a^{\dagger}_{\tau}(\bmk)\,\Big|\, 0\, \Big\rangle
= s \left(\frac{E}{m_t}\right)^n\,
\end{eqnarray*}
for the spin-averaged forward scattering amplitude.

The contribution of the forward scattering operators to the Lagrangian
reads
\begin{eqnarray}
\CL_{\rm fsc} & = & \sum\limits_n\tilde C_V^{(n)} \tilde {\cal
  O}_V^{(n)} + \tilde C_A^{(n)} \tilde {\cal O}_A^{(n)} \,,
\label{fsc}
\end{eqnarray}
where the $\tilde C_{V/A}^{(n)}(\Lambda,\nu)$ are the Wilson coefficients. They
depend on the renormalization velocity scale $\nu$, and they have a dependence
on the selection cuts, generically denoted by $\Lambda$. Frequently we will use
the shorter notation $\tilde C_{V/A}\equiv\tilde C_{V/A}^{(0)}$ for the 
coefficients of the dominant energy-independent forward scattering operators 
$\tilde{\cal O}_{V/A}$. For the examinations in this work we
consider the operators $\tilde{\cal O}_{V/A}$ and $\tilde{\cal O}_{V/A}^{(1)}$.

\vskip 5mm
\noindent
{\bf \underline {QCD factorization formula:}} \\[2mm]
For the inclusive cross section of $t\bar t$ production
close to threshold accounting for the phase space matching contributions up to
N${}^3$LL order and for the QCD and other electroweak and finite lifetime
effects at NNLL order we have the factorization 
formula~\cite{Hoang:2001mm,Hoang:2006pd,Hoang:2004tg}
\begin{eqnarray}
\label{crosssection}
\sigma_{{\rm incl}}(\Lambda) & = & \frac{1}{s}L^{lk}\, \im
\Bigg[
\Big(C_{V,1}(\Lambda,\nu)^2 + C_{A,1}(\Lambda,\nu)^2\Big) \CA_{1}^{lk} 
\nonumber\\[2mm]
&&\qquad\quad{}  +  \Big(2C_{V,1}(\Lambda,\nu)C_{V,1}^{(1)}(\nu) +
       2C_{A,1}(\Lambda,\nu)C_{A,1}^{(1)}(\nu)\Big)({E/m_t}) \CA_{1}^{lk}
\nonumber\\[2mm]
&&\qquad\quad{}  +  \Big(2C_{V,1}(\Lambda,\nu)C_{V,2}(\nu) +
       2C_{A,1}(\Lambda,\nu)C_{A,2}(\nu)\Big) \CA_{2}^{lk}
\nonumber\\[2mm]
&&\qquad\quad{}+ \Big(C_{V,3}(\nu)^2 + C_{A,3}(\nu)^2\Big) \CA_{3}^{lk}  \Bigg]
\nonumber\\
&&{}\quad+\sum_{n=0}^1 (E/m_t)^n\,\im\Big[\,\tilde C_V^{(n)}(\Lambda,\nu) 
+{\tilde
    C_A^{(n)}(\Lambda,\nu)}\,\Big]
\,.
\end{eqnarray}
The spin-averaged lepton tensor reads
\begin{eqnarray}
L^{lk} &=&
       \frac{1}{4}\,\sum\limits_{\tau,\tau^\prime}\,
       \Big[\,\bar v_{\tau^\prime}(\bmkp)\,\gamma^l\,(\gamma_5)\,u_{\tau}(\bmk)\,\Big]
       \,\Big[\,\bar
         u_{\tau}(\bmk)\,\gamma^k\,(\gamma_5)\,v_{\tau^\prime}(\bmkp)\,\Big] 
\nonumber
\\[2mm]
       &=& \frac{1}{2}\,(k+k^\prime)^2\,(\delta^{lk}-\hat e_z^l  \hat e_z^k)\,,
\end{eqnarray}
with the definitions of electron/positron momenta given in
Eqs.~(\ref{electronmomenta}). 
The quantities $\CA_i^{lk}$ are time-ordered products of the
$t\bar t$ production and annihilation currents defined in
Eq.~(\ref{currents}).
Note that the electron and positron field 
operators, from which the operators in Eqs.~(\ref{eettop}),
(\ref{eettop1}) and~(\ref{eeeeop}) are composed, only pick out the
initial and final  
$e^+e^-$ states and do not affect the correlators $\CA_i^{lk}$ in any
way. The explicit expressions for the $\CA_i^{lk}$ are
\begin{eqnarray}
{\cal A}_{1}^{lk} &=& 
i\, \sum\limits_{\bmp,\bmpp} \int\! d^4x\: 
e^{-i\hat q \cdot x}\:
\Big\langle\, 0\,\Big|\,T\,
{{\cal O}_{\bmp,1}^l}^{\!\!\!\dagger} (0)\, 
{\cal O}_{\bmpp,1}^k (x)\Big|\,0\,\Big\rangle\,,
\nonumber\\[2mm]
{\cal A}_2^{lk} &=& 
\frac{i}{2}\, \sum\limits_{\bmp,\bmpp} \int\! d^4x\: 
e^{-i\hat q \cdot x}\:
\Big\langle\, 0\,\Big|\,T\,\Big[
{{\cal O}_{\bmp,1}^l}^{\!\!\!\dagger} (0)\, 
{\cal O}_{\bmpp,2}^k (x) + 
{{\cal O}_{\bmp,2}^l}^{\!\!\!\dagger} (0)\, 
{\cal O}_{\bmpp,1}^k (x) \Big]\Big|\,0\,\Big\rangle\,,
\nonumber\\[2mm]
{\cal A}_3^{lk} &=& 
i\, \sum\limits_{\bmp,\bmpp} \int\! d^4x\: 
e^{-i\hat q \cdot x}\:
\Big\langle\, 0\,\Big|\,T\,
{{\cal O}_{\bmp,3}^l}^{\!\!\!\dagger} (0)\, 
{\cal O}_{\bmpp,3}^k (x)\Big|\,0\,\Big\rangle\,,
\end{eqnarray}
where $\hat{q}\equiv(\sqrt{s}-2m_t,0)$.
For the $v^2$-suppressed electroweak and finite lifetime matching 
coefficients $C_{V/A,1}^{\rm 1loop}$ and $iC_{V/A,1}^{\rm int}$ contained in
$C_{V/A,1}$ 
it is sufficient to use the LL current correlator $\CA_{1,\rm LL}$. 
All terms in the second, third and fourth lines of Eq.~(\ref{crosssection}) are 
$v^2$-suppressed and therefore contribute at NNLL order. This suppression
originates from  factors $E/m_t$, $\bmp^2/m_t^2$ appearing in $\CO^k_{\bmp,2}$
and two factors of $\bmsigma\cdot \bmp/m_t$ appearing in $\CO^k_{\bmp,3}$,
respectively.  The terms in the first four lines also appear in pure QCD for
stable heavy quarks see e.g.\ Refs.~\cite{Hoang:2000ib,Hoang:2001mm,Pineda:2006ri}.
The fifth line contains the phase space matching corrections related to the
$(e^+e^-)(e^+e^-)$ forward scattering operators.
As shown in Sec.~\ref{subsectionphasespacematchingNRQCD} the coefficients
$\tilde C^{(0)}_{V/A}=\tilde C_{V/A}$ start contributing at NLL, and the terms
$E/m_t \tilde C^{(1)}_{V/A}$ are N${}^3$LL corrections. 
We note that all terms shown in Eq.~(\ref{crosssection}) are understood
as finite $\msb$-renormalized quantities.

We can write ${\cal A}_i^{lk}=\delta^{lk}/3 {\cal A}_i$ after tracing the
sigma matrices of the currents in 3 dimensions. The correlators ${\cal A}_i$ can
then be expressed in terms of contributions to the  
zero-distance $S$-wave and $P$-wave Green functions of the two-body
Schr\"odinger equation:
\begin{eqnarray}
\CA_1(v,m_t,\nu) & = & 6N_c \Big[ G^c(a,v,m_t,\nu) +
  \left(\CV_2^{(s)}(\nu) + 2\CV_s^{(s)}(\nu) \right)
  G^\delta(a,v,m_t,\nu) \nonumber\\[2mm]
&&{}\qquad + \CV_r^{(s)}(\nu) G^r(a,v,m_t,\nu) +
  \CV_k^{(s)}(\nu) G^k(a,v,m_t,\nu) \nonumber\\[2mm]
&&{}\qquad + G^{\rm
    kin}(a,v,m_t,\nu) + G^{\rm dil}(a,v,m_t,\nu)\Big]\,,\nonumber\\[2mm]
\CA_3(v,m_t,\nu) &=& \frac{4N_c}{m_t^2} G^1(a,v,m_t,\nu)\,.
\label{a1correlator}
\end{eqnarray}
Here, the terms $\CV_i^{(s)}(\nu)$ ($i=2,s,r,k$) are the Wilson coefficients of
the $v^2$-suppressed potentials~\cite{Hoang:2002yy}.
The correlator  $\CA_2$ can be related to $\CA_1$
by the heavy quark equation of motion giving 
$\CA_2(v,m_t,\nu) = v^2 \CA_1(v,m_t,\nu)$.
Thus only the LL terms in $\CA_1$ are necessary to obtain
the NNLL order contributions of $\CA_2$. The function $G^c$ is the Coulomb Green
function. The LL approximation for $G^c$ is known analytically and has been
displayed in Eq.~(\ref{deltaGCoul}).  At NLL and
NNLL order, related to the ${\cal O}(\alpha_s)$~\cite{Fischler:1977yf} and  
${\cal O}(\alpha_s^2)$~\cite{Schroder:1998vy,Peter:1996ig,Peter:1997me} 
corrections to the Coulomb potential, we use the numerical results obtained in
Refs.~\cite{Hoang:2000ib,Hoang:2001mm}. They are based on an exact
solution of the corresponding Schr\"odinger equation using computational
techniques developed in Refs.~\cite{Strassler:1990nw,Jezabek:1992np}. 
For the analytic formula for $G^1$, see
Ref.~\cite{Hoang:2001mm}.
All the relativistic corrections to the Green function, 
$G^{\delta,r,k,{\rm kin}}$, are available in analytic form, see
Refs.~\cite{Hoang:2001mm,Hoang:2003ns}. They are computed from insertions of the
$v^2$-suppressed potentials and the kinetic energy corrections. The Green
function correction $G^{\rm dil}$ arises from an insertion of the lifetime
dilation correction to the bilinear quark field operators shown in
Eq.~(\ref{Lke}). The expressions for $G^{\rm dil}$ reads~\cite{Hoang:2004tg} 
\begin{eqnarray}
G^{\rm dil} 
& = &
-i\,\frac{\Gamma_t}{2m_t}\,\bigg[\,
1+\frac{v}{2}\frac{\partial}{\partial v} + a\frac{\partial}{\partial a}
\,\bigg]\,G^0(a,v,m_t,\nu)
\label{greendilfirst}
\,.
\end{eqnarray}

\vskip 5mm
\noindent
{\bf \underline {Phase space divergences and renormalization group
evolution:}}\\[2mm] 
Using the unrenormalized current correlators $\CA_i^{lk}$ in 
the factorization formula~(\ref{crosssection}) leads to the ultraviolet
$1/\epsilon$ phase space divergences either from insertions of $v^2$-suppressed 
operators or insertions of Wilson coefficient corrections describing finite
lifetime corrections. 
The divergences are absorbed by the counterterms associated to the
$(e^+e^-)(e^+e^-)$ forward scattering operators $\tilde{\cal O}_{V/A}$ 
given in Eq.~(\ref{eeeeop}) and 
are treated with the usual renormalization techniques 
known from effective theories. However, it is a novel feature that the phase
space divergences and the anomalous dimension of the operators 
$\tilde{\cal O}_{V/A}$ are purely imaginary. 
In the $\msb$ scheme  the NNLL
counterterms of the renormalized $\tilde{\cal O}_{V/A}$ 
operators have the form~\cite{Hoang:2004tg}
\begin{eqnarray*}
\delta \tilde C_{V/A} & = &
i\,\frac{N_c m_t^2}{32 \pi^2 \epsilon}\,
\bigg[ (C_{V/A,1}^{\rm Born})^2 \frac{\Gamma_t}{m_t} 
  +2 C_{V/A,1}^{\rm Born} C_{V/A,1}^{\rm int}  
\bigg]\,
{\cal V}_c^{(s)}(\nu) 
\nonumber \\[2mm] & & 
+\,i\,\frac{N_c m_t^2}{32 \pi^2 \epsilon}\,
(C_{V/A,1}^{\rm Born})^2 \frac{\Gamma_t}{m_t} \,
\Big[ \Big( 2c_2(\nu) -1\Big){\cal V}_c^{(s)}(\nu) + {\cal V}_r^{(s)}(\nu)  
\Big]
\nonumber \\[2mm] & & 
+\,i\,\frac{N_c m_t^2}{48 \pi^2 \epsilon}\,
(C_{V/A,3}^{\rm Born})^2 \frac{\Gamma_t}{m_t} \,{\cal V}_c^{(s)}(\nu)
\,.
\label{countereeee}
\end{eqnarray*}
Solving the resulting renormalization group equations for the $\tilde C_{V/A}$
one obtains 
\begin{eqnarray}
\tilde C_{V/A}(\Lambda,\nu)& = & \tilde C_{V/A}(\Lambda,1) + 
i\,\frac{2 N_c m_t^2 C_F}{3\beta_0}\,\bigg\{
\bigg[ \Big( (C_{V/A,1}^{\rm Born})^2+ (C_{V/A,3}^{\rm Born})^2 
\Big)\,\frac{\Gamma_t}{m_t}
\nonumber\\[2mm] & &\quad
 +\, 3 C_{V/A,1}^{\rm Born} C_{V/A,1}^{\rm int }\bigg]\,\ln(z)
 - \frac{4C_F}{\beta_0}\,\frac{\Gamma_t}{m_t}\,(C_{V/A,1}^{\rm
   Born})^2\ln^2(z)
\nonumber\\[2mm] & &\quad
 +\, \frac{4(C_A+2C_F)}{\beta_0}\,\frac{\Gamma_t}{m_t} \,(C_{V/A,1}^{\rm 
Born})^2\rho(z)
\bigg\}
\,,
\label{tildeCav}
\end{eqnarray}
where
\begin{eqnarray}
\rho(z) & = & \frac{\pi^2}{12}-\frac{1}{2}\ln^22 + \ln2 \ln(z) - {\rm
  Li}_2\left(\frac{z}{2}\right)
\,,
\nonumber\\[2mm]
z & \equiv & \frac{\alpha_s(m_t \nu)}{\alpha_s(m_t)}
\,.
\end{eqnarray}
Here we have introduced the $\Lambda$-dependent hard scale ($\nu =
1$) matching conditions
$\tilde C_{V/A}(\Lambda,1)$.\footnote{Frequently we drop the argument `$\nu=1$'
in the coefficients $\tilde C_{V/A}(\Lambda,1)$ to simplify the notation.}
They are determined by the phase space
matching procedure as described in the following sections and incorporate the information
on the experimental selection cuts as well as the contributions from background diagrams.
The phase space logarithms resummed in Eq.~(\ref{tildeCav}) correspond to
logarithmic terms involving ratios of the hard scales $m_t, \Lambda$ and
nonrelativistic kinematic scales. They contribute 
at order $\alpha^3\sim v^6$ in the inclusive cross section. So compared to the
LL cross section, which counts as $\alpha^2 v\sim v^5$, the phase space logs 
constitute NLL contributions. This is expected since the phase space divergences 
arise from matrix elements contributing at NNLL order. In the following sections
we determine the matching  
conditions $\tilde C_{V/A}(\Lambda,1)$, which contain the details of the
selection cuts that are applied for the definition of the cross section. 
As explained in the introduction, for inclusive cross section definitions the
information on these selection cuts represents hard effects within the NRQCD
framework.

\section{Concepts of Phase Space Matching} 
\label{sectionphasespacematching}

In this section we discuss the main concepts that go into the phase space
matching on the basis of an explicit computation of the phase space matching
conditions for a cut on the invariant masses of the reconstructed top and
antitop quarks. For simplicity we neglect in this section the effects of the
strong interactions, i.e.\ we set $\alpha_s=0$. A number of additional issues
arise if QCD effects are included. QCD corrections to the phase space matching
conditions are computed and discussed in Sec.~\ref{sectionQCD}.

\subsection{Basic Setup}
\label{sec:basic}

We define what we call the invariant masses of the top and antitop quarks
through the reconstructed masses of  
the $b W^+$ and $\bar{b}W^-$ systems coming from the top and antitop decays: 
\begin{align}
& M_t^2  =p_t^2 = (p_b+p_{W^+})^2 \,,
& M_{\bar{t}}^2 = p^2_{\bar{t}}=(p_{\bar{b}}+p_{W^-})^2 \,.
\label{eq:invariantmassdef}
\end{align}
Without QCD effects related to jet and soft particle emission this definition is
unambiguous, and we neglect combinatorial background in the following. The
latter would have to be determined from Monte-Carlo studies in addition to the
examinations carried out in this work.   
We consider selection cuts on the top and antitop invariant masses of the form
\begin{align}
(m_t -\Delta M_t) &  \le M_{t,\bar{t}} \le (m_t +\Delta M_t)\,.
\label{eq:lambdac}
\end{align}
We note that to keep Eq.~(\ref{eq:lambdac}) ambiguity-free once QCD effects are
accounted for, it is necessary to employ a short-distance top quark mass
definition for $m_t$ that is suitable for reconstruction. Such mass
definitions have a low infrared cutoff scale $R\sim \Gamma_t$ to avoid the pole
mass ${\cal O}(\Lambda_{\rm QCD})$ renormalon~\cite{Hoang:2008yj}. An example is
the jet mass
introduced in Refs.~\cite{Fleming:2007xt,Fleming:2007qr}. The $\msb$ mass does not
belong to this class of short-distance masses since it has $R\sim m_t$.

\begin{figure}[t]
  \begin{center}
  \includegraphics[width=0.4\textwidth]{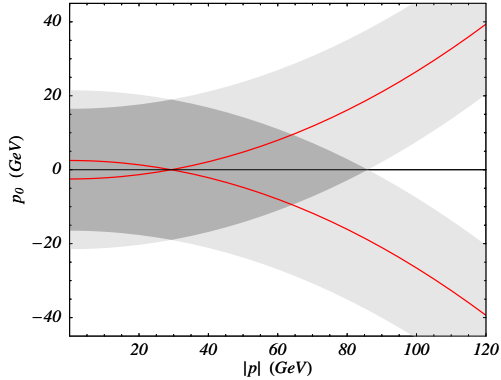}
  \caption{Phase space integration region spanned in the $p^\mu=(p_0,\bmp)$ 
variables
    for $t\bar t$ pair production. In terms of $p^\mu$ and the external energy
    $E=\sqrt{s}-2m_t$ the top and antitop momenta read $p_{t,\bar t}^\mu =
    (m_t+E/2,0)\pm p^\mu$.   
  The red solid lines correspond to the top and antitop nonrelativistic on-shell
  conditions; in the limit $\Gamma_t =  0$ 
  the phase-space shrinks to the point where the two on-shell lines intersect,
  $(p_0,|\bmp|)=(0,\sqrt{mE})$. 
  Gray, light-gray and white areas correspond to the double-resonant,
  single-resonant and non-resonant regions, respectively. 
  The gray area represents the phase space region compatible with the invariant
  mass constraints.
  We have chosen $E=5\,$GeV and 
  $\Delta M_t =20\,$GeV for this picture. 
  }
  \label{fig4}
  \end{center}
\end{figure}

The constraint on the top and antitop invariant masses $M_{t,\bar t}$
can be translated into a condition on the off-shellness $p_{t,\bar t}^2-m_t^2$
appearing in the top and antitop propagators, see Eq.~(\ref{eq:prop}). At LL
order in the nonrelativistic expansion the constraint on the nonrelativistic
off-shellness in the NRQCD propagator has the form 
\begin{align}
\label{eq:constraintNR}
-\Delta M_t & \, \le \, 
p_{t,\bar t,0}- m_t - \frac{\bmp_{t,\bar t}^2}{2m_t}  \, \le \,
\Delta M_t
\,.
\end{align}
The relativistic NNLL order corrections to these constraints are given in
the second part of Sec.~\ref{subsectionphasespacematchingNRQCD}. Through
momentum conservation Eq.~(\ref{eq:constraintNR}) leads to a constraint on the
phase space integrations for the cross section as illustrated in
Fig.~\ref{fig4}.    
From Eq.~(\ref{eq:constraintNR}) we see that the ultrasoft and the soft phase
space momentum integrations are limited by the scales $\Delta M_t$ and
$\sqrt{2m_t\Delta M_t}$, respectively.
In this work we consider moderate invariant mass cuts with $\Delta 
M_t\sim 15-35$~GeV. Thus we have
\begin{align}
& m_t v^2 \, \lsim \, \Delta M_t \sim  15-35~\mbox{GeV}\,,
& m_t v \, \lsim \, \sqrt{2 m_t \Delta M_t} \sim 70-110~\mbox{GeV} \,,
\end{align}
and the upper bounds of integration are substantially above the generic
$v$-scaling of the soft and ultrasoft momentum components. 
Since $\sqrt{2 m_t \Delta M_t}$ is parametrically of order $m_t$ we
use for our bookkeeping the counting $\sqrt{2 m_t \Delta M_t}
\sim m_t$. This also implies that $\Delta M_t \sim m_t$. Within the
vNRQCD framework this counting scheme is natural since, due to the pull-up
mechanism~\cite{Hoang:2002yy,Manohar:2006nz}, the ultrasoft scale is directly
connected to the hard matching scale via RG evolution without an additional
matching at the soft scale. 
In this counting scheme the phase space constraints are incorporated through 
the NRQCD Wilson coefficients. On the other hand, numerically   
the scales $\Delta M_t$ and $\sqrt{2m_t\Delta M_t}$ are sufficiently below the
top mass scale such that all $t\bar t$ phase space configurations that pass the
invariant mass constraint can still be adequately described by NRQCD. This fact
is crucial for the phase space matching method we describe in the following.

\begin{figure}[t]
  \begin{center}
  \includegraphics[width=0.9\textwidth]{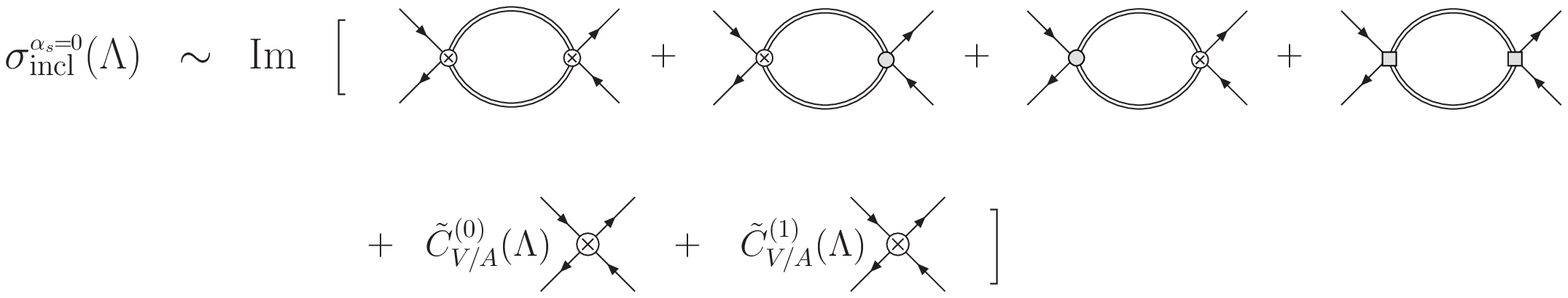}
  \caption{Graphical illustration of the factorization
    formula~(\ref{crosssection}) for $\alpha_s=0$. The $t\bar{t}$ production 
operators
    ${\cal O}_{V/A,\bmp,\sigma}^{(n)}$ for $\sigma=1,2,3$ are represented by a 
crossed circle, 
    gray circle and gray box, respectively. Diagrams in the second line 
correspond to the
    contributions from the forward scattering operators.}
  \label{fig:factheoas0}
  \end{center}
\end{figure}

\begin{figure}[t]
  \begin{center}
  \includegraphics[width=0.9\textwidth]{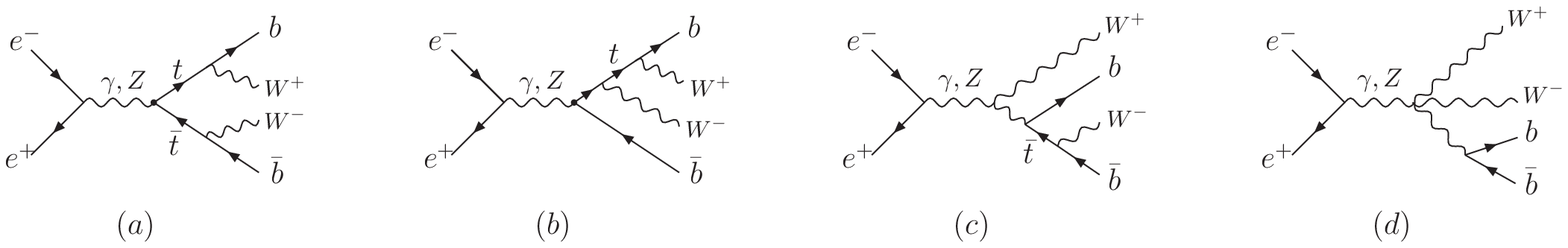}
  \caption{(a) Full theory diagram for $e^+e^-\to t\bar t \to b\bar b W^+W^-$.
   (b,c) Typical single-resonant full theory diagram for  
  $e^+e^-\to t\bar b W^- \to b\bar b W^+W^-$. 
   (d) Typical full theory diagram for $e^+e^-\to b\bar b W^+W^-$ without top or
   antitop quarks as intermediate states. 
  \label{fig:eebbWW}
}
  \end{center}
\end{figure}

When QCD effects are neglected the factorization formula in
Eq.~(\ref{crosssection}) can be illustrated graphically as in
Fig.~\ref{fig:factheoas0}. 
For $\alpha_s=0$ and in the absence of potentials the phase space matching
conditions can only contribute to the Wilson coefficients $\tilde C_{V/A}^{(n)}$
of the $(e^+e^-)(e^+e^-)$ forward scattering operators 
$\tilde{\cal O}_{V/A}^{(n)}$. This is because off-shell (anti)top phase space
contributions (corresponding to the light-gray and white areas in
Fig.~\ref{fig4}) do not match to the operator structure of the 
$(e^+e^-)(t\bar t)$ currents. From Eq.~(\ref{crosssection}) and Fig.~\ref{fig4}
we see that for the determination of the Wilson coefficients 
$\tilde C_{V/A}^{(n)}$ we need to know the result for the inclusive cross
section $\sigma_{\rm incl}^{\alpha_s=0}(\Lambda)$ with invariant mass
constraints. In the common approach to matching computations  
$\sigma_{\rm incl}^{\alpha_s=0}(\Lambda)$ is computed in the full relativistic
theory. After the result is expanded nonrelativistically using the counting of
Eq.~(\ref{powercounting}) one can identify the 
pieces belonging to the Wilson coefficients $\tilde C_{V/A}^{(n)}$. 
On the other hand, as mentioned above, the $t\bar t$ phase space regions passing
the invariant mass cuts can be determined within the nonrelativistic
expansion. We therefore write the expression for the inclusive cross section as
a sum of two terms,
\begin{align}
& \sigma_{\rm incl}^{\alpha_s=0}(\Lambda) \, = \, 
\sigma_{\rm NRQCD}^{\alpha_s=0}(\Lambda)  + 
\sigma_{\rm rem}^{\alpha_s=0}(\Lambda)
\,.
\end{align}
Here, $\sigma_{\rm NRQCD}^{\alpha_s=0}$ is the cross section computed from NRQCD
Feynman rules with the (anti)top invariant mass constraints being applied for
the phase space integration. The parameter $\Lambda$ is related to the invariant mass
cut $\Delta M_t$ and we use the formal counting $\Lambda\sim m_t$ according to
the discussion above. The exact definition of $\Lambda$ will be discussed
below. In this computation the $(e^+e^-)(e^+e^-)$ forward 
scattering operators do not contribute, and the resulting expressions are just
the nonrelativistic expansions of full theory squared matrix elements containing
the square of the double resonant diagram $e^+e^-\to t\bar t \to b\bar b W^+W^-$
(see Fig.~\ref{fig:eebbWW}a)
and the interference of the double resonant diagram with the diagrams for 
$e^+e^-\to b\bar b W^+W^-$ having only either the top or the antitop in
intermediate stages (see Fig.~\ref{fig:eebbWW}b and c for typical diagrams). 
As we show in Sec.~\ref{subsectionphasespacematchingNRQCD}, the
contributions to the Wilson coefficients $\tilde C_{V/A}^{(n)}$ that result from 
$\sigma_{\rm NRQCD}^{\alpha_s=0}(\Lambda)$ are local (i.e. energy-independent)
and only depend on powers of $\Gamma_t/m_t$ and $\Lambda/m_t$. 
While $\Gamma_t/m_t\sim v^2$, which obeys the natural NRQCD counting, the
$\Lambda/m_t$ term is of order unity and can - as we show in
Sec.~\ref{subsectionphasespacematchingNRQCD} - lead to power
counting breaking contributions for insertions of operators that are higher
order in the nonrelativistic expansion. 
However, we find that the numerical effects of the power-counting breaking
contributions are very small and do not spoil the nonrelativistic
expansion. This is partly due to the fact that the phase space cutoff $\Lambda$
is sufficiently smaller than the convergence 
radius of the nonrelativistic expansion. 
We refer to this feature as ``mild'' power-counting breaking. 

The remainder contribution of the inclusive cross section, 
$\sigma_{\rm rem}^{\alpha_s=0}(\Lambda)$ accounts for all other 
contributions to the full theory matrix element. This includes for example pure
background $e^+e^-\to b\bar b W^+W^-$ diagrams, see Fig.~\ref{fig:eebbWW}d for a
typical diagram, and also the square of the single-top diagrams in
Figs.~\ref{fig:eebbWW}b and c. In Sec.~\ref{subsectionfulltheory} we determine
the remainder contribution from a numerical analysis 
using MadEvent~\cite{Alwall:2007st}. We demonstrate that the remainder contribution is very
small and can be neglected in view of the experimental precision expected at a
future linear collider (see Sec.~\ref{sectionintroduction}).
Restricting the MadEvent amplitude to the diagrams with an intermediate 
$t\bar t$ pair we also show the excellent approximation that is provided by 
the nonrelativistic computations in $\sigma_{\rm NRQCD}^{\alpha_s=0}$.

\subsection{NRQCD Phase Space Matching}
\label{subsectionphasespacematchingNRQCD}

In this section we compute $\sigma_{\rm NRQCD}^{\alpha_s=0}$, the NRQCD cross
section for $\alpha_s=0$ with a cut 
$\Delta M_t$ on the invariant masses $M_{t,\bar t}$, see
Eq.~(\ref{eq:lambdac}). As explained above, we treat $\Delta M_t$ and $\Lambda$
as a hard scale. 

\vskip 5mm \noindent
{\bf \underline {Leading order diagram:}} \\[2mm]
We start with the leading order
NRQCD diagram to set up the notation and explain our method of
computation. Technically, the least involved method to determine the form of the
phase space integral and the proper normalization factors is to use the 
factorization theorem of Eq.~(\ref{crosssection}) and apply the cutting rules on
the (anti)top propagators in the NRQCD current-current correlators. After
identifying the top and antitop momenta according to Eq.~(\ref{eq:invariantmassdef}) 
it is then straightforward to derive the expression for the phase
space integral and the phase space boundaries compatible with the invariant 
mass constraints. 

\begin{figure}[t] %
\begin{center}
\includegraphics[width=.5\textwidth]{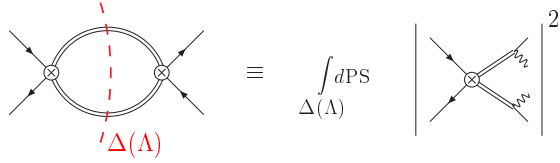}
\end{center}
\caption{Leading order diagram for the phase space matching computation.
The red dashed line
in the diagram on the LHS means that we cut through the (unstable) top lines
and integrate over the phase space region defined
by $\Delta(\Lambda)$. This is equivalent to the 
full theory cross section computation from the tree-level
matrix element for $e^+e^-\to t\bar t \to b\bar b W^+W^-$ (shown on the RHS) 
in the nonrelativistic limit. The average over 
the electron and positron spin states is implicit.  \label{LOdiagram}}
\end{figure}

We start from the form of the leading order current correlator
\begin{align}
\label{A1LO}
{\cal A}_1^{0,\alpha_s=0}(v,m_t,\nu) & \, = \, 6 N_c G^0(a=0,v,m_t,\nu) 
\nonumber\\[2mm]
& \,=\,  6 N_c\,i
\int \!\!\! \frac{d^4p}{(2\pi)^4}
 \frac{i}{\left(\frac{E}{2} + p_0 - \frac{\bmp^2}{2m_t} +
                         i\frac{\Gamma_t}{2}\right) }
 \frac{i}{\left(\frac{E}{2} - p_0 - \frac{\bmp^2}{2m_t} +
                         i\frac{\Gamma_t}{2}\right) }
\,.
\end{align}
It is easy to identify $p^\mu_{t,\bar t}=(m_t+E/2,0)\pm p^\mu$ as the top and
antitop four-momenta. Using the cutting rule for the unstable (anti)top
propagators 
\begin{align}
\frac{i}{\frac{E}{2} \pm p_0 - \frac{\bmp^2}{2m_t} + i\frac{\Gamma_t}{2}} \to
-2\,\mbox{Im}\,\left[ \frac{1}{\frac{E}{2} \pm p_0 - \frac{\bmp^2}{2m_t}+
    i\frac{\Gamma_t}{2}}\right]
\label{propcuttingrule}
\end{align}
and recalling the form of the invariant mass constraints in the nonrelativistic
limit given in Eq.~(\ref{eq:constraintNR}) we obtain
\begin{align}
\label{eq:sigmaNRQCDLOas0}
\sigma_{\rm NRQCD}^{0,\alpha_s=0}(\Lambda)  \, = \,&
N_c \, \big((C_{V,1}^{\rm Born})^2 + (C_{A,1}^{\rm Born})^2 \big)
\int \limits_{\Delta(\Lambda)} \!\!\!
                         \frac{d^4p}{(2\pi)^4}
 \frac{\Gamma_t}{\left(\frac{E}{2} + p_0 - \frac{\bmp^2}{2m_t} +
                         i\frac{\Gamma_t}{2} \right)
                 \left(\frac{E}{2} + p_0 - \frac{\bmp^2}{2m_t} -
                         i\frac{\Gamma_t}{2} \right)}\nonumber\\
             & \qquad\qquad \times
 \frac{\Gamma_t}{\left(\frac{E}{2} - p_0 - \frac{\bmp^2}{2m_t} +
                         i\frac{\Gamma_t}{2} \right)
                 \left(\frac{E}{2} - p_0 - \frac{\bmp^2}{2m_t} -
                         i\frac{\Gamma_t}{2} \right)}
\,,
\end{align}
where $\Delta(\Lambda)$ stands for the phase space constraint
\begin{align}
\label{eq:constraintNR2} 
\Big|\, \frac{E}{2} \pm p_0 - \frac{\bmp^2}{2m_t}\,\Big|  \, \le \,
\Delta M_t
\,
\end{align}
as derived before in Eq.~(\ref{eq:constraintNR}).
The form of the allowed region in the $(p_0,\bmp)$-plane 
is shown in Fig.~\ref{fig4} and a graphical illustration of the
computation in Eq.~(\ref{eq:sigmaNRQCDLOas0}) is depicted in 
Fig.~\ref{LOdiagram}. 
The expression in
Eq.~(\ref{eq:sigmaNRQCDLOas0}) agrees with the result of a full theory cross
section computation from the tree-level matrix element for   
$e^+e^-\to t\bar t \to b\bar b W^+W^-$ in the nonrelativistic limit. In this
computation one has to first carry out the $b W^+$ and $\bar b W^-$ phase space
integrals as a function of the top and antitop invariant masses, respectively.
The proper nonrelativistic limit is obtained by setting the invariant
masses of the $b W^+$ and $\bar b W^-$ systems to $m_t$ (which gives two factors
of the on-shell width $\Gamma_t$) and by taking the nonrelativistic limit of all
other remaining terms. Upon contraction of all indices in Dirac space and
resumming the width terms into the top propagators - which is required from the
counting in the double resonant kinematic region -  one
arrives at Eq.~(\ref{eq:sigmaNRQCDLOas0}).

\begin{figure}[t]
  \begin{center}
  \includegraphics[width=0.30\textwidth]{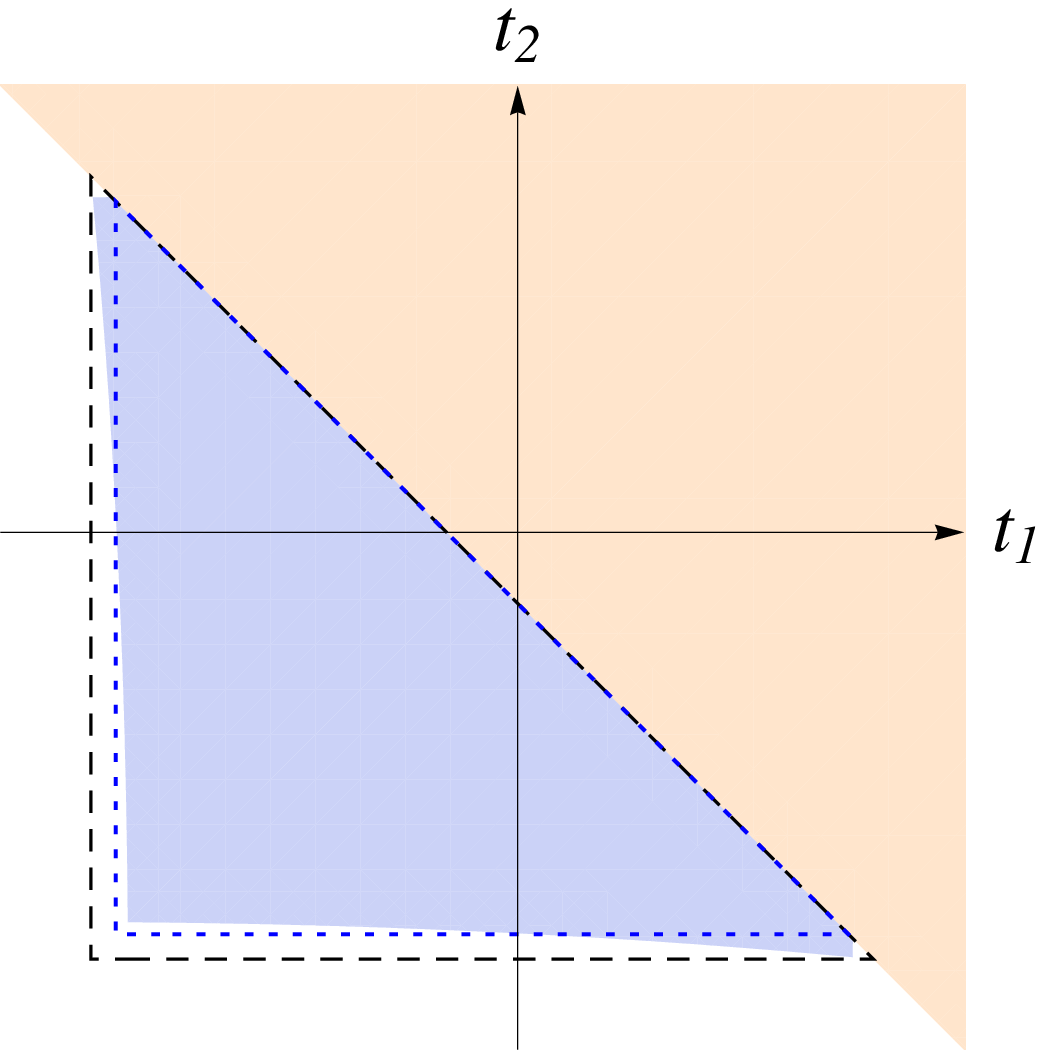}
  \hspace*{2cm}
  \includegraphics[width=0.30\textwidth]{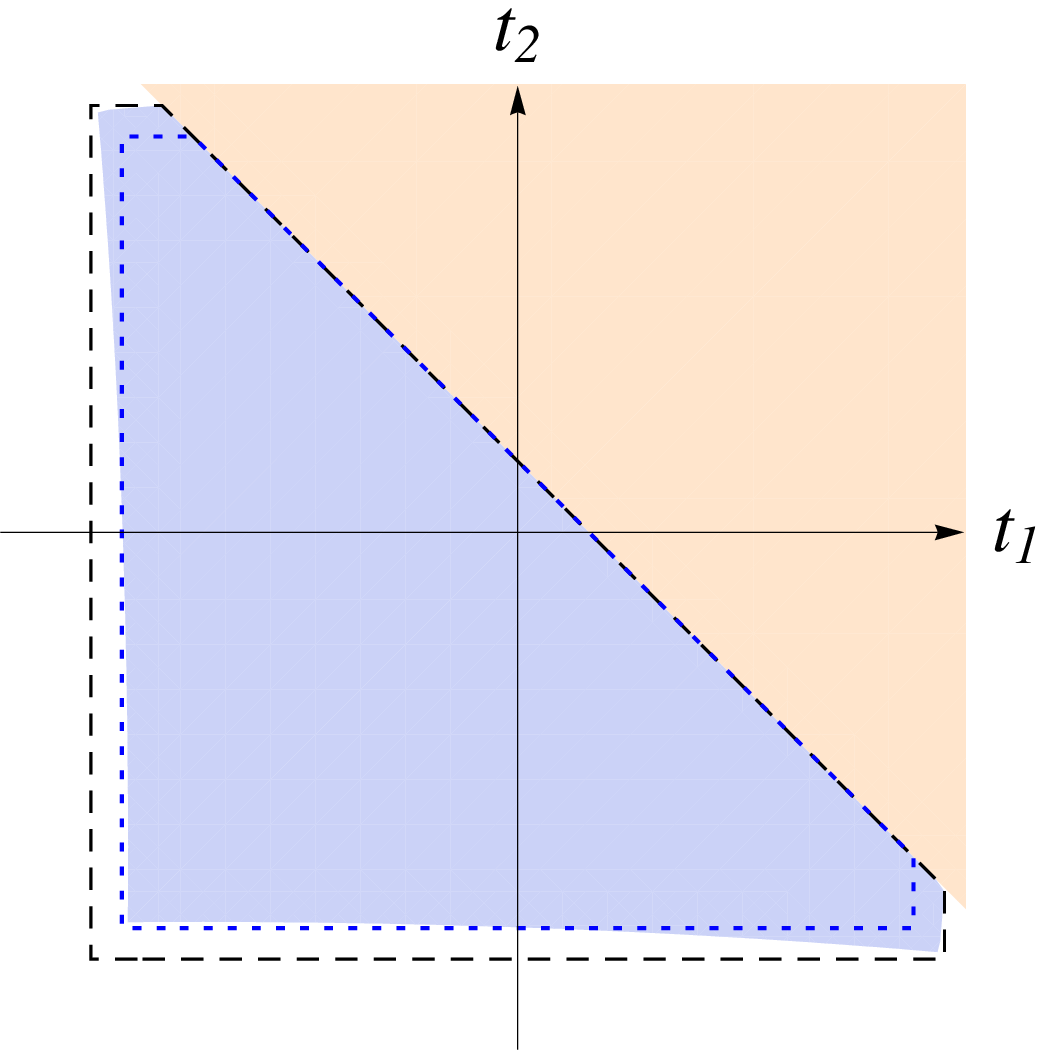}
  \caption{Allowed phase-space region in the $t_1-t_2$ plane. The left and right 
panels show the cases $E<0$ and $E>0$, respectively.
The triangular-shaped blue shaded regions corresponds to the exact relativistic
phase-space region allowed by the kinematics and the invariant mass cuts in
Eq.~(\ref{eq:lambdac}). The orange region located on the upper right of each
panel is forbidden since $|\bmp|$ must be real. 
The area inside the dashed lines corresponds to the points that pass the
conditions~(\ref{t1t2limits}). When the relativistic corrections
to the cut $\Lambda$ are included according to Eq.~(\ref{cutrelcor}), the
allowed region is the one indicated by the dotted lines. }
  \label{fig:t1t2boundaries}
  \end{center}
\end{figure}

\begin{figure}[t] %
\begin{center}
\includegraphics[width=.80\textwidth]{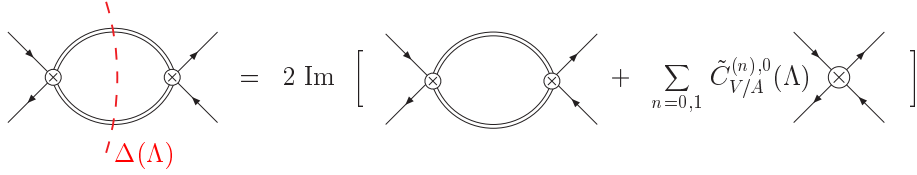}
\end{center}
\caption{
One-loop renormalization condition for $\tilde C_{V/A}^{(n),0}(\Lambda)$. 
 \label{figopeone} }
\end{figure}

A more compact representation of the $t\bar t$ phase space integral in
Eq.~(\ref{eq:sigmaNRQCDLOas0}) is obtained by switching variables to the
nonrelativistic invariant mass variables $t_1$ and $t_2$ defined by
\begin{align}
t_{1,2}=2m_t\Big( \frac{E}{2} \pm p_0 - \frac{\bmp^2}{2m_t} \Big)
\,.
\label{t12def}
\end{align}
Inverting the relations gives
\begin{align}
& p_0 \, = \, \frac{t_1-t_2}{4 m_t}\,,
\qquad
\bmp^2 \, = \, E m_t - \frac{t_1+t_2}{2}
\,.
\end{align}
Using the Jacobian $dp_0d^3\bmp = \pi/(2m_t)\times\sqrt{m_tE-\frac{1}{2}(t_1+t_2)}\,
dt_1dt_2$ the expression in 
Eq.~(\ref{eq:sigmaNRQCDLOas0}) takes the form
\begin{align}
\label{eq:sigmaNRQCDas0generic}
\sigma_{\rm NRQCD}^{i,\alpha_s=0}(\Lambda)  \, = \,&
N_c \, \big((C_{V,1}^{\rm Born})^2 + (C_{A,1}^{\rm Born})^2 \big) 
\frac{m_t^3\Gamma_t^2}{2\pi^3} \int \limits_{\tilde
  \Delta(\Lambda)} \!\! dt_1 dt_2 
\frac{\sqrt{m_t E - \frac12(t_1+t_2)}}{(t_1^2
  + m_t^2\Gamma_t^2)(t_2^2 + m_t^2 \Gamma_t^2)}\, \Delta^i(t_1,t_2)\,
\,,
\end{align}
with 
\begin{align}
\Delta^0(t_1,t_2) \, = \, 1 \quad 
\mbox{for}\quad
\sigma_{\rm NRQCD}^{0,\alpha_s=0}
\,.
\end{align}
The integration region in $(t_1,t_2)$-space has the form
\begin{align}
\label{t1t2limits}
\tilde \Delta (\Lambda) = \Big\{(t_1,t_2) \in \mathbb{R}^2:
\Big(|t_{1,2}| < \Lambda^2 \Big) \wedge 
\Big(0 < m_t E - \frac12(t_1+t_2) \Big) \Big\}\,,
\end{align}
where $\Lambda^2\equiv 2m_t\Delta M_t$. In Fig.~\ref{fig:t1t2boundaries} the
allowed region in the $t_1$-$t_2$ plane is illustrated by the area within the
dotted lines for $E<0$ (left panel) and $E>0$ (right panel). 
The second condition in Eq.~(\ref{t1t2limits}) for $t_{1,2}$ is from kinematics
to ensure real values of the momentum 
$\bmag p= \sqrt{m_t E -(t_1+t_2)/2}$. Beyond the nonrelativistic approximation  
the relation between $\Lambda^2$ and $\Delta M_t$ receives additional
relativistic corrections which are discussed below. 
We have to evaluate Eq.~(\ref{eq:sigmaNRQCDas0generic}) in a nonrelativistic
expansion for
\begin{align}
m_t E\,,m_t \Gamma_t\, \sim \,m_t^2 v^2 \ll \Lambda^2 \,.
\end{align}
To this end, we carry out an asymptotic expansion based on the four regions 
$(t_1, t_2)\sim (m_t^2 v^2,m_t^2 v^2)$ (double resonant), 
$(t_1, t_2)\sim (m_t^2 v^2,\Lambda^2)$ and $(t_1, t_2)\sim (\Lambda^2, 
m_t^2v^2)$
(single resonant), and
$(t_1, t_2)\sim (\Lambda^2,\Lambda^2)$ (hard). We obtain
\begin{align}
\sigma_{\rm NRQCD}^{0,\alpha_s=0}(\Lambda)
    & \, = \, 2 N_c \, \big((C_{V,1}^{\rm Born})^2 + (C_{A,1}^{\rm Born})^2 
\big) \,\frac{m_t^2}{4\pi}\bigg(
    \mbox{Im}\,(iv)  -
    \frac{2\sqrt{2}}{\pi}\,\frac{\Gamma_t}{\Lambda}\nonumber
+ \frac{4+2\sqrt{2}\arsinh
    (1)}{3\,\pi^2}\,\frac{m_t\Gamma_t^2}{\Lambda^3}\nonumber\\
&\qquad\quad{}-\frac{2\sqrt{2}}{3\,\pi}\, 
    \frac{m_t E\, \Gamma_t}{\Lambda^3}+
    \Ord{v^6\,\frac{m_t^5}{\Lambda^5}}\bigg)
\,,
\label{LOresult}
\end{align}
where $v = \sqrt{(E+i\Gamma_t)/m_t}$.
The first term in the parenthesis is the well known NRQCD Born cross section
obtained from the unrestricted phase space integration. It constitutes the
leading order cross section and is 
${\cal O}(v)$, see Eq.~(\ref{deltaGCoul}) for $\alpha_s=0$. The second term in
Eq.~(\ref{LOresult}) proportional to $\Gamma_t/\Lambda$ is the 
dominant phase space correction and of order $v^2$, i.e.\ it
contributes at NLL order. We also see terms proportional to
$m_t\Gamma_t^2/\Lambda^3$ and $m_t E\Gamma_t/\Lambda^3$ which contribute at
N${}^3$LL order.
These NLL and N${}^3$LL corrections are subtracting the phase space
contributions that do not pass the invariant mass constraints, represented
by the white regions in Figs.~\ref{fig:t1t2boundaries}. One may wonder why these
corrections are polynomial in $E$ and $\Gamma_t$ and not a non-trivial function
of $E/\Gamma_t$ given that this region also has
single resonant regions where either $t_1$ or $t_2$ are resonant, i.e.\ of order
$m_t^2 v^2$. This can be understood from the form of the integrand in
Eq.~(\ref{eq:sigmaNRQCDas0generic}) which shows that it is impossible to
generate a nontrivial $E/\Gamma_t$ dependence if $t_1\gg t_2\sim m_t^2 v^2$ or
$t_2\gg t_1\sim m_t^2 v^2$. 
Thus the contributions that arise from the invariant mass constraints, and in
fact all selection criteria that do not cut into the double resonant phase space
region $(t_1, t_2)\sim (m_t^2 v^2,m_t^2 v^2)$ represent hard contributions that
can be matched onto the local $(e^+e^-)(e^+e^-)$ forward scattering operators 
$\tilde{\cal O}_{V/A}^{(n)}$. Accounting for the form of Eq.~(\ref{crosssection})
it is then straightforward to identify the contributions
to the Wilson coefficients $\tilde C_{V/A}(\Lambda,1)$ and 
$\tilde C_{V/A}^{(1)}(\Lambda,1)$. A graphical illustration of the procedure
is shown in Fig.~\ref{figopeone}. We obtain the results
\begin{align}
& {\tilde C_{V/A}^{0}}(\Lambda,1) \, = \,
 2\,i\,N_c\,\big(C_{V/A,1}^{\rm Born}\big)^2\,
\frac{m_t^2}{4\pi}\,
\Big(-\frac{2\sqrt{2}}{\pi}\,\frac{\Gamma_t}{\Lambda}
+\frac{4+2\sqrt{2}\arsinh
    (1)}{3\,\pi^2}\,\frac{m_t\Gamma_t^2}{\Lambda^3}\Big)
\,,
\label{tildeCVA0}
\end{align}
and
\begin{align}
& {\tilde C_{V/A}^{(1),0}}(\Lambda,1) \, = \,
 -\,2\,i\,N_c\,\big(C_{V/A,1}^{\rm Born}\big)^2\,
\frac{m_t^2}{4\pi}\,\frac{2\sqrt{2}}{3\,\pi}\, 
    \frac{m_t^2\, \Gamma_t}{\Lambda^3}
\,.
\label{tildeCVA01}
\end{align}
The coefficients ${\tilde C_{V/A}^{0}}(\Lambda,1)$ contribute at NLL order and 
the ${\tilde C_{V/A}^{(1),0}}(\Lambda,1)$ at N${}^3$LL order. The
superscripts~``0'' are a reminder that QCD corrections are neglected.

\vskip 5mm \noindent
{\bf \underline {Relativistic corrections:}} \\[2mm]
We now determine the phase space matching contributions from
${\cal O}(v^2)$ relativistic corrections to the NRQCD cross section. Since they come
with one additional factor 
of $\bmp^2/m_t^2$, they are suppressed by $v^2$ according to the
counting $\bmp \sim m_t v$. However, since the imposed invariant mass
constraints act like a hard momentum-cutoff for the NRQCD phase space
integration, power-like dependences involving the cut parameter $\Lambda$ 
appear as a consequence of employing the nonrelativistic expansion.  
This occurs whenever the nonrelativistic expansion leads to integrals that are
divergent for $|\bmp|\to\infty$. In effective field theory computations where a
hard cutoff regularization is used, this means that once power-counting breaking
occurs from the insertion of a higher dimensional operator ${\cal O}$, Wilson
coefficients contributing to a lower order than the operator ${\cal O}$ might
have to be modified during the matching procedure to the full theory. This
happens e.g.\ in lattice computations and, while it is admittedly not
esthetically elegant, it only represents a technical subtlety from the field
theoretic point of view.   
For the method we use for the phase space matching, however, the occurrence of
power-counting breaking terms that are numerically large could be disastrous
since our method relies on the assumption that the $t\bar t$ phase space passing
the selection cuts can be computed reliably in the nonrelativistic
expansion. This restricts the size of the invariant mass cut $\Delta M_t$ we 
can handle within our method for the phase space matching procedure.  We show in
Sec.~\ref{subsectionfulltheory} that for the invariant mass cuts $\Delta
M_t$ between $15$ and $35$~GeV which we consider in 
this work power-counting breaking effects are very small and do not cause any
problems. 

There is an important relativistic effect related to the invariant mass
constraint given in Eq.~(\ref{eq:constraintNR2}). These corrections modify the
bounds on the $t_{1,2}$ integrations 
in Eq.~(\ref{t1t2limits}). To determine these
corrections one needs to solve the equations
\begin{align}
\Big(\frac{E}{2}+m_t\pm p_0\Big)^2-\bmp^2-\Big(m_t\mp\Delta M_t\Big)^2 & = \, 0
\,,\nonumber \\
\Big(\frac{E}{2}+m_t\pm p_0\Big)^2-\bmp^2-\Big(m_t\pm\Delta M_t\Big)^2 & = \, 0
\,,
\label{boundsrelcor}
\end{align}
where $E=\sqrt{s}-2m_t$. The condition $m_t E-\frac{1}{2}(t_1+t_2)>0$ remains
unchanged. The allowed phase space region in the $t_1-t_2$ plane accounting for
all relativistic effects is illustrated by the blue shaded regions in 
Fig.~\ref{fig:t1t2boundaries}. Since an exact analytic integration within 
the relativistic phase space boundary is not required  for the precision
aimed for in this work, we derive the following approximation.
The main effect of accounting for the exact boundaries 
is a change in the lower boundaries of the $t_{1,2}$ integrations. The 
approximation can be made because the single resonant
regions where either $t_2$ or $t_1$ are zero ({\it i.e.} the regions close to
the negative $x$- or $y$-axes inside the blue shaded areas in
Fig.~\ref{fig:t1t2boundaries}) give the largest contributions. For the
determination of the new lower boundary for the 
$t_{1,2}$ integration one can take, for example, the first of the relations in
Eqs.~(\ref{boundsrelcor}). Switching to the $t$-variables and setting for
example $t_2=0$, it is straightforward to find the relation
\begin{align}
|t_{1,2}| \, < \,
\Lambda^2 \, = \, 2m_t \Delta M_t 
- \frac{3}{4}(\Delta M_t)^2
- \frac{1}{2} E\Delta M_t
+ \, \ldots
\label{cutrelcor}
\end{align}
The associated region in the
$t_1$-$t_2$ plane is indicated by the dotted lines in Fig.~\ref{fig:t1t2boundaries}. 
The omitted terms in Eq.~(\ref{cutrelcor}) indicate: (a) terms involving the
energy $E$, which contribute beyond N$^{3}$LL order to the cross section and thus can be
safely dropped for the range of energies $E$ considered
in our analysis, and (b) terms which
give power-counting breaking contributions similar to those arising from the  N${}^4$LL order kinetic energy corrections,
that are shown below to be numerically negligible.
Apart from the neglected terms in Eq.~(\ref{cutrelcor}), 
the difference between the exact relativistic 
phase space region (with curved boundaries) and the approximation
given by the dotted lines in Fig.~\ref{fig:t1t2boundaries}, can also be shown to yield terms beyond N${}^3$LL, or terms of type (b).

The other contributions from ${\cal O}(v^2)$ relativistic corrections to
$\sigma_{\rm NRQCD}^{\alpha_s=0}(\Lambda)$ can be
cast into the form of Eq.~(\ref{eq:sigmaNRQCDas0generic}), where
the functions $\Delta^i(t_1,t_2)$ read
\begin{align}
\Delta^{\rm kin,0} &\, = \, -
\frac{\bmp^4}{2m_t^2}\left(\frac{t_1}{t_1^2+m_t^2\Gamma_t^2} +
\frac{t_2}{t_2^2 + m_t^2\Gamma_t^2}\right)\,,
\nonumber\\
\Delta^{\rm dil,0} &\, = \,
\frac{\bmp^2}{m_t^2}\, \frac{m_t^4\Gamma_t^4 -
  t_1^2t_2^2}{(t_1^2+m_t^2\Gamma_t^2) (t_2^2 + m_t^2\Gamma_t^2)}\,,
\nonumber\\
\Delta^{v^2,0} &\, = \, \frac{2C_{V,1}^{\rm Born}C_{V,2}^{\rm 
Born}+2C_{A,1}^{\rm
    Born}C_{A,2}^{\rm Born}}{(C_{V,1}^{\rm Born})^2 + (C_{A,1}^{\rm
    Born})^2} \,
\frac{\bmp^2}{m_t^2} \,,
\nonumber\\
\Delta^{\text{$P$-wave,0}}  &\, = \frac{2}{3}\,\frac{(C_{V,3}^{\rm
    Born})^2 + (C_{A,3}^{\rm  Born})^2}{(C_{V,1}^{\rm 
    Born})^2 + (C_{A,1}^{\rm Born})^2}\, 
\frac{\bmp^2}{m_t^2} \,,
\nonumber\\
\Delta^{\rm int,0} &\, = \,
- \frac{C_{V,1}^{\rm Born}C_{V,1}^{\rm int}+C_{A,1}^{\rm
    Born}C_{A,1}^{\rm int}}{(C_{V,1}^{\rm Born})^2 + (C_{A,1}^{\rm
    Born})^2} \, \frac{t_1+t_2}{m_t\Gamma_t}
\,.
\label{Deltav2def}
\end{align}
Here $\Delta^{\rm kin,0}$ comes from the insertion of the  
kinetic energy correction $\bmp^4/(8m_t^3)$ and $ \Delta^{\rm dil,0}$ from the
time dilation correction $-i\,\Gamma_t\,\bmp^2/(4m_t^2)$ contained in the quark
bilinear Lagrangian Eq.~(\ref{Lke}). The term
$\Delta^{v^2,0}$ comes from the insertions of the $v^2$-suppressed $S$-wave
current ${\cal O}_{\bmp,2}$ and $\Delta^{\text{$P$-wave,0}}$ from the $P$-wave
current ${\cal O}_{\bmp,3}$, see Eqs.~(\ref{currents}). Finally, the function 
$\Delta^{\rm int,0}$ arises from interference contributions of the double
resonant amplitudes $e^+e^-\to t\bar t\to b\bar b W^+W^-$ with those where only
either the top or the antitop appear at intermediate stages, see
Eqs.~(\ref{CintV}) and (\ref{CintA}). Using the methods of the previous
subsection to compute the corrections to the cross section with invariant mass
cuts we obtain the following expressions for the contributions to the Wilson
coefficients $\tilde C_{V/A}(\Lambda,1)$ and 
$\tilde C_{V/A}^{(1)}(\Lambda,1)$:
\begin{align}
{\tilde C_{V/A}^{\rm kin,0}}(\Lambda,1) \, & = \, 
2\,i\,N_c\,\big(C_{V/A,1}^{\rm Born}\big)^2\,
    \frac{5m_t^2}{32\pi}\,
    \Big(\frac{9}{5\sqrt{2}\,\pi}\,\frac{\Gamma_t\Lambda}{m_t^2} +
    \frac{-86+105\sqrt{2}\arsinh 
    (1)}{30\pi^2}\,\frac{\Gamma_t^2}{m_t\Lambda}\Big)
\,,
\nonumber \\[2mm]
{\tilde C_{V/A}^{\rm dil,0}}(\Lambda,1) \, & = \, 
2\,i\,N_c\,\big(C_{V/A,1}^{\rm Born}\big)^2\,
    \frac{3m_t^2}{16\pi}\,
    \Big(-\frac{2\sqrt{2}}{3\,\pi}\,\frac{\Gamma_t\Lambda}{m_t^2} +
    \frac{4\,(2-3\sqrt{2}\arsinh(1))}{3\,\pi^2} 
    \,\frac{\Gamma_t^2}{m_t\Lambda}\Big)
\,,
\nonumber\\[2mm]
{\tilde C_{V/A}^{v^2,0}}(\Lambda,1) \, & = \, 
2\,i\,N_c\,2\,C_{V/A,1}^{\rm Born}C_{V/A,2}^{\rm Born}\,
\frac{m_t^2}{4\pi}
\Big(\frac{\sqrt{2}}{\pi}\,\frac{\Gamma_t\Lambda}{m_t^2} +
\frac{-2+3\sqrt{2}\arsinh(1)}{\pi^2}
\,\frac{\Gamma_t^2}{m_t\Lambda}\Big)
\,,
\nonumber\\[2mm]
{\tilde C_{V/A}^{\text{$P$-wave},0}}(\Lambda,1) \, & = \, 
 \frac{4}{3}\,i\,N_c\,\left(C_{V/A,3}^{\rm Born}\right)^2\,
\frac{m_t^2}{4\pi}\,
\Big(\frac{\sqrt{2}}{\pi}\,\frac{\Gamma_t\Lambda}{m_t^2} +
\frac{-2+3\sqrt{2}\arsinh(1)}{\pi^2}
\,\frac{\Gamma_t^2}{m_t\Lambda}\Big)
\,,
\nonumber\\[2mm]
{\tilde C_{V/A}^{\rm int,0}}(\Lambda,1) \, & = \, 
2\,i\,N_c\,2\,C_{V/A,1}^{\rm Born}C_{V/A,1}^{\rm int}
\Big(\frac{m_t\Lambda}{2\sqrt{2}\,\pi^2}
+\frac{-2+3\sqrt{2}\arsinh(1)}{4\pi^3}\,\frac{m_t^2
  \Gamma_t}{\Lambda}\Big)   
\,,
\label{relativisticresults1}
\end{align}
and
\begin{align}
{\tilde C_{V/A}^{(1),\rm kin,0}}(\Lambda,1) \, & = \, 
 -\,2\,i\,N_c\,\big(C_{V/A,1}^{\rm Born}\big)^2\,
    \frac{5m_t^2}{32\pi}\,
    \frac{7}{\sqrt{2}\,\pi}\,
    \frac{\Gamma_t}{\Lambda}
\,,
\nonumber\\[2mm]
{\tilde C_{V/A}^{(1),\rm dil,0}}(\Lambda,1) \, & = \, 
2\,i\,N_c\,\big(C_{V/A,1}^{\rm Born}\big)^2\,
    \frac{3m_t^2}{16\pi}\,
    \frac{2\sqrt{2}}{\pi}\,\frac{\Gamma_t}{\Lambda}
\,,
\nonumber\\[2mm]
{\tilde C_{V/A}^{(1),v^2,0}}(\Lambda,1) \, & = \, 
-2\,i\,N_c\,2\,C_{V/A,1}^{\rm Born}C_{V/A,2}^{\rm Born}\,
\frac{m_t^2}{4\pi} \frac{3\sqrt{2}}{\pi}\,\frac{\Gamma_t}{\Lambda}
\,,
\nonumber\\[2mm]
{\tilde C_{V/A}^{(1),\text{$P$-wave},0}}(\Lambda,1) \, & = \, 
- \frac{4}{3}\,i\,N_c\,\left(C_{V/A,3}^{\rm Born}\right)^2\,
\frac{m_t^2}{4\pi}\,
\frac{3\sqrt{2}}{\pi}\, 
    \frac{\Gamma_t}{\Lambda}
\,,
\nonumber\\[2mm]
{\tilde C_{V/A}^{(1),\rm int,0}}(\Lambda,1) \, & = \, 
-2\,i\,N_c\,2\,C_{V/A,1}^{\rm Born}C_{V/A,1}^{\rm int}
\frac{1}{2\sqrt{2}\pi^2}\,\frac{m_t^3}{\Lambda}
\,,
\label{relativisticresults2}
\end{align}
where $C_{V/A,2}^{\rm Born} = -1/6\, C_{V/A,1}^{\rm Born}$.
As anticipated for the results contributing to $\tilde C_{V/A}(\Lambda,1)$,
apart from the N${}^3$LL terms $\propto \Gamma_t^2/m_t^2$, 
there are in Eq.~(\ref{relativisticresults1}) terms $\propto\Gamma_t\Lambda/m_t$
that are power-counting breaking 
since they contribute at NLL order. 
These contributions feature a relative factor $\Lambda^2/m_t^2=2\Delta
M_t/m_t=0.15-0.4$ with 
respect to the NLL order terms in $\tilde C_{V/A}^{0}$ of
Eq.\,(\ref{tildeCVA0}) that come from the cross section in the nonrelativistic
limit. For the interference contributions this statement also applies because
$C_{V/A,1}^{\rm int} \sim C_{V/A,1}^{\rm Born} \Gamma_t/m_t$.  
The small size of this factor is one reason why these
power-counting contributions do not spoil the quality of the nonrelativistic
expansion.

\begin{figure}[t]
  \begin{center}
  \includegraphics[width=0.49\textwidth]{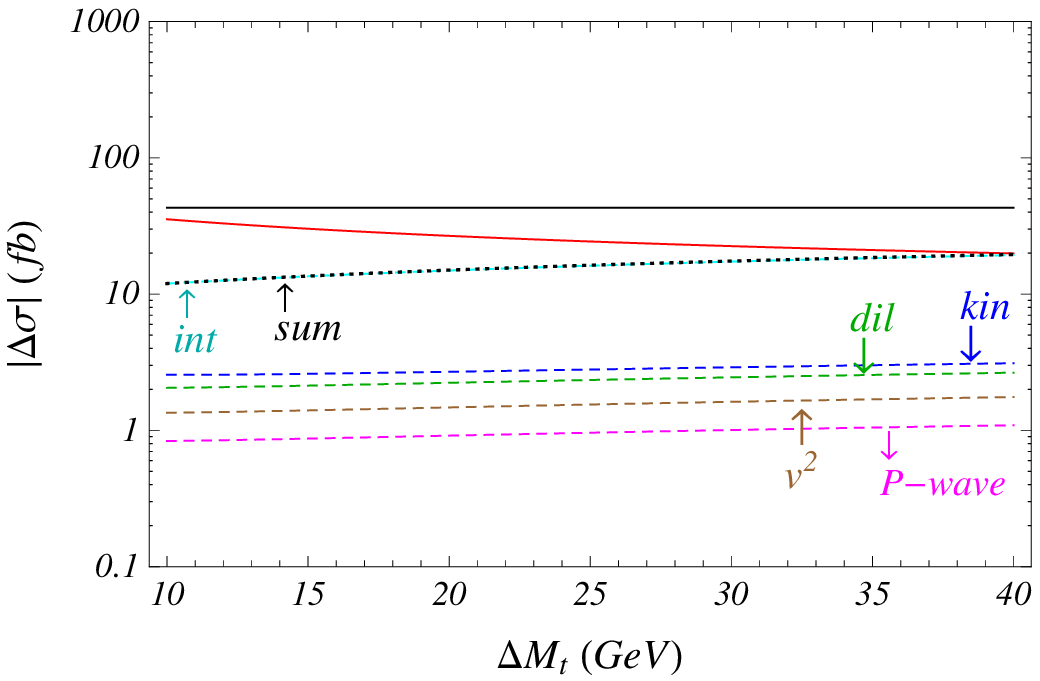}\;\;
  \includegraphics[width=0.49\textwidth]{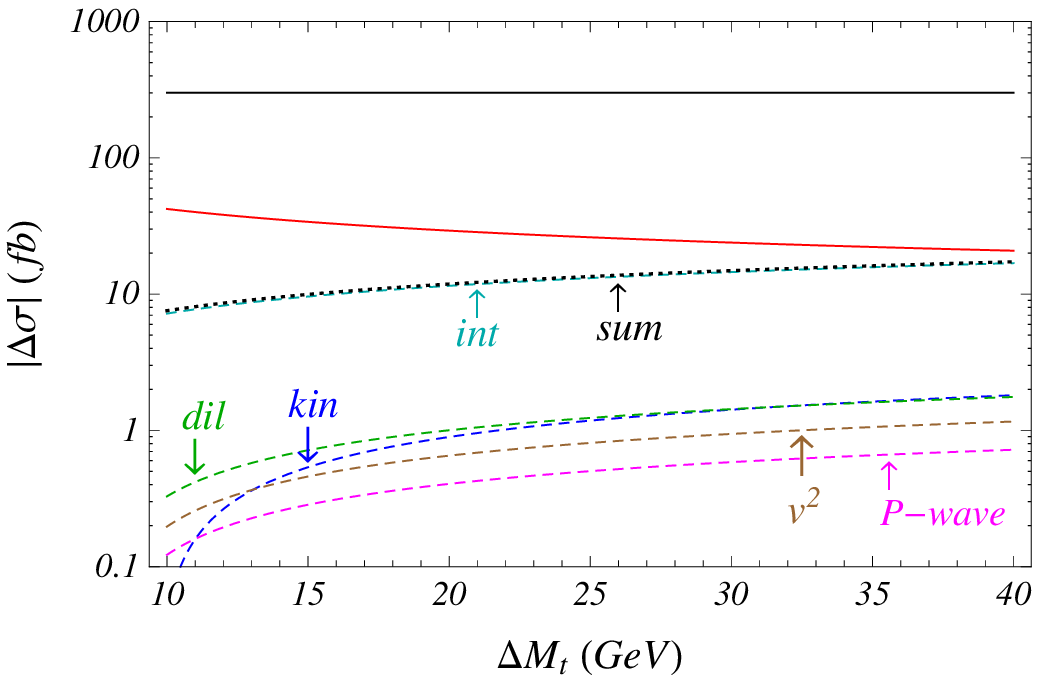}
  \caption{Absolute values of the contributions to the inclusive cross section
from the leading order diagram phase space
matching coefficients,
${\tilde C_{V/A}^{0}}(\Lambda,1) + E/m_t {\tilde
  C_{V/A}^{(1),0}}(\Lambda,1)$, found in Eqs.~(\ref{tildeCVA0}),
and~(\ref{tildeCVA01})
(red solid line), and those from the different NNLL relativistic corrections, 
${\tilde C_{V/A}^{i,0}}(\Lambda,1) + E/m_t {\tilde
  C_{V/A}^{(1),i,0}}(\Lambda,1)$ ($i=\mbox{kin},
\mbox{dil},v^2,\mbox{$P$-wave}, \mbox{int}$) given in
Eqs.~(\ref{relativisticresults1}) and (\ref{relativisticresults2})
(blue, green, brown, magenta and cyan dashed lines,
respectively). For comparison we also display the corresponding contributions
from the tree-level NRQCD cross
section without phase space matching contributions, 
$\sigma^{0,\alpha_s=0}_{\rm NRQCD}(\Lambda=\infty) =
2N_c\,((C_{V,1}^{\rm Born})^2+(C_{A,1}^{\rm
  Born})^2)\,m_t^2/(4\pi)\,\mbox{Im}(iv)$ (black horizontal lines).
The left panel corresponds to $E=-5$ while for the right we have
$E=5$~GeV. The values chosen for the input parameters can be found in
Eq.~(\ref{parameters}). We have used the energy-independent form  of the coefficients 
$C_{V/A,i}^{\rm Born}$ and $C_{V/A,i}^{(1),\rm Born}$ ($i=1,2,3$) given 
by Eqs.~(\ref{treematching}). 
}
  \label{fig:relatcor}
  \end{center}
\end{figure}

For visualization we show in Fig.~\ref{fig:relatcor}, as a function of
$\Delta M_t$, the
numerical contributions to the inclusive cross section from the NLL phase space
matching coefficients ${\tilde C_{V/A}^{0}}(\Lambda,1) + E/m_t {\tilde
  C_{V/A}^{(1),0}}(\Lambda,1)$
 in Eqs.~(\ref{tildeCVA0}) and Eqs.~(\ref{tildeCVA01})
(red solid line), and those from the different NNLL relativistic corrections, 
${\tilde C_{V/A}^{i,0}}(\Lambda,1)+E/m_t {\tilde
  C_{V/A}^{(1),i,0}}(\Lambda,1)$ ($i=\mbox{kin},
\mbox{dil},v^2,\mbox{$P$-wave}, \mbox{int}$) given in
Eqs.~(\ref{relativisticresults1}) and (\ref{relativisticresults2})
(blue, green, brown, magenta and cyan dashed lines,
respectively). The difference of dashed lines between left and
right panels illustrates the size of N${}^3$LL effects in the $E/m_t {\tilde
  C_{V/A}^{(1),i,0}}(\Lambda,1)$ terms.
Note that the various contributions have different signs and that
in Fig.~\ref{fig:relatcor} only their absolute value is displayed. For
comparison, the solid black line shows the NRQCD cross section without phase
space cuts. The sum of all the contributions in 
Eqs.~(\ref{relativisticresults1}) and (\ref{relativisticresults2}) is displayed as the dotted black line,
which almost overlaps with the interference contributions.
We see
that except for the interference contributions all phase space matching
corrections in Eqs.~(\ref{relativisticresults1}) are an order of magnitude
smaller than those obtained from the leading order NRQCD cross section in 
Eq.~(\ref{tildeCVA0}). This 
shows that the power-counting breaking terms $\propto \Gamma_t\Lambda/m_t^2$ are
small and do not spoil the nonrelativistic expansion. 
Interestingly, in the sum these phase space matching corrections also
cancel to a large extent due to their different signs. In Fig.~\ref{fig:relatcor}
we also see that the contributions from the interference coefficients, 
${\tilde C_{V/A}^{{\rm int},0}}(\Lambda,1)$, are the by far largest terms that
come from NNLL order relativistic insertions. For $\Delta M_t>35$~GeV they are
comparable to the contributions of ${\tilde C_{V/A}^{0}}(\Lambda,1)$.
Since we have numerically that 
$2 C_{V,1}^{\rm Born} C_{V,1}^{\rm int} + 2 C_{A,1}^{\rm Born}
C_{A,1}^{\rm int} = -4.7\, ((C_{V,1}^{\rm Born})^2 + (C_{A,1}^{\rm Born})^2)
\,\Gamma_t/m_t$, we see, however, that the large size of the interference terms
comes from the size of the coefficients $C^{\rm int}_{V/A,1}$ and is not related
to power-counting breaking effects. As far as the size of higher order
relativistic corrections are concerned we thus also expect a good perturbative 
behavior for the interference effects.

\begin{figure}[t]
  \begin{center}
  \includegraphics[width=0.49\textwidth]{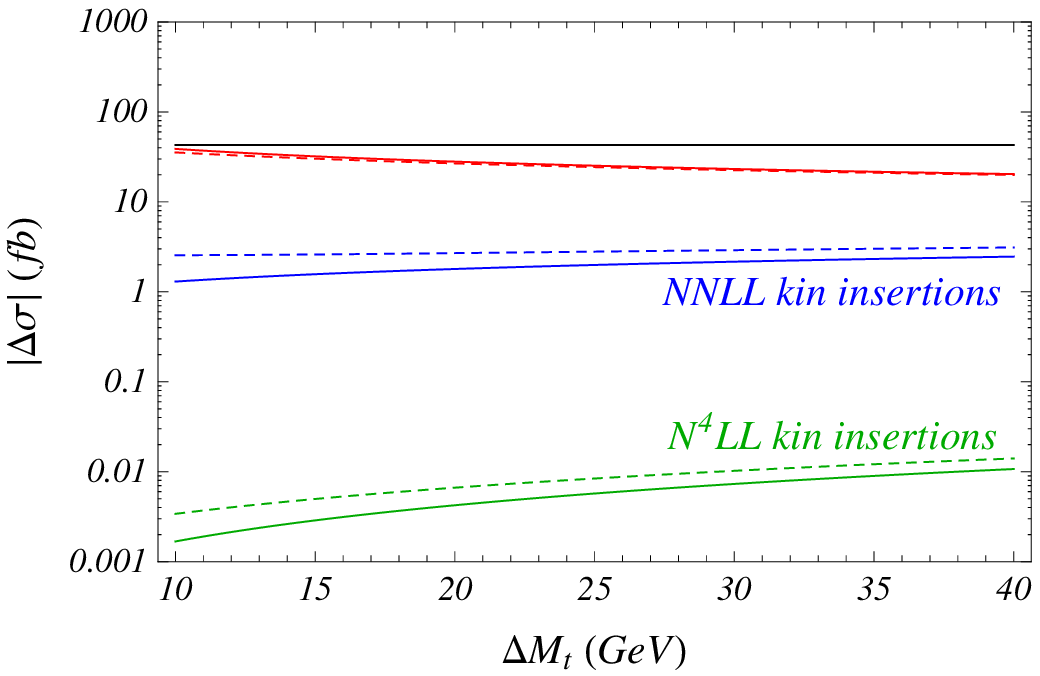}\;\;
  \includegraphics[width=0.49\textwidth]{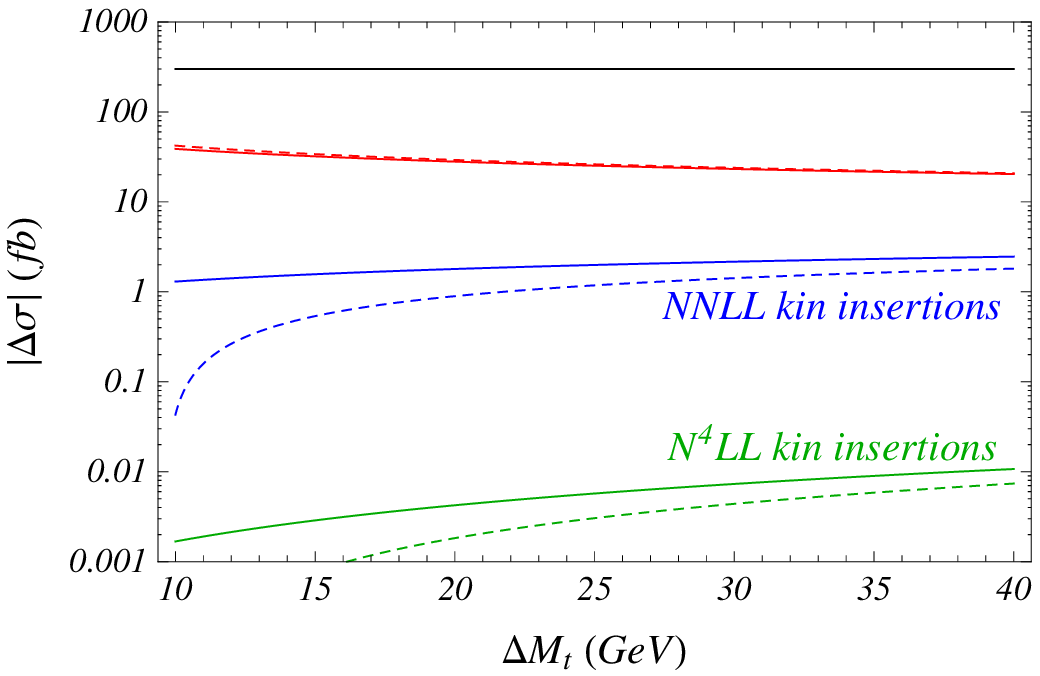}
  \caption{Absolute values of the phase space matching contributions to the
    inclusive cross section from the NNLL (from
    Eqs.~(\ref{relativisticresults1}, \ref{relativisticresults2}), blue lines),
    and N${}^4$LL insertions of the kinetic energy operators
    (Eq.~(\ref{tildeCVAkinhigh1}, \ref{tildeCVAkinhigh2}), green lines), as a
    function of $\Delta M_t$ for $E=-5$ (left panel) and $5$~GeV (right
    panel). The solid lines show the contributions from $\tilde
    C_{V/A}(\Lambda,1)$ and the dashed lines show the contributions for $\tilde
    C_{V/A}(\Lambda,1) + E/m_t \tilde C_{V/A}^{(1)}(\Lambda,1)$. For comparison
    we also display the corresponding contributions from the tree-level NRQCD
    cross section without phase space matching contributions (black horizontal
    lines), and 
    the dominant phase space matching corrections from
    Eq.~(\ref{tildeCVA0}) and Eq.~(\ref{tildeCVA01}) (red lines). We have used the energy-independent form  of the coefficients 
$C_{V/A,i}^{\rm Born}$ and $C_{V/A,i}^{(1),\rm Born}$ ($i=1,2,3$) given 
by Eqs.~(\ref{treematching}). 
}
  \label{fig:kinetic}
  \end{center}
\end{figure}

To examine in an example the behavior of power-counting breaking effects from
operator insertions beyond NNLL order we now consider the phase space matching 
contributions arising  from the N${}^4$LL order 
kinetic energy corrections. These emerge either from two
insertions of the ${\cal O}(v^4)$ kinetic energy operator $\bmp^4/(8m_t)^3$ or
from one insertion of the ${\cal O}(v^6)$ subleading kinetic energy operator 
$-\bmp^6/(16m_t)^5$. The respective expressions for the functions
$\Delta^i(t_1,t_2)$ read
\begin{align}
\Delta^{\rm kin,0,2\times \bmp^4/(8m_t^3)} &\, = \,
\frac{\bmp^8}{32m_t^4}
\left(\frac{3t_1^2 - 5m_t^2\Gamma_t^2}{(t_1^2+m_t^2\Gamma_t^2)^2} +
  \frac{3t_2^2 - 5m_t^2\Gamma_t^2}{(t_2^2+m_t^2\Gamma_t^2)^2}
+\frac{3(t_1^2+t_2^2)+8 t_1 t_2+6 m_t^2 \Gamma_t^2}{ (t_1^2+m_t^2 \Gamma_t
      ^2)(t_2^2+ m_t^2 \Gamma_t^2)}\right)\,,\nonumber\\[2mm] 
\Delta^{\rm kin,0,1\times \bmp^6/(16m_t^5)} &\, = \,
-\frac{\bmp^2}{2m_t^2}\Delta^{\rm kin,0}\,,
\end{align}
and the corresponding contributions to the Wilson coefficients 
$\tilde C_{V/A}(\Lambda,1)$ and 
$\tilde C_{V/A}^{(1)}(\Lambda,1)$ are
\begin{align}
{\tilde C_{V/A}^{\rm kin,[2\times\bmp^4/(8m_t^3)],0}}(\Lambda,1) \, & = \, 
2\,i\,N_c\,\big(C_{V/A,1}^{\rm Born}\big)^2\,
\frac{63m_t^2}{512\pi}\, \Big(
       \frac{53}{252\sqrt{2}\,\pi}\frac{\Gamma_t\Lambda^3}{m_t^4}
\nonumber\\[2mm]
&\quad+\frac{5\,(754-483\sqrt{2}\arsinh
       (1))}{1176\pi^2}\frac{\Gamma_t^2\Lambda}{m_t^3}\Big)
\,,
\nonumber \\[2mm]
{\tilde C_{V/A}^{\rm kin,[1\times\bmp^6/(16m_t^5)],0}}(\Lambda,1) \, & = \, 
2\,i\,N_c\,\big(C_{V/A,1}^{\rm Born}\big)^2\,
\frac{-7m_t^2}{64\pi}\,
\Big(\frac{11}{42\sqrt{2}\,\pi}\frac{\Gamma_t\Lambda^3}{m_t^4}
\nonumber\\[2mm]
&\quad+\frac{514-315\sqrt{2}\arsinh(1)}{140\pi^2}\frac{\Gamma_t^2\Lambda}{m_t^3} 
\Big)
\,,
\label{tildeCVAkinhigh1}
\end{align}
and
\begin{align}
{\tilde C_{V/A}^{(1),[2\times\bmp^4/(8m_t^3)],0}}(\Lambda,1) \, & = \, 
2\,i\,N_c\,
    \big(C_{V/A,1}^{\rm Born}\big)^2\,
    \frac{63m_t^2}{512\pi}\,
    \frac{115}{28\sqrt{2}\,\pi}\frac{\Gamma_t\Lambda}{m_t^2}
\,,
\nonumber \\[2mm]
{\tilde C_{V/A}^{(1),[1\times\bmp^6/(16m_t^5)],0}}(\Lambda,1) \, & = \, 
-2\,i\,N_c\,\big(C_{V/A,1}^{\rm Born}\big)^2\,
\frac{7m_t^2}{64\pi}\,
\frac{9}{2\sqrt{2}\,\pi}\frac{\Gamma_t\Lambda}{m_t^2}
\,.
\label{tildeCVAkinhigh2}
\end{align}
In Eqs.~(\ref{tildeCVAkinhigh1}) and (\ref{tildeCVAkinhigh2}) we have displayed
all terms that contribute at NLL and N${}^3$LL order. They all are
proportional to powers of $\Lambda/m_t$ and power-counting breaking. 
In Fig.~\ref{fig:kinetic}
the numerical contributions of the kinetic energy corrections to the inclusive
cross section from the NNLL (from Eqs.~(\ref{relativisticresults1},
\ref{relativisticresults2}), blue lines) and 
N${}^4$LL insertions (Eq.~(\ref{tildeCVAkinhigh1}, \ref{tildeCVAkinhigh2}), 
green lines) of the kinetic energy
operators are displayed over $\Delta M_t$ for $E=-5$ and $5$~GeV. The respective
solid lines show the contributions to $\tilde C_{V/A}(\Lambda,1)$ and the
dashed lines show the contributions to  
$\tilde C_{V/A}(\Lambda,1) + E/m_t \tilde C_{V/A}^{(1)}(\Lambda,1)$.
For comparison we also display the corresponding contributions from the tree-level NRQCD 
cross
section without phase space matching contribution, 
$\sigma^{0,\alpha_s=0}_{\rm NRQCD}(\Lambda=\infty) =
2N_c\,((C_{V,1}^{\rm Born})^2+(C_{A,1}^{\rm
  Born})^2)\,m_t^2/(4\pi)\,\mbox{Re}(v)$ (black lines),  
and the dominant phase space matching corrections
$\tilde C_{V/A}^0(\Lambda,1)$ from Eq.~(\ref{tildeCVA0}) (red lines).
The red dashed lines account for the energy
dependent terms in Eq.~(\ref{tildeCVA01}), that also
originate from the leading order diagram.
The results show again that
despite the existence of contributions that are power-counting
breaking the numerical impact of these terms is very small. The phase space
matching 
corrections from the NNLL kinetic energy insertions never exceed the level of a
few fb and those from the N${}^4$LL insertions, which are purely power-counting
breaking at the order we consider, are of order $0.01$~fb or smaller.
Morever we see that the contributions  to $\tilde C_{V/A}(\Lambda,1)$ and 
to $E/m_t \tilde C_{V/A}^{(1)}(\Lambda,1)$ coming from  NNLL 
kinetic energy insertions are similar in size, as would be 
expected in the absence of power-counting breaking terms.
This confirms that power-counting breaking effects are not of any concern. 
In particular we see that the quality of the
nonrelativistic expansion is excellent and the corrections from insertions of
higher dimensional relativistic operators are numerically compatible with the 
nonrelativistic $v$-counting.

\subsection{Full Theory Analysis}
\label{subsectionfulltheory}

In this section we analyze results for the inclusive cross section obtained from 
relativistic amplitudes in the full theory for $\alpha_s=0$. In the first part
we compare the NRQCD prediction using the factorization formula of
Eq.~(\ref{crosssection}) at NNLL 
order with the corresponding inclusive cross section from the full theory 
accounting only for the amplitude $e^+e^-\to t\bar t \to  b\bar b W^+W^-$.
In this case the remainder contribution $\sigma^{\alpha_s=0}_{\rm rem}(\Lambda)$
vanishes since this amplitude is fully accounted for in the usual NRQCD matrix
element and matching computations. We show that the NRQCD factorization formula
provides an 
excellent approximation to the full theory prediction for invariant mass cuts
$\Delta M_t=15$--$35$~GeV. We also demonstrate that the
NRQCD prediction in the $\msb$ scheme without the phase space matching
contributions\footnote{
All NRQCD predictions for top pair threshold
production that can be found in the previous literature were carried out in
this approximation.}  
overestimates by far the full theory predictions. This shows the importance
of the phase space matching procedure and also explains why
the NRQCD contributions turn out to be the dominant
terms in the phase space matching corrections. 
In the second part of this section we compare the full Standard Model prediction
for  $e^+e^-\to b\bar b W^+W^-$ with invariant mass cuts with the NNLL
NRQCD prediction including the NRQCD phase space matching contribution. 
We demonstrate that, as indicated before, the remainder cross section
$\sigma^{\alpha_s=0}_{\rm rem}(\Lambda)$ is small and only amounts to at most
several femptobarn. For the precision expected for threshold cross section
measurements at a future linear collider the remainder contributions can
therefore be safely neglected. For the full theory computations carried out in
this section we use 
MadGraph and MadEvent~\cite{Alwall:2007st}. At this point we note that the
amplitudes generated by MadGraph are at the tree-level and use the fixed-width
scheme for the top quark propagator,
$i (\slash{p}+m_t)/(q^2-m_t^2+i m_t\Gamma_t)$. 
This expression for the top propagator is the correct
form in the resonance 
limit and compatible with the nonrelativistic expansion of NRQCD at the order we
are working. We note that a discussion on invariant mass cuts for the threshold
production of a $W^+W^-$ pair in $e^+e^-$ collisions based on
full theory calculations similar to the ones given here was presented in
Ref.~\cite{Actis:2008rb}. We also refer to Refs.~\cite{Beneke:2003xh,Beneke:2004km}.

\vskip 5mm \noindent
{\bf \underline {Analysis for $e^+e^-\to t\bar t \to b\bar b W^+W^-$:}} \\[2mm]
Since the amplitude for 
$e^+e^-\to \gamma, Z\to t\bar t \to b\bar b W^+W^-$ contains a 
$t\bar t$ intermediate state, its contribution in the $t\bar t$ threshold region
is fully described in NRQCD. Thus as long as invariant
mass cuts are applied in the region where the nonrelativistic expansion is 
valid,
the phase space matching contributions can be computed entirely within NRQCD and
the remainder contribution to the cross section, 
$\sigma^{\alpha_s=0}_{\rm rem}(\Lambda)$ is zero. Thus the comparison of the
NRQCD prediction of the inclusive cross section with invariant mass cuts based
on Eq.~(\ref{crosssection}) with a fully relativistic tree-level computation
based on the same amplitude serves as an important numerical check of the
nonrelativistic expansion and of the NRQCD phase space matching method. In this
section we carry out this check for the tree-level full theory cross section
(i.e.\ without accounting for one-loop electroweak corrections) and at NNLL
order in the nonrelativistic 
$v$-expansion. We note that the amplitude for $e^+e^-\to b\bar b
W^+W^-$ is not gauge-invariant if one includes only the diagrams with a $t\bar
t$ intermediate state.\footnote{
In the nonrelativistic expansion the gauge-dependence of the results based 
only on the amplitude for 
$e^+e^-\to \gamma, Z\to t\bar t \to b\bar b W^+W^-$ starts at NNLL order. 
} 
To be definite we therefore pick the unitary gauge for
all the calculations in this section.

For the relativistic calculations we employ MadGraph for the amplitude
generation and MadEvent with $10^4$ events to numerically compute the cross
section with cuts on the top and antitop invariant masses. For the NRQCD
calculation one employs the factorization formula of Eq.~(\ref{crosssection})
neglecting all QCD effects from matching coefficients and potentials and the
real one-loop electroweak corrections to the Wilson coefficients 
$C_{V/A,1}^{\rm 1loop}$ of the $(e^+e^-)(t\bar t)$ current operators. One also
needs to neglect the imaginary interference contributions $i C_{V/A,1}^{\rm
  int}$ to the Wilson coefficients of the $(e^+e^-)(t\bar t)$ current operators
(see Eqs.~(\ref{CintV}) and (\ref{CintA})) because we do not account for
diagrams with either only a 
top or an antitop at intermediate stages. However, there are off-shell 
corrections to the (anti)top decay from the full theory relativistic amplitude
which in the matching procedure for $i C_{V/A,1}^{\rm int}$ are related to the
imaginary part of the top wave-function renormalization constant in unitary
gauge. These imaginary matching contributions have the form~\cite{Hoang:2004tg}
\begin{align}
(i C_{V,1}^{\rm int})^{Z_t,\rm uni} & \, = \, -i \, \frac{\alpha ^2 \pi 
  |V_{tb}|^2}{32 x m_t^2 s_w^4 (4 c_w^2-x)} \Big[3 Q_e  Q_t s_w^2 
(4-x+4 x^2-9 x^3+2 
x^4)-4 Q_e s_w^2(1-x^3) \nonumber\\
& \quad +3  Q_t s_w^2 (1+x^2-2 x^3) - 1+x^3\Big]\,,
\nonumber\\
(i C_{A,1}^{\rm int})^{Z_t,\rm uni} & \, = \, i \,\frac{\alpha ^2 \pi
  |V_{tb}|^2}{32 x m_t^2 s_w^4
  (4 c_w^2-x)} \Big[3 Q_t s_w^2(1+x^2-2 x^3) -1+x^3\Big]\,,
\end{align} 
and need to be 
included. The resulting expression for the NRQCD cross section at
NNLL order is remarkably simple and has the form
\begin{align}
\label{sigmaNRQCDas0NNLL}
\sigma_{\rm NRQCD,t\bar t}^{\alpha_s=0,\rm NNLL}(\Lambda) & \, = \,
N_c\, \Big(\big((C_{V,1}^{\rm Born})^2+(C_{A,1}^{\rm
  Born})^2\big)\Big(\frac{m_t^2}{2\pi}\mbox{Im}(iv)-\frac{\sqrt{2}
  m_t^2\Gamma_t}{\pi^2 \Lambda}-\frac{13 \Gamma_t
  \Lambda}{48\sqrt{2}\pi^2} 
\nonumber\\
& {}\quad-\frac{3 m_t\Gamma_t}{8\pi}\mbox{Re}(iv) + \frac{7
  m_t^2}{48\pi} \mbox{Im}(iv^3) \Big) \nonumber\\
& {}\quad+ \big((C_{V,3}^{\rm Born})^2+(C_{A,3}^{\rm
  Born})^2\big) \Big( \frac{\sqrt{2}\Gamma_t
    \Lambda}{3\pi^2} +
\frac{m_t^2}{3\pi}\mbox{Im}(i v^3) \Big)\nonumber\\
&{}\quad+\big(C_{V,1}^{\rm Born}(C_{V,1}^{\rm int})^{Z_t,\rm uni}+C_{A,1}^{\rm Born}(C_{A,1}^{\rm int})^{Z_t,\rm uni}\big)\Big(
\frac{\sqrt{2} m_t \Lambda}{\pi^2} +\frac{m_t^2}{\pi}\mbox{Re}(iv)\Big)
 \Big)
\,,
\end{align}
where the subscript $t\bar t$ is a reminder that for the amplitude
only diagrams with a $t\bar t$ pair in the intermediate state are
considered and where we use the $s$-dependent convention for the
current coefficients $C_{V/A,i}^{\rm Born}$ given in Eq.~(\ref{treematchingv2}). 
The corresponding LL cross section is just the well known nonrelativistic lowest
order expression which does not contain phase space matching
corrections
\begin{align}
\label{sigmaNRQCDas0LL}
\sigma_{\rm NRQCD,t\bar t}^{\alpha_s=0,\rm LL} & \, = \,
2N_c\,\big((C_{V,1}^{\rm Born})^2+(C_{A,1}^{\rm
  Born})^2\big)\,\frac{m_t^2}{4\pi}\,\mbox{Im}\,(iv)\,,
\end{align}
and the NLL cross section reads
\begin{align}
\label{sigmaNRQCDas0NLL}
\sigma_{\rm NRQCD,t\bar t}^{\alpha_s=0,\rm NLL}(\Lambda) & \, = \,
N_c\, \Big(\big((C_{V,1}^{\rm Born})^2+(C_{A,1}^{\rm
  Born})^2\big)\Big(\frac{m_t^2}{2\pi}\mbox{Im}(iv)-\frac{\sqrt{2}
  m_t^2\Gamma_t}{\pi^2 \Lambda}-\frac{13 \Gamma_t
  \Lambda}{48\sqrt{2}\pi^2}  \Big) \nonumber\\
& {}\quad+ \big((C_{V,3}^{\rm Born})^2+(C_{A,3}^{\rm
  Born})^2\big) \frac{\sqrt{2}\Gamma_t
    \Lambda}{3\pi^2} \nonumber\\
&{}\quad+\big(C_{V,1}^{\rm Born}(C_{V,1}^{\rm int})^{Z_t,\rm uni}+C_{A,1}^{\rm Born}(C_{A,1}^{\rm int})^{Z_t,\rm uni}\big)
\frac{\sqrt{2} m_t \Lambda}{\pi^2}
 \Big)\,.
\end{align}

\begin{figure}[t]
  \begin{center}
  \includegraphics[width=0.48\textwidth]{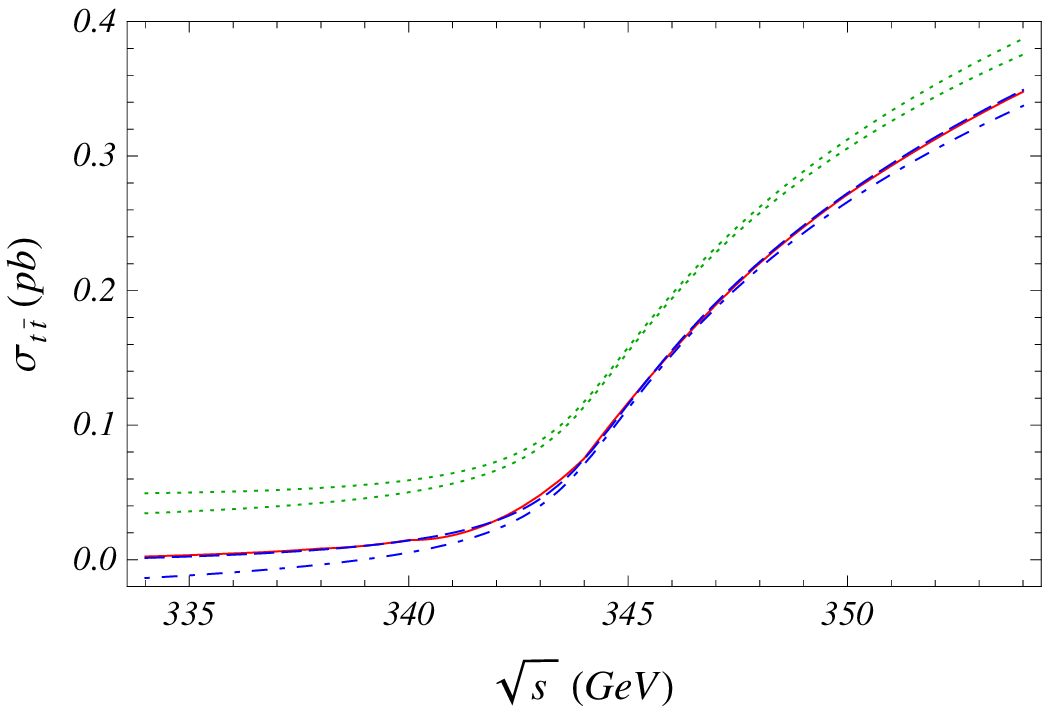}\quad\;
  \includegraphics[width=0.48\textwidth]{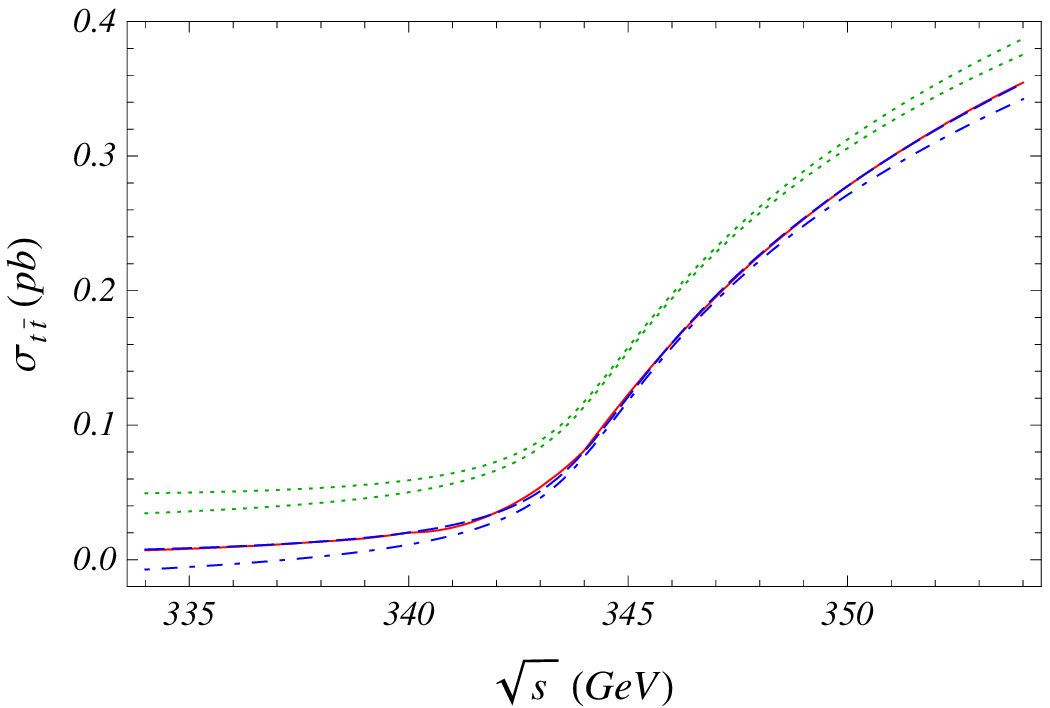}\\
  \includegraphics[width=0.48\textwidth]{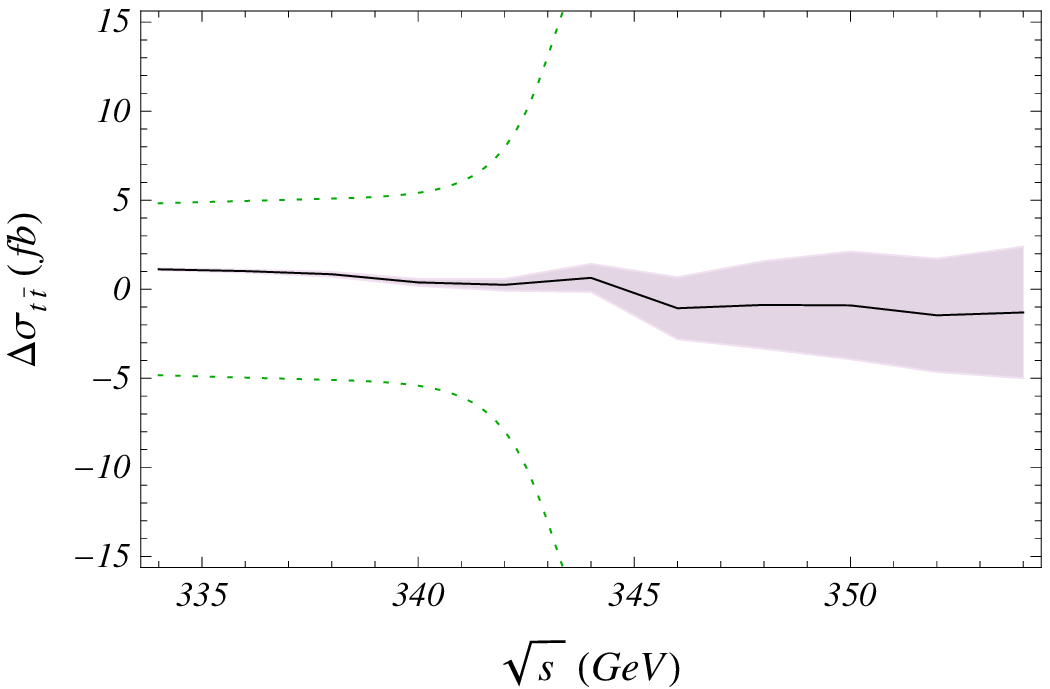}\quad\;
  \includegraphics[width=0.48\textwidth]{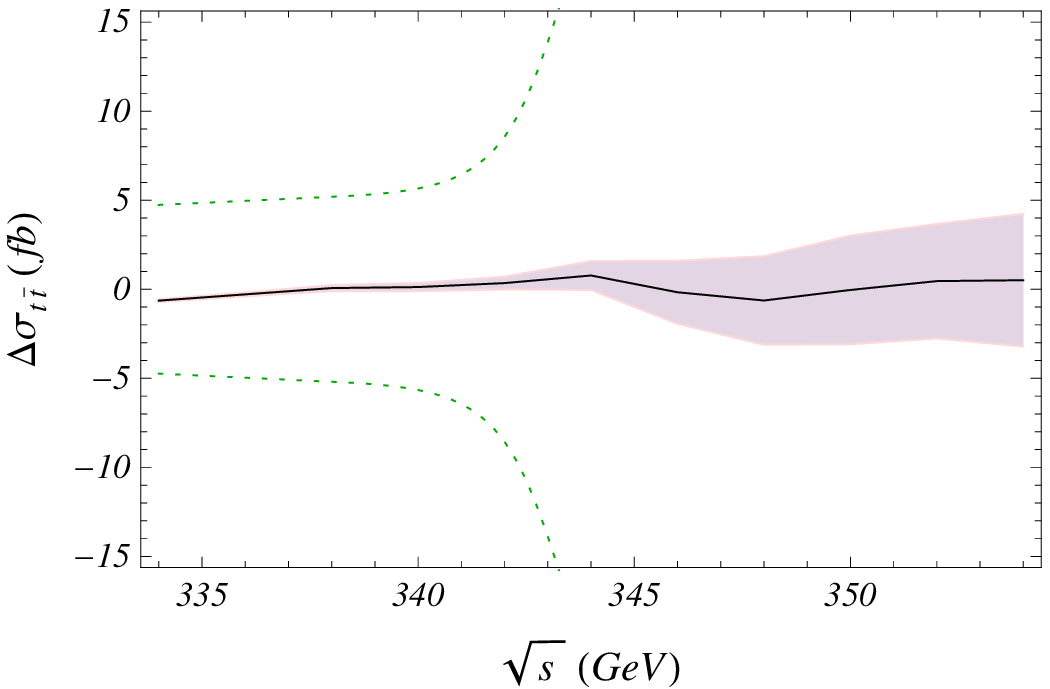}
  \caption{Upper panels: the full relativistic cross section 
$\sigma_{t\bar t}^{\alpha_s=0}(\Lambda)$ obtained from MadEvent (red lines) and
the LL (lower green dotted line), 
NLL (blue dash-dotted lines) and 
NNLL order (blue dashed lines) NRQCD cross sections.
The uncut NNLL NRQCD cross section is also shown (upper green-dotted lines).
Lower panels: difference of the relativistic and the
NNLL NRQCD cross section  (solid line), 
shaded bands represent the statistical uncertainty of the MadEvent
integrations while the green dotted lines delimit the theoretical precision goal.
We have chosen $\Delta M_t=15$ for all panels on the left, and $\Delta M_t=35$~GeV 
for those on the right.
  \label{fig:sigttbar}}
  \end{center}
\end{figure}

In the upper panels of Fig.~\ref{fig:sigttbar} the full relativistic cross 
section 
$\sigma_{t\bar t}^{\alpha_s=0}(\Lambda)$ obtained from MadEvent (red lines) and
the LL (lower green dotted line), 
NLL (blue dash-dotted lines) and 
NNLL order (blue dashed lines) NRQCD cross sections of 
Eqs.~(\ref{sigmaNRQCDas0LL}), (\ref{sigmaNRQCDas0NLL})
and (\ref{sigmaNRQCDas0NNLL}),
are shown for $\Delta M_t=15$ (left panels) and $35$~GeV (right panels). We use
the following set of parameters for the computations:
\begin{align}
\label{parameters}
\begin{array}[b]{rclrcl}
m_t &=& 172.00\, \mbox{GeV}\,,\quad & \Gamma_t &=& 1.4614 \,
\mbox{GeV}\,, \\[2mm]
M_W &=& 80.419 \, \mbox{GeV}\,, & M_Z &=& 91.188 \,
\mbox{GeV}\,, \\[2mm]
c_w &=& M_W/M_Z\,, & \alpha &=& 1/132.51
\,.
\end{array}
\end{align}
We see that the quality of the nonrelativistic expansion is excellent. 
Moreover, at the scale of the figures no differences between the full 
relativistic and the NNLL order NRQCD cross section are visible.
The corresponding lower panels show the difference of the relativistic and the
NNLL NRQCD cross section, $\Delta\sigma_{t\bar t}(\Lambda)=
\sigma_{t\bar t}^{\alpha_s=0}(\Lambda) -
\sigma_{\rm NRQCD,t\bar t}^{\alpha_s=0,\rm NNLL}(\Lambda)$, which illustrates
the small size of 
corrections coming from beyond NNLL order in the NRQCD computation. Here the
shaded bands represent the statistical uncertainty of the MadEvent
integrations. (The statistical error is proportional to the cross section value
and is therefore increasing for $\sqrt{s}>2m_t$.) We see that
$\Delta\sigma_{t\bar t}(\Lambda)$ is at the level of $1-2$~fb for $\sqrt{s}<2
m_t$ and compatible with zero within the statistical uncertainty of the MadEvent
results for $\sqrt{s}>2m_t$. The greed dotted lines show the theoretical
precision goal we have formulated in view of the experimental uncertainties
expected at a future linear collider. The lines are determined from the
quadratic sum of an energy-independent error of $5$~fb and a $2$\% relative
uncertainty with respect to the full NRQCD inclusive cross section prediction
(including QCD effects) presented in Sec.~\ref{sectionanalysis}.
Overall, we find that $\Delta\sigma_{t\bar t}(\Lambda)$ is
much smaller than the theoretical precision goal indicated by the green dotted
lines and compatible with the size of
${\cal O}(v^3)$ relativistic corrections. This demonstrates the excellent
quality of the NRQCD phase space matching procedure we propose in this work and
it should be adequate for the expected precision achievable at a future linear
collider. 

In the upper panels of Fig.~\ref{fig:sigttbar} we have also shown the NNLL NRQCD
cross section without the phase space matching contributions\footnote{
For $\alpha_s=0$ and without phase space constraints there are no NLL order
corrections, and the LL and NLL NRQCD cross sections agree.
} (upper green dotted lines). Except for the wave function contribution in 
$i (C_{V/A,1}^{\rm int})^{Z_t,\rm uni}$, which contributes less than $15$~fb for
the displayed energy range, this represents the 
approximation which has been used in all previous literature on the top
pair threshold cross section where the top quark instability was 
accounted for by the replacement rule $E\to E+i~\Gamma_t$. 
We see that without phase space matching contributions the cross section is 
overestimated by about $30-50$~fb at all energies. The relative discrepancy to
the correct answer is particularly large for $\sqrt{s}<2m_t$ where the
predictions that only account for the complex shift $E\to E+i\Gamma_t$ do not
vanish sufficiently fast. This unphysical feature is related to the fact that
the top propagator of Eq.~(\ref{eq:prop}) leads to an overestimate of the
contributions from NRQCD phase space regions with large top and antitop
invariant masses. Thus the phase space matching
corrections are essential to reach the intended precision goal for the
theoretical predictions.

\begin{figure}[t]
  \begin{center}
  \includegraphics[width=0.46\textwidth]{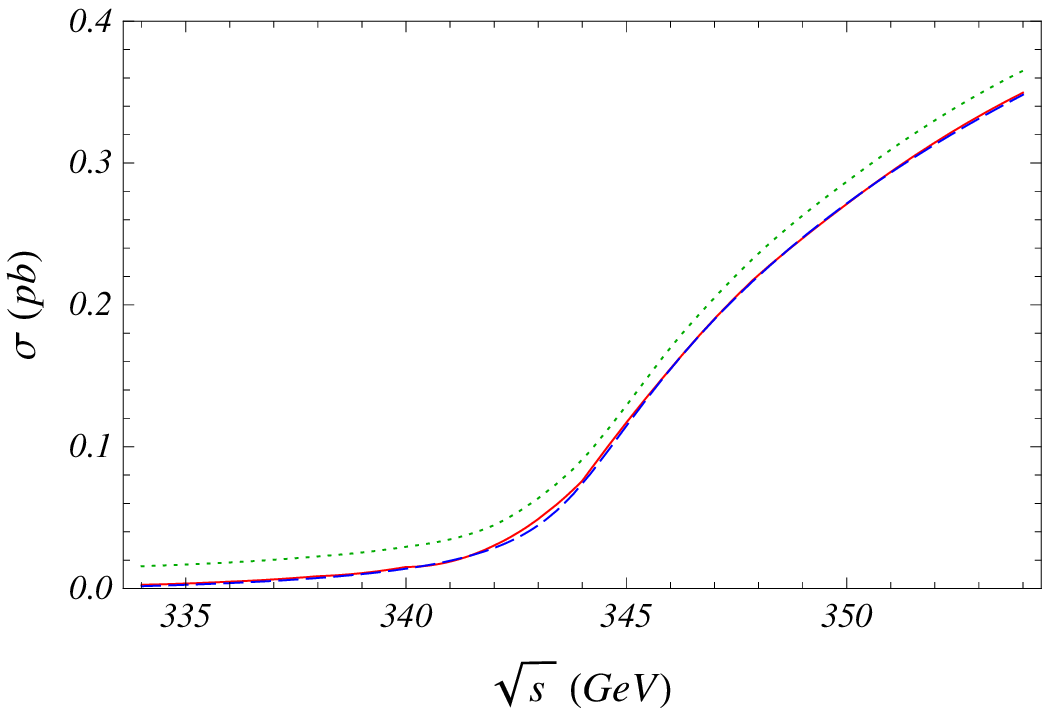}\qquad
  \includegraphics[width=0.46\textwidth]{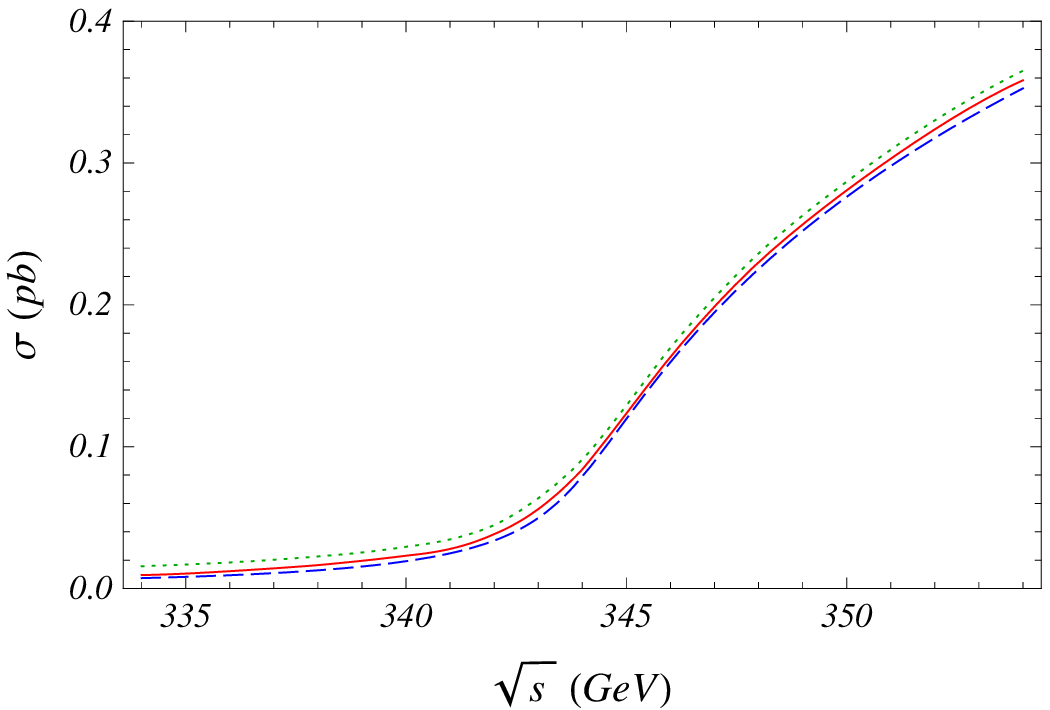}\\
  \includegraphics[width=0.46\textwidth]{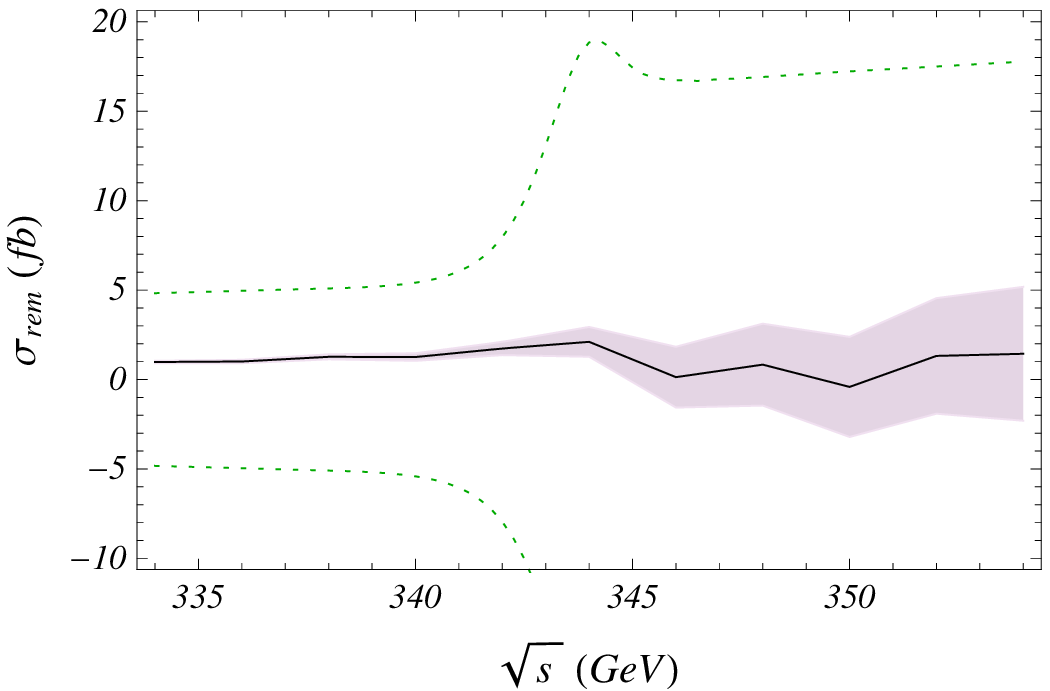}\qquad
  \includegraphics[width=0.46\textwidth]{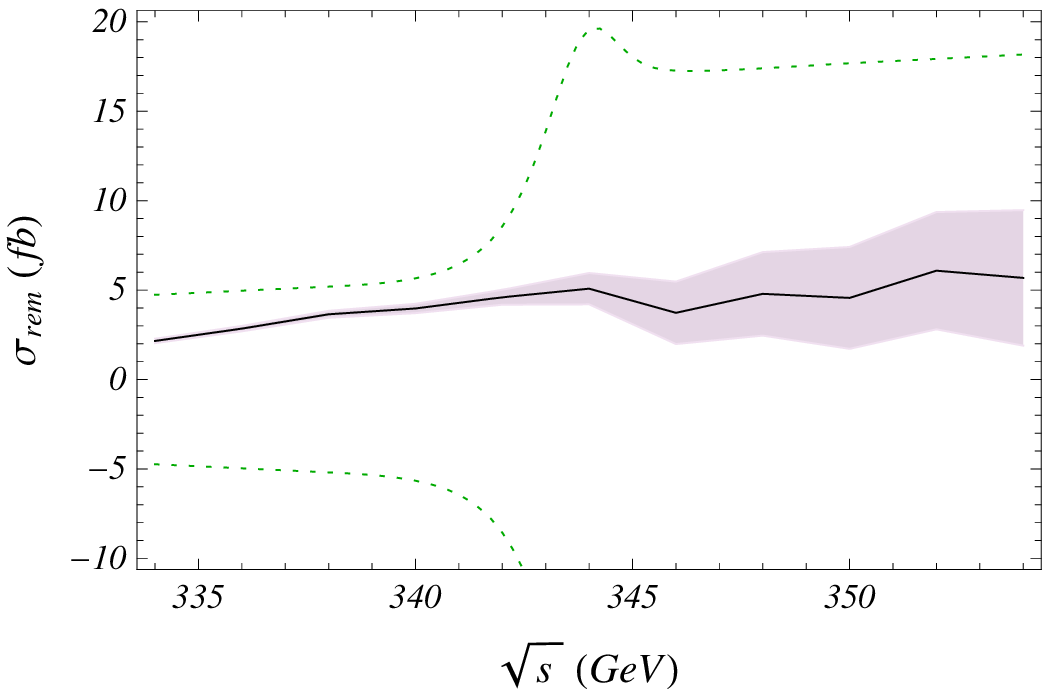}\\
  \includegraphics[width=0.46\textwidth]{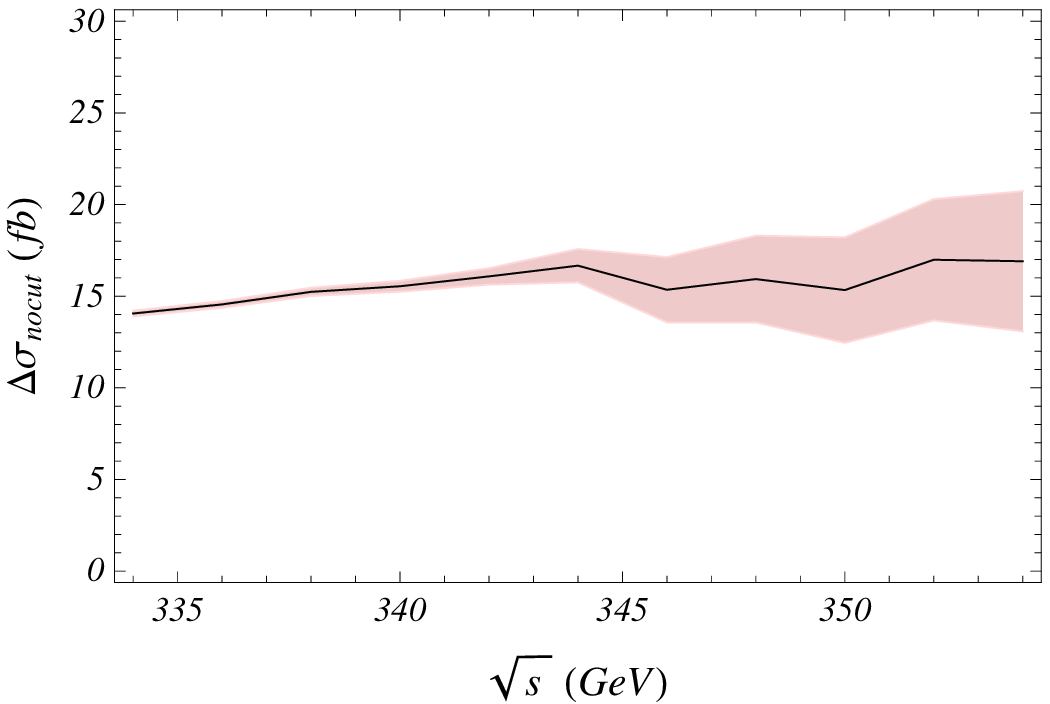}\qquad
  \includegraphics[width=0.46\textwidth]{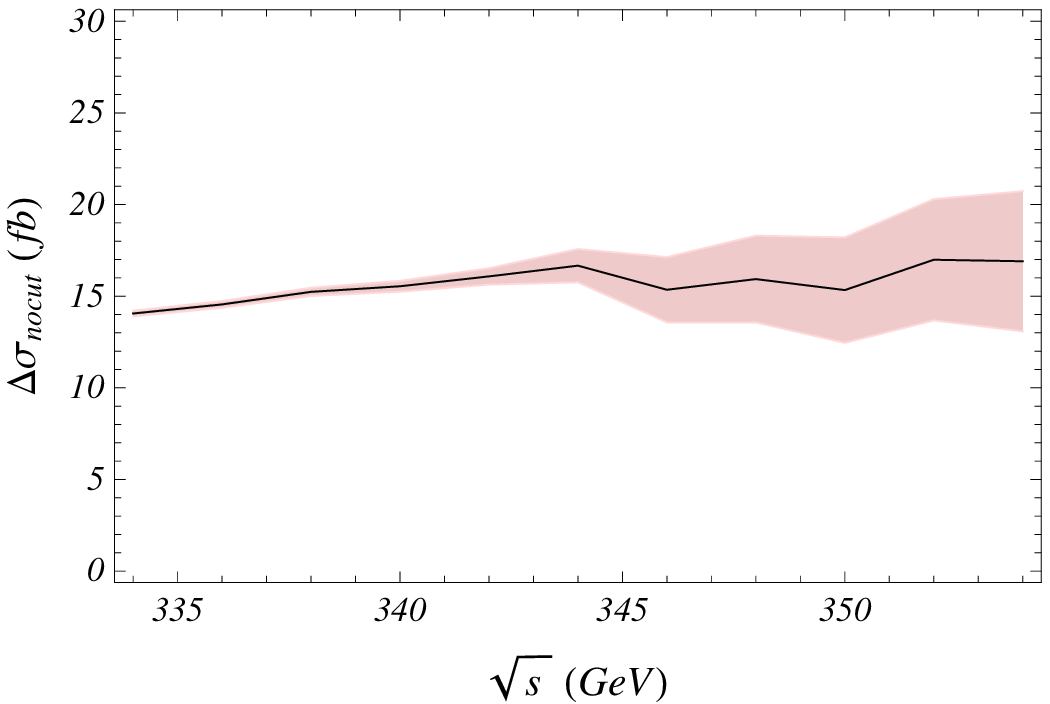}
  \caption{Upper panel: fully relativistic  $b\bar b W^+W^-$ cross section  
$\sigma^{\alpha_s=0}(\Lambda)$ obtained from MadEvent with cut (red line) and
without cut (green dotted lines), and NNLL
order NRQCD cross section 
$\sigma_{\rm NRQCD,t\bar t}^{\alpha_s=0,\rm NNLL}(\Lambda)$ (blue dashed lines).
Middle panel: $\sigma_{\rm rem}^{\alpha_s=0}(\Lambda)$ in fb~units 
with the statistical
uncertainties of the MadEvent computation (shaded region), 
and  the theoretical precision goal (green dotted lines).
Lower panel: $\Delta\sigma_{\rm nocut}(\Lambda) =
\sigma^{\alpha_s=0}_{{\rm nocut}}-
\sigma_{\rm NRQCD}^{\alpha_s=0,{\rm NNLL}}(\Lambda)$ (black solid lines),
 and 
the statistical uncertainties from MadEvent (shaded regions).
We have chosen $\Delta M_t=15$ for all panels on the left, and $\Delta M_t=35$~GeV 
for those on the right.
  \label{fig:sigall}}
  \end{center}
\end{figure}
\vskip 5mm \noindent
{\bf \underline {Analysis for the full amplitude 
$e^+e^-\to b\bar b W^+W^-$:}} \\[2mm]
We now consider the fully relativistic Standard Model amplitude for
$e^+e^-\to b\bar b W^+W^-$ at tree-level, 
$\sigma^{\alpha_s=0}(\Lambda)$.
The results allow us to determine the remainder parts of the cross section,
$\sigma_{\rm rem}^{\alpha_s=0}(\Lambda)$. In the upper panel of
Fig.~\ref{fig:sigall} the fully relativistic cross section  
$\sigma^{\alpha_s=0}(\Lambda)$ obtained from MadEvent (red line) and the NNLL
order NRQCD cross section 
$\sigma_{\rm NRQCD,t\bar t}^{\alpha_s=0,\rm NNLL}(\Lambda)$ (blue dashed lines)
are shown for $\Delta M_t=15$ (left panels) and $35$~GeV (right panels), where 
the parameter set of Eq.~(\ref{parameters}) has been employed and $10^4$ events
have been used for integrations. 
The analytic formula for 
$\sigma_{\rm NRQCD,t\bar t}^{\alpha_s=0,\rm NNLL}(\Lambda)$ is just 
Eq.~(\ref{sigmaNRQCDas0NNLL}) with the electroweak top wave function
contributions  $(i C_{V/A,1}^{\rm int})^{Z_t,\rm uni}$ being replaced by the full
interference coefficients $i C_{V/A,1}^{\rm int}$ from Eqs.~(\ref{CintV}) and
(\ref{CintA}). Up to relativistic corrections beyond NNLL order, which we
neglect in the following discussion, the difference between the relativistic and
the NNLL order NRQCD results is just $\sigma_{\rm rem}^{\alpha_s=0}(\Lambda)$,
which we use to determine the remainder contributions to the phase space 
matching coefficients $\tilde C_{V/A}^0(\Lambda)$ and 
$\tilde C_{V/A}^{(1),0}(\Lambda)$. 
For the invariant mass cuts $\Delta M_t=15$ and $35$~GeV
$\sigma_{\rm rem}^{\alpha_s=0}(\Lambda)$ is displayed in fb~units in the middle
panel of Fig.~\ref{fig:sigall}. The shaded region represents the statistical
uncertainties of the MadEvent computation. For $\Delta M_t=35$~GeV
$\sigma_{\rm rem}^{\alpha_s=0}(\Lambda)$ ranges from $0$ to $5$~fb and increases
with the c.m.\ energy. For 
$\Delta M_t=15$~GeV it is almost energy-independent and below $1$~fb for 
$\sqrt{s}<2m_t$ and compatible with zero within the statistical MadEvent errors
for $\sqrt{s}>2m_t$. The results are fully consistent with a linear dependence
in $E$, which confirms the structure and NRQCD counting of the
$(e^+e^-)(e^+e^-)$ forward scattering operators. Overall, the size of the
remainder contributions  
is much smaller than the theoretical precision goal explained above
and indicated by the green dotted lines. So the remainder part can be neglected
for the theoretical 
predictions. With Eq.~(\ref{crosssection}) it is straightforward to fit for the
remainder contributions to the phase space matching coefficients 
$\tilde C_{V/A}^0$ and $\tilde C_{V/A}^{(1),0}$. We use an analysis for
$\Delta M_t=15, 20, 25, 30$ and $35$~GeV, and assuming a linear dependence on 
$\Delta M_t$ the result of this fit reads
\begin{align}
\Big(\tilde C_V^0(\Lambda) + \tilde C_A^0(\Lambda)\Big)^{\rm rem} & \, = \, 
i\,\bigg[\,(-1.2\pm 0.9)\, {\rm fb}\,+ \, ( 28\pm 6){\Delta M_t \over m_t}\, {\rm fb}\,\bigg]\,,
\nonumber \\
\Big(\tilde C_V^{(1),0}(\Lambda) + \tilde C_A^{(1),0}(\Lambda)\Big)^{\rm rem} & 
 \, = \,  
i\,\bigg[\,(-14\pm 25)\, {\rm fb}\, + \, ( 260\pm 170) \,{\Delta M_t \over m_t} \,{\rm fb} 
\,\bigg]
\,.
\label{phsspfit1}
\end{align}
For the fit we added an energy-independent error of $1$~fb to
$\sigma_{\rm rem}^{\alpha_s=0}(\Lambda)$ to 
account for higher order relativistic corrections $\propto
(E/m_t)^n$ with $n>1$.

\vskip 5mm \noindent
{\bf \underline {Total cross section without cuts:}} \\[2mm]
Rather than predicting the inclusive cross section with cuts on the (anti)top
invariant masses it is also natural to ask about the total cross section
for the case that no such cuts are applied on the $b\bar b W^+W^-$ final
state. The phase space matching concept can be also applied in this case since
the difference between the full theory phase space for the cases with and without
invariant mass cuts constitutes again a hard NRQCD effect, much like the
difference between applying cuts of different sizes. This can be easily understood
because even without invariant mass cuts the phase space is constrained by 
kinematic bounds that in the $t\bar t$ threshold region ($\sqrt{s}\approx 2m_t$)
also represent hard scales of order $m_t$.
In the upper panels of
Fig.~\ref{fig:sigall} the green dotted lines represent the full theory $b\bar b
W^+W^-$ cross section without any cuts from MadEvent. 
The difference $\Delta\sigma_{\rm nocut}(\Lambda) =
\sigma^{\alpha=0}_{{\rm nocut}}-
\sigma_{\rm NRQCD}^{\alpha_s=0,{\rm NNLL}}(\Lambda)$
is displayed by the black solid lines in the lowest panels 
of Fig.~\ref{fig:sigall} for $\Delta M_t=15$ (left) and $35$~GeV (right). 
In the figure the statistical uncertainties from MadEvent are again represented
by the gray shaded regions. In contrast to predictions for the cross section
with invariant mass cuts, here, $\Lambda$ acts simply as a UV cutoff for the
NRQCD phase space integration. Since the cross section is independent of
the cutoff $\Lambda$, $\Delta\sigma_{\rm nocut}(\Lambda)$ just compensates the
$\Lambda$-dependence of the NRQCD cross section.  
Applying the same fitting
procedure as for Eq.~(\ref{phsspfit1}) we obtain the following results for
the remainder contributions to the phase space matching coefficients
$\tilde C_{V/A}^0$ and $\tilde C_{V/A}^{(1),0}$:
\begin{align}
\Big(\tilde C_V^0(\infty) + \tilde C_A^0(\infty)\Big)^{\rm rem} & \, = \, 
i\,\bigg[\,(-18.9\pm 0.9)\,{\rm fb} \,+ \, ( -42\pm 6){\Delta M_t \over m_t}\,{\rm fb}\,\bigg]  \,,
\nonumber \\
\Big(\tilde C_V^{(1),0}(\infty) + \tilde C_A^{(1),0}(\infty)\Big)^{\rm rem} & 
 \, = \,  
i\,\bigg[\,(21\pm 26)\, {\rm fb}\, + \, ( 90\pm 170) \,{\Delta M_t \over m_t} \, {\rm fb}
\,\bigg]
\,.
\label{phsspfit2}
\end{align}
Although we do not need the phase space matching coefficients 
$\tilde C_{V/A}^0(\infty)$ and $\tilde C_{V/A}^{(1),0}(\infty)$
for the predictions intended in this work, we give the results for future
reference.

\section{Phase Space Matching with QCD Effects} 
\label{sectionQCD}

In the previous sections we have discussed the concepts of the phase space
matching procedure and computed the phase space matching coefficients neglecting
QCD effects. We have demonstrated that the phase space matching coefficients can
be reliably computed within NRQCD and that the remainder contributions which
contain all effects that need the evaluation of full theory diagrams are small.
For the precision goal we need to achieve, these remainder contributions can be
neglected. We have also shown that for invariant mass cuts $\Delta
M_t=15$--$35$~GeV, power-counting breaking terms that arise from insertions of
higher order relativistic operators do not spoil the nonrelativistic
expansion. 

In this section we extend the phase space matching procedure to account for QCD
corrections. Due to the additional QCD-induced interactions the required
operator structure is more complicated. Up to NNLL order
only the $(e^+e^-)(e^+e^-)$ forward scattering operators already
discussed in the previous sections are required. At N${}^3$LL order we also need
to account for imaginary phase space matching contributions to the Wilson
coefficients of the $(e^+e^-)(t\bar t)$ top pair production operators to account
for non-analytic energy-dependent terms in the matching relations.
As for the previous sections we
compute the phase space matching coefficients within NRQCD. Since for
$\alpha_s=0$ we found the remainder contributions to the coefficients to be
negligible and since no kinematic or dynamical enhancement is expected for the
QCD corrections, we ignore the QCD corrections to the remainder contributions
in our analysis. For the determination of the QCD corrections we have to account
for the QCD potentials and for the exchange of ultrasoft gluons among the top
and antitop quarks and their decay products. Both types of corrections are
analyzed in the following subsections.

\subsection{Ultrasoft Gluon Exchange and Nonperturbative Effects}
\label{sec:QCDusoft}

\begin{figure}[t]
  \begin{center}
  \includegraphics[width=0.9\textwidth]{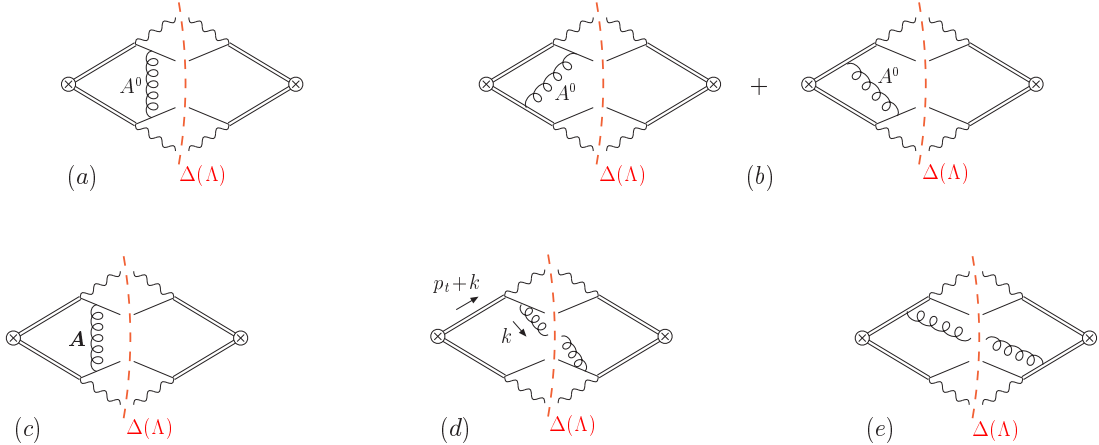}
  \caption{${\cal O}(\alpha_s)$ QCD radiative corrections to 
$e^+e^-\to\bar{t}t\to bW^-\bar{b}W^+$ originating from $(a)$ time-like gluon 
exchange 
 between $b\bar{b}$-quarks, $(b)$ time-like gluon exchange between $t\bar{b}$ 
and $\bar{t}b$-quarks,
 $(c)$ space-like gluon exchange between $b\bar{b}$, and $(d)$ interference 
between gluon radiation
from $b$ and $\bar{b}$. Contribution $(e)$ involving gluon radiation off the 
nonrelativistic 
top/antitop lines is suppressed with respect to the previous ones.
The conjugated diagrams have to be added for each case. The definition of
the top momenta $p_t$ in terms of the nonrelativistic momenta
can be found in Fig.~\ref{fig4}. }
  \label{fig6}
  \end{center}
\end{figure}

In the threshold region ultrasoft gluons, which carry momenta of order
$m_tv^2\sim\Gamma_t$, can propagate and lead to interactions among the top and
the antitop quark and their decay products. These ultrasoft interactions are
sometimes called ``QCD interference effects'' since the corresponding Feynman
diagrams and their interference contributions all have comparable numerical
size. The exchange of ultrasoft gluons starts to contribute at NLL order; the
relevant diagrams at ${\cal O}(\alpha_s)$ in Coulomb gauge are displayed in
Fig.~\ref{fig6}. It is known that the QCD interference effects cancel in the
total cross section at NLL~\cite{Fadin:1993dz,Melnikov:1993np} and NNLL
order~\cite{Hoang:2004tg} if no cuts are 
imposed on the NRQCD phase space.  Some of the interference effects have shown
to cancel at NLL also for other inclusive observables where the top energy is
integrated out, for example for the top quark three-momentum
distribution~\cite{Peter:1997rk}. For the top and antitop invariant masses $M_t$
and $M_{\bar t}$, on the other hand, the effects of ultrasoft gluon exchange are
essential because they directly affect the form of the distribution. The
invariant mass distribution is further affected - at leading order - by
nonperturbative effects. The precise way how both effects enter the distribution
depends on the details of the reconstruction prescription, see Refs.~\cite{Fleming:2007xt,Fleming:2007qr}
for a detailed treatment at large c.m.\ energies $\sqrt{s}\gg 2m_t$. The
situation is, however, considerably simpler for the predictions of the inclusive
cross section with the invariant mass cuts of
Eq.~(\ref{eq:invarmasscuts}) for $\Delta M_t\gg \Gamma_t$. Because the inclusive
cross section is an integral over $M_t$ and $M_{\bar t}$ in the resonance
region, the main effects of the ultrasoft gluon exchange and of the
nonperturbative contributions effectively correspond to a shift in $\Delta
M_t$ of order $\Gamma_t$ and $\Lambda_{\rm QCD}$, respectively. This allows us
to estimate that the size of the ultrasoft and the nonperturbative corrections
to the phase space matching coefficients are of order $\alpha_s
m_t\Gamma_t\Lambda^2$ and $m_t\Lambda_{\rm QCD}/\Lambda^2$, respectively, with
respect to the dominant NLL contributions of Eq.~(\ref{tildeCVA0}), which are
proportional to $\Gamma_t/\Lambda$. The form of the nonperturbative corrections
can also be understood from the point of view that there is an operator product
expansion for the inclusive cross section.
For the estimate for the ultrasoft gluons we
have included a factor of $\alpha_s$ from the coupling of the gluon to the
quarks. With $\Lambda_{\rm QCD}\approx \alpha_s\Gamma_t$ and $\Lambda\sim m_t$
both effects represent N${}^4$LL corrections and are beyond the N${}^3$LL order
level we consider in this work. Numerically they are suppressed by two orders of
magnitude 
with respect to the NLL order phase space matching contributions.
To confirm our argumentation for the ultrasoft corrections we
explicitly compute in the appendix the phase space matching corrections to the 
$(e^+e^-)(e^+e^-)$ forward scattering operators that arise from the diagrams in
Fig.~\ref{fig6}. We use the invariant mass definition of
Eq.~(\ref{eq:invariantmassdef}) and assume that the ultrasoft gluon can be
resolved. While being not entirely realistic from the experimental
point of view, this
prescriptions should nevertheless give the typical size of the
ultrasoft corrections. As shown in the appendix, we find that the ultrasoft
phase space matching corrections indeed have the scaling behavior estimated
above. In addition, they have a strong numerical suppression that make them
irrelevant. For the rest of this paper we therefore ignore ultrasoft and
nonperturbative phase space matching corrections.

\subsection{Potential Interactions}
\label{sec:potentials}

We now consider the phase space matching corrections originating from the
potential interactions. Since the phase space matching procedure shares the
properties of common loop graph matching computations, we find that once loop
diagrams are involved, we also have to carry out phase space matching for
subdiagrams in order to obtain local results for the matching coefficients. 

\begin{figure}[t] %
\begin{center}
\includegraphics[width=.95\textwidth]{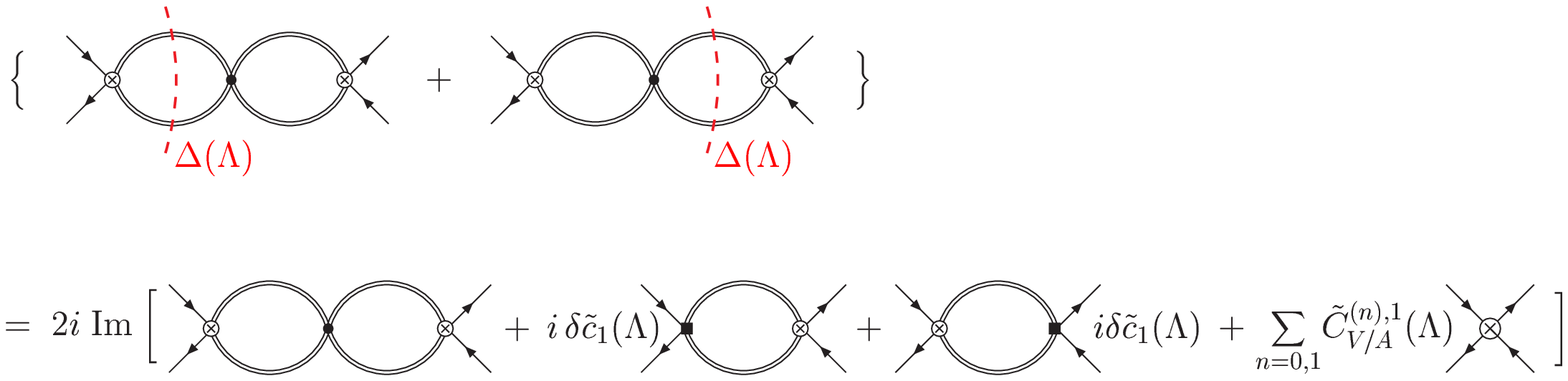}
\end{center}
\caption{
$\Ord{\alpha_s}$ matching condition for the Wilson coefficients of 
$e^+e^-$ forward scattering operators.
\label{figopetwo}}
\end{figure}

We start with the ${\cal O}(\alpha_s)$ contribution to the NRQCD cross section
that arises from the insertion of one Coulomb potential. The 
diagrams to be considered are depicted on the LHS of the equality shown in
Fig.~\ref{figopetwo}. Following the
approach of Sec.~\ref{subsectionphasespacematchingNRQCD}, the result can be cast
into the form
\begin{align}
\label{eq:sigmaNRQCDas1generic}
\sigma_{\rm NRQCD}^{i,{\cal O}(\alpha_s)}  \, = \,&
N_c \, \big((C_{V,1}^{\rm Born})^2 + (C_{A,1}^{\rm Born})^2 \big) 
\frac{m_t^3\Gamma_t^2}{2\pi^3}\!\!\! 
\int \limits_{\tilde
  \Delta(\Lambda)} \!\! dt_1 dt_2 
\frac{\sqrt{m_t E - \frac12(t_1+t_2)}}{(t_1^2
  + m_t^2\Gamma_t^2)(t_2^2 + m_t^2 \Gamma_t^2)}\, \Delta^i(t_1,t_2)\,
+\, \mbox{c.c.}
\,,
\end{align}
where in comparison to Eq.~(\ref{eq:sigmaNRQCDas0generic}) we have to add the
complex conjugate expressions because the computation involves a loop subgraph
either to the left or to the right of the cut. For the Coulomb potential it is
straightforward to derive 
\begin{align}
\Delta^1(t_1,t_2) \, = \, {\cal D}\bigg(\sqrt{m_t E-\frac{t_1+t_2}{2}}\bigg)
\,,
\end{align}
where ($q=|\bmq|$, $\CV_c^{(s)}=-4\pi C_F\alpha_s\equiv -4\pi a$)
\begin{align}
{\cal D}(q) & = 
\tilde\mu^{2\epsilon}\int\!\!\frac{d^d\bmr}{(2\pi)^d}\,
\frac{i}{\frac{E}{2}+ r_0 - \frac{{\mathbf r}^{2}}{2m_t} +
  i\frac{\Gamma_t}{2}}\,
\frac{i}{\frac{E}{2}- r_0 - \frac{\bmr^2}{2m_t} +
  i\frac{\Gamma_t}{2}}\, 
\frac{-i\,\CV_c^{(s)}}{(\bmr-\bmq)^2}
\nonumber\\&=
\,i\,a\,\frac{m_t}{2q}\,
  \ln \frac{m_t v+ q}{m_t v- q}\,.
\label{Dc}
\end{align}
The term ${\cal D}$ is just the vertex diagram for $t\bar t$ production with one
insertion of the Coulomb potential. Using Eq.~(\ref{t1t2limits}) for the
boundaries of the phase space integration we obtain the result
\begin{align}
\sigma_{\rm NRQCD}^{i,{\cal O}(\alpha_s)}  \, = \,&
2\,N_c \, \big((C_{V,1}^{\rm Born})^2 + (C_{A,1}^{\rm Born})^2 \big) \, \frac{m_t^2}{4\pi}\,a\,\bigg[- \mbox{Im}
          \big[\ln(-i\,v)\,\big]  
-2 \frac{m_t\Gamma_t}{\Lambda^2} 
\nonumber\\[2mm] &{}\qquad 
-  \frac{8\sqrt{2}}{3\pi}\,\frac{m_t^2\Gamma_t
          }{\Lambda^3}\, \mbox{Re}\big[\,i\,v\,\big]
+  \Ord{v^4\frac{m_t^4}{\Lambda^4}} 
\bigg]\,.
\label{eq:lowestcoulomb}
\end{align}

\begin{figure}[t]
  \begin{center}
  \includegraphics[width=0.49\textwidth]{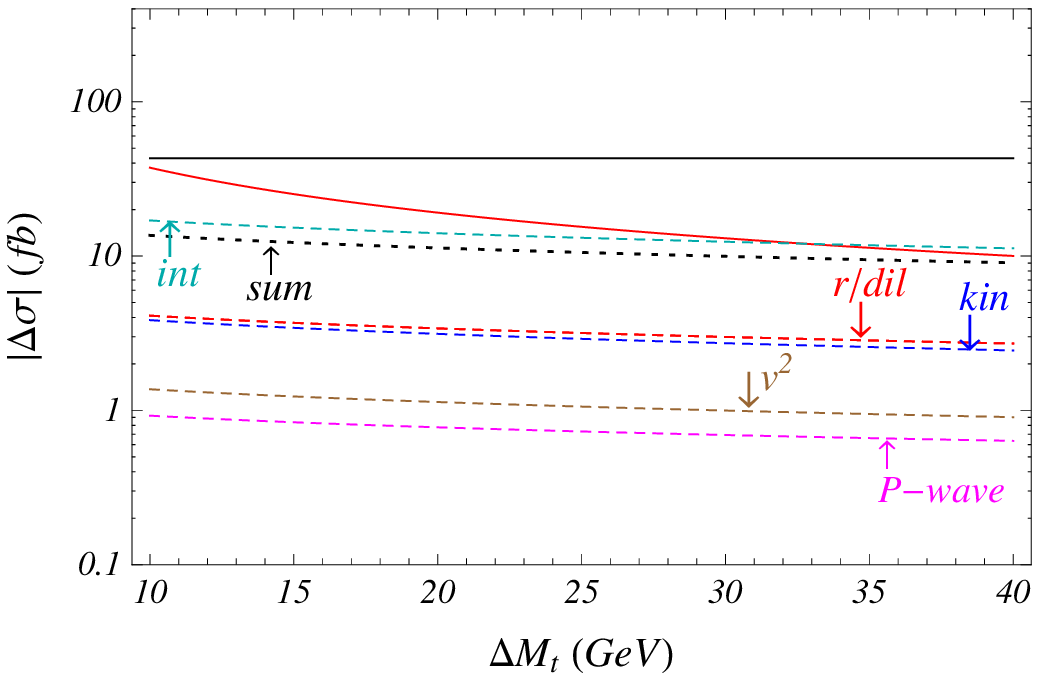}\;\;
  \includegraphics[width=0.49\textwidth]{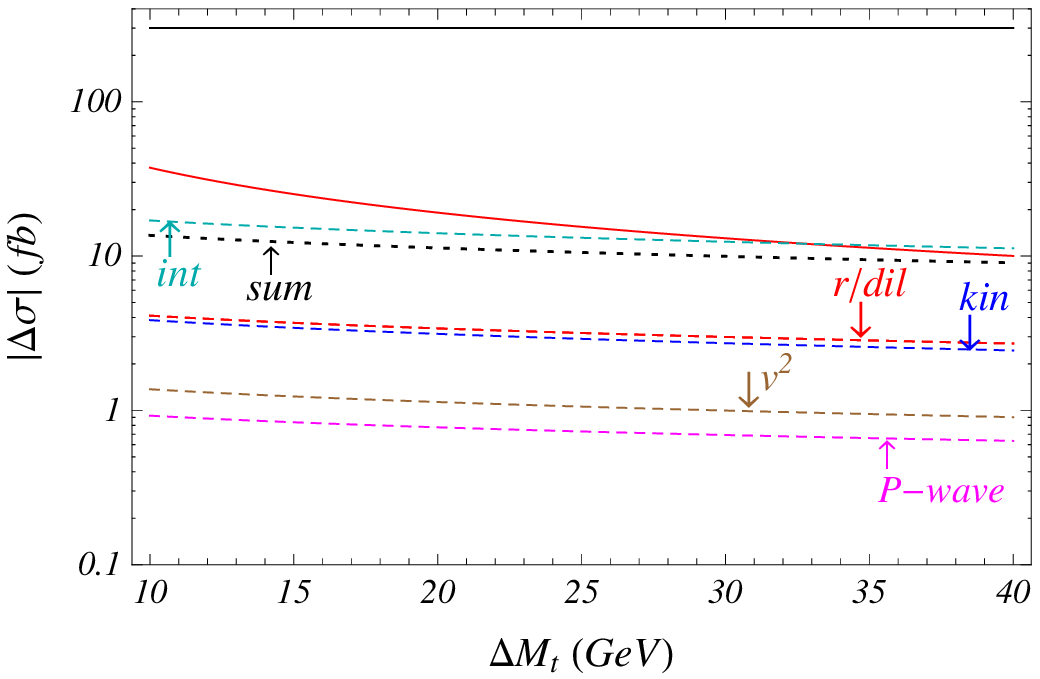}
  \caption{Absolute values of the contributions to the inclusive cross section
from the ${\cal O}(\alpha_s)$ phase space matching corrections 
$\tilde C_{V/A}^1(\Lambda)$  of Eq.~(\ref{tildeCVA1}) (red lines)
and 
${\tilde C_{V/A}^{i,1}}(\Lambda)$,
$i=\mbox{r/dil},\mbox{kin},v^2,\mbox{$P$-wave}, \mbox{int}$
(red,blue, brown, magenta and cyan dashed lines,
respectively),
listed in Eq.~(\ref{tildeCVA1v2}). The sum of the ${\tilde C_{V/A}^{i,1}}(\Lambda)$
contributions is shown as the dotted black line.
For comparison we also display 
the tree-level NRQCD cross
section without phase space matching contributions (black lines).
The left panel corresponds to $E=-5$ while for the right
$E=5$~GeV.
 The values chosen for the input parameters can be found in
Eq.~(\ref{parameters}), and $\alpha_s=0.1077$. We have used the energy-independent form  of the coefficients 
$C_{V/A,i}^{\rm Born}$ and $C_{V/A,i}^{(1),\rm Born}$ ($i=1,2,3$) given 
by Eqs.~(\ref{treematching}). }
  \label{fig:relatcoras}
  \end{center}
\end{figure}

\vskip 5mm \noindent
{\bf \underline {${\cal O}(\alpha_s)$ Matching for the forward
scattering operators:}} \\[2mm]
The first $\Lambda$-independent and non-analytic term in the brackets in
Eq.~(\ref{eq:lowestcoulomb}) is from the
well known ${\cal O}(\alpha_s)$ contribution of the nonrelativistic Coulomb
Green function, which contributes to the LL NRQCD cross section without cuts,
see Eq.~(\ref{deltaGCoul}). The second term $\propto \alpha_s
m_t\Gamma_t/\Lambda^2$ is local and contributes at NNLL order. Through the phase
space matching procedure using Eq.~(\ref{crosssection}), the second term
contributes to the imaginary Wilson coefficient $\tilde C_{V/A}$ of the
$(e^+e^-)(e^+e^-)$ forward scattering operators $\tilde{\cal O}_{V/A}$.
The result reads 
\begin{align}
{\tilde C_{V/A}^1}(\Lambda) & = \,
 2\,i\,N_c\,\big(C_{V/A,1}^{\rm Born}\big)^2\,
\frac{m_t^2}{4\pi}\,
\Big[-2 \,a\,\frac{m_t\Gamma_t}{\Lambda^2}
\Big]\,,
\label{tildeCVA1}
\end{align}
and constitutes an ${\cal O}(\alpha_s)$ correction to the NLL matching
contributions of Eq.~(\ref{tildeCVA0}). The determination of the ${\cal
  O}(\alpha_s)$ contributions to ${\tilde C_{V/A}^1}$ arising from insertions of
higher-dimension operators and potentials describing relativistic
$v^2$-corrections is carried out in analogy. Here we just quote the final
results and refer to Ref.~\cite{ReisserPhd} for details:
\begin{align}
{\tilde C_{V/A}^{r,1}}(\Lambda)
 & = \, 2\,i\,N_c\,
    \big(C_{V/A,1}^{\rm Born}\big)^2\,
    \frac{m_t^2}{4\pi}\,
    \left[-a\left(\ln\frac{m_t}{\Lambda} +\frac12
      +\frac12\ln2\right) \frac{\Gamma_t}{m_t} \right]\,,
\nonumber\\[2mm]
{\tilde C_{V/A}^{s,1}}(\Lambda) & = \, 0\,,
\nonumber\\[2mm]
{\tilde C_{V/A}^{\rm kin,1}}(\Lambda)
& = \, 2\,i\,N_c\,
    \big(C_{V/A,1}^{\rm Born}\big)^2\,
    \frac{m_t^2}{4\pi}\,
    \left[-a\left(\ln\frac{m_t}{\Lambda} +\frac38
      +\frac12\ln2\right) \frac{\Gamma_t}{m_t}
\right]\,,
\nonumber\\[4mm]
{\tilde C_{V/A}^{\rm dil,1}}(\Lambda)
& = \, 2\,i\,N_c\,
    \big(C_{V/A,1}^{\rm Born}\big)^2\,
    \frac{m_t^2}{4\pi}\,
    \left[\,a\left(\ln\frac{m_t}{\Lambda} +\frac12
      +\frac12\ln2\right) \frac{\Gamma_t}{m_t}
\right]\,,
\nonumber\\[4mm]
{\tilde C_{V/A}^{v^2,1}}(\Lambda)
& = \, 2\,i\,N_c\,
    2C_{V/A,1}^{\rm Born}C_{V/A,2}^{\rm Born}\,
    \frac{m_t^2}{4\pi}\,
    \left[-a\left(\ln\frac{m_t}{\Lambda} +\frac12
      +\frac12\ln2\right) \frac{\Gamma_t}{m_t}
\right]\,,
\nonumber\\[4mm]
{\tilde C_{V/A}^{\text{$P$-wave},1}}(\Lambda)
& = \, \frac{4}{3}\,i\,N_c\,\left(C_{V/A,3}^{\rm Born}\right)^2\,
    \frac{m_t^2}{4\pi}\,
    \left[-a\left(\ln\frac{m_t}{\Lambda} +\frac23
      +\frac12\ln2\right) \frac{\Gamma_t}{m_t}
\right]\,,
\nonumber\\[4mm]
{\tilde C_{V/A}^{\rm int,1}}(\Lambda)  
& = \,   2\,i\,N_c\,2\,C_{V/A,1}^{\rm Born}C_{V/A,1}^{\rm int}
    \frac{m_t^2}{4\pi}\,
    \left[-a\left(\ln\frac{m_t}{\Lambda} +\frac12
      +\frac12\ln2\right)
\right]\,, 
\label{tildeCVA1v2}
\end{align}
where the definitions of the coefficients
$C_{V/A,i}^{\rm Born}$ are given in Eqs.~(\ref{treematching}) and
(\ref{treematchingv2}). The coefficients  
${\tilde C_{V/A}^{r,1}}$ and ${\tilde C_{V/A}^{s,1}}$ come from one insertion of
the potentials ($\bmk = \bmp - \bmpp$)
\begin{align}
& \frac{\CV_r^{(s)}(\bmp^2+\bmpp^2)}{2m_t^2\bmk^2}
\quad \mbox{and}\quad
  \frac{\CV_s^{(s)}}{m_t^2}\,,
\label{v2potentials}
\end{align}
where we used  $\CV_r^{(s)}(\nu=1)= -4\pi a$ and 
${\cal V}_s^{(s)}(\nu=1)=4\pi a/3$,
respectively. The origin of the other coefficients arises from insertions of
the kinetic energy correction (kin), the time dilation correction (dil), the
$v^2$-suppressed $S$-wave current ${\cal O}_{\bmp,2}$ ($v^2$), the $P$-wave current
${\cal O}_{\bmp,3}$ ($P$-wave) and the interference coefficient (int), as
explained already after Eq.~(\ref{Deltav2def}).
The spin-dependent, momentum-independent potential $\CV_s^{(s)}/m_t^2$ does not
lead to a contribution at the order we consider.%
\footnote{Since the potential $\CV_s^{(s)}/m_t^2$ is momentum-independent, its
  insertion leads to a factorized expression for the correction to the Green
  function that can only contribute to the phase space matching
  coefficient $\delta\tilde c_1^{s,1}$ of the $S$-wave $t\bar t$ current 
  discussed in the following subsection.}
In analogy to our examinations for $\alpha_s=0$ we find that the phase space
matching contributions originating from the relativistic insertions also
contribute at NNLL order and are power-counting breaking. They also feature a 
relative factor $\Lambda^2/m_t^2$ with respect to the terms in 
${\tilde C_{V/A}^1}(\Lambda)$ of Eq.~(\ref{tildeCVA1}).
Through the numerical analysis of the ${\tilde C_{V/A}^{i,1}}(\Lambda)$ coefficients
which we carry out in Fig.~\ref{fig:relatcoras}, we find a similar
situation as for the case of the corresponding tree-level matching coefficients
${\tilde C_{V/A}^{i,0}}(\Lambda)$ studied in Sec.~\ref{subsectionphasespacematchingNRQCD}:
The phase space matching contributions in 
${\tilde C_{V/A}^1}(\Lambda)$ of Eq.~(\ref{tildeCVA1}) are about an order of
magnitude larger than the ones from the relativistic corrections in
Eqs.~(\ref{tildeCVA1v2}), which also
cancel each other partly due to their different signs. 
The cancellation is, however, less effective as for $\alpha_s=0$ due to the
additional contribution coming from the ${\cal V}_r$ potential.
Again, this does not apply to the contributions from the interference
corrections in ${\tilde C_{V/A}^{\rm int,1}}(\Lambda)$, which are comparable 
to ${\tilde C_{V/A}^1}(\Lambda)$ for $\Delta M_t\gsim 25$~GeV. 
As explained in
Sec.~\ref{subsectionphasespacematchingNRQCD} before, this behavior is related to the large size of the interference
coefficients $C_{V/A,1}^{\rm int}$ and not to power-counting breaking effects.
As far as the size of higher order relativistic corrections are concerned we
thus also find a good perturbative behavior for the ${\cal O}(\alpha_s)$
corrections to the phase space matching coefficients 
$\tilde C_{V/A}(\Lambda)$.

\vskip 5mm \noindent
{\bf \underline {${\cal O}(\alpha_s)$ Matching for the 
$t\bar t$ currents:}}\\[2mm] 
\noindent
Let us now discuss the third term in the brackets on the RHS of
Eq.~(\ref{eq:lowestcoulomb}) which is 
$\propto\mbox{Re}[iv]$. 
It is non-analytic in the energy and can therefore
not contribute to the phase space matching coefficients of the
$(e^+e^-)(e^+e^-)$ forward scattering operators $\tilde{\cal
  O}_{V/A}^{(n)}$. This term contributes at 
N${}^3$LL order and is also of theoretical interest as it illustrates how  
the phase space matching procedure is carried out at higher orders. We recall
that the phase space matching 
follows the common rules of matching computations, where the matching involving
diagrams with a higher number of loops first requires the matching of operators 
relevant for subdiagrams in order to deal with all terms that are non-analytic in
the external momenta. To deal with the non-analytic term
$\propto\mbox{Re}[iv]$ in Eq.~(\ref{eq:lowestcoulomb}) we need to determine the
imaginary phase space matching 
term $i\,\delta\tilde c_1(\Lambda)$ contained in the Wilson coefficient 
$C_{V/A,1}$ of the dominant $S$-wave $t\bar t$ current operators ${\cal
  O}_{V/A,\bmp,1}$, see Eqs.~(\ref{wilsoncoeff}). 

\begin{figure}[t] %
\begin{center}
\includegraphics[width=.8\textwidth]{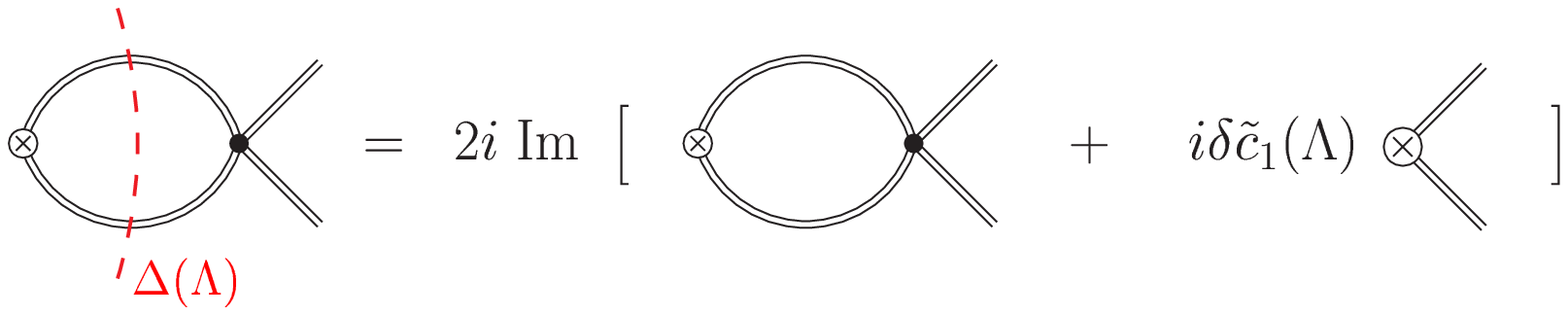}
\end{center}
\caption{
$\Ord{\alpha_s}$ matching condition for the Wilson coefficient of a
$t\bar t$ production current. The black dot represents an insertion of a
potential operator.
\label{opencondition}}
\end{figure}

In the following we compute the contribution to the phase space matching term
$i\,\delta\tilde c_1(\Lambda)$ generated by an insertion of one Coulomb
potential, as illustrated 
in Fig~\ref{opencondition}. Suppressing the external (anti)top quark
spinors and the lepton current the amplitude for the cut diagram reads 
\begin{eqnarray}
{\cal A}^{\rm cur}(\Lambda) & = & \int \limits_{\Delta(\Lambda)} \!\!\!
                         \frac{d^4p}{(2\pi)^4}
 \frac{\Gamma_t}{\left(\frac{E}{2} + p_0 - \frac{\bmp^2}{2m_t} +
                         i\frac{\Gamma_t}{2} \right)
                 \left(\frac{E}{2} + p_0 - \frac{\bmp^2}{2m_t} -
                         i\frac{\Gamma_t}{2} \right)}\nonumber\\
             &&{}\times
 \frac{\Gamma_t}{\left(\frac{E}{2} - p_0 - \frac{\bmp^2}{2m_t} +
                         i\frac{\Gamma_t}{2} \right)
                 \left(\frac{E}{2} - p_0 - \frac{\bmp^2}{2m_t} -
                         i\frac{\Gamma_t}{2} \right)}
  \, \frac{-i\,\CV_c^{(s)}}{(\bmp-\bmpp)^2}\,,\hspace{1cm}
\label{currentabc}
\end{eqnarray}
where $\pm\bmpp$ is the 3-momentum of the top and the antitop
quark in the c.\,m.\ frame, respectively,
and $\CV_c^{(s)}=-4\pi C_F\alpha_s\equiv -4\pi a$.
For the determination of the matching condition for the $t\bar t$ current 
it is sufficient to consider on-shell external top quarks, and we therefore set
$\bmpp^2 = m_t E$, $E>0$. 
By carrying out the angular integration and changing variables we
obtain an intermediate result of the form~(\ref{eq:sigmaNRQCDas0generic}) with
($|\bmp|=(m_t E -\tfrac{1}{2}(t_1+t_2))^{1/2}$)
\begin{align}
\Delta^{\rm cur}(t_1,t_2) & = \, -i\,\CV_c\,\frac{2\pi}{|\bmp|\sqrt{m_tE}}\,\ln
\left|\frac{\sqrt{m_tE}+|\bmp|}{\sqrt{m_tE}-|\bmp|}\right|
\,.
\end{align} 
The evaluation of the remaining integral with the invariant mass 
restrictions~(\ref{t1t2limits}) yields
\begin{align}
{\cal A}^{\rm cur}(\Lambda)&\, = \,  i\,a\,\Bigg[
      \sqrt{\frac{m_t}{E}}\,\mbox{Re}\left[\ln\frac{m_tv+\sqrt{m_tE}} 
                                              {m_tv-\sqrt{m_tE}}\right] 
       -\frac{8\sqrt{2}}{3\pi}\frac{m_t^2\Gamma_t}{\Lambda^3}
+ \Ord{v^4\frac{m_t^5}{\Lambda^5}}
      \Bigg]\,.
\end{align}
The first term in the brackets is just twice the imaginary part of the vertex diagram
${\cal D}$ defined in Eq.~(\ref{Dc}) without phase space restrictions 
\begin{align}
{\cal A}^{\rm cur}(\infty) & = \,
2\,i\,\im\,\Big[ {\cal D}\left(\sqrt{m_tE}\right)\Big]\,,
\label{idopt}
\end{align}
and
corresponds to the first term on the RHS of the matching relation in
Fig.~\ref{opencondition}. The second term, on the other hand, gives a
contribution to the matching term $i \delta\tilde c_1(\Lambda)$, and we obtain
\begin{align}
i\,\delta\tilde c_1^1(\Lambda) &=\,
\frac12\left({\cal A}^{\rm cur}(\Lambda)- 
{\cal A}^{\rm cur}(\infty) \right)
\, = \, -i\,a
\frac{4\sqrt{2}}{3\pi}\frac{m_t^2\Gamma_t}{\Lambda^3} \,,
\label{leadingswavecurrent}
\end{align}
where the factor $1/2$ is to compensate the factor $2$ that appears in the
optical theorem, see Fig.~\ref{opencondition}. It is now straightforward to
check that using the matching relation of Fig.~\ref{figopetwo} the phase space
matching term $i\,\delta\tilde c_1^1(\Lambda)$ in the Wilson coefficient of the
$t\bar t$ current operators  ${\cal O}_{V/A,\bmp,1}$ indeed accounts for
the non-analytic term $\propto\mbox{Re}[iv]$ in Eq.~(\ref{eq:lowestcoulomb}). 
This demonstrates the consistency of the phase space matching procedure at the
loop level. We note that the same coefficient is obtained
for the respective annihilation currents. Thus the imaginary phase space
matching coefficients are not affected by the hermitian conjugation of the
current operators in Eq.~(\ref{currentlagrangian}).

The determination of the ${\cal
  O}(\alpha_s)$ contributions to 
$i\,\delta\tilde c_1(\Lambda)$ arising from insertions of
higher-dimension operators and potentials describing relativistic
$v^2$-corrections is carried out in analogy. Here we just quote the final
results and refer again to Ref.~\cite{ReisserPhd} for details:
\begin{align}
i\,\delta \tilde c_1^{r,1}(\Lambda)&=
-i\,a \frac{\sqrt{2}}{\pi}\frac{\Gamma_t}{\Lambda}  
\,,&
i\,\delta \tilde c_1^{s,1}(\Lambda)&=
i\,a \frac{4\sqrt{2}}{3\pi}\frac{\Gamma_t}{\Lambda}  
\,,&
i\,\delta \tilde c_1^{\rm kin,1}(\Lambda)&=
-i\,a \frac{7}{4\sqrt{2}\,\pi}\frac{\Gamma_t}{\Lambda} 
\nonumber\\[2mm]
i\,\delta \tilde c_1^{\rm dil,1}(\Lambda)&=
i\,a\frac{\sqrt{2}}{\pi}\frac{\Gamma_t}{\Lambda}  
\,,&
i\,\delta \tilde c_1^{v^2,1}(\Lambda)&=
i\,a\frac{\sqrt{2}}{3\pi}\frac{\Gamma_t}{\Lambda}   
\,,&
i\,\delta\tilde c_1^{\rm int,1}(\Lambda)&=
-i\,a \frac{2\sqrt{2}}{\pi}\frac{m_t}{\Lambda} 
\,. 
\label{deltacint1}
\end{align}
Note that the insertion of the current operators
$\CO_{V/A,\bmp,2}$ gives the mixing term $i\delta \tilde c_1^{v^2,1}$
contributing to the matching condition of 
$\CO_{V/A,\bmp,1}$. An analogous mixing does not arise from insertions of the
$P$-wave current $\CO_{V/A,\bmp,3}$ due to angular momentum conservation.
All terms shown in Eq.~(\ref{deltacint1}) are of relative order
$\Lambda^2/m_t^2$ with respect to the one for the dominant $S$-wave
current in Eq.~(\ref{leadingswavecurrent}).
In the case of the interference correction this is also true because 
$C_{V/A,1}^{\rm int} \sim C_{V/A,1}^{\rm Born} \Gamma_t/m_t$.

\subsection{Perturbative Expansion and Higher Order Corrections}
\label{sec:expansiontest}

In Secs.~\ref{subsectionphasespacematchingNRQCD} and \ref{sec:potentials} we
have computed the ${\cal O}(\alpha_s^0)$ and ${\cal O}(\alpha_s)$ phase space
matching contributions to the $(e^+e^-)(e^+e^-)$ forward scattering
operators. We have completed all contributions at
NNLL order, and we have demonstrated that power-counting breaking terms do not
spoil the nonrelativistic expansion. For the ${\cal O}(\alpha_s)$ contributions
we have also shown that the Wilson coefficients of the $(e^+e^-)(t\bar t)$
current operators receive phase space matching terms, which contribute
at N${}^3$LL order. In this section we examine the $\alpha_s$-series of the
phase space matching contributions. 
To render the analysis more transparent it is useful to distinguish the phase
space matching contributions according to where they originate in the computation
of the NRQCD cross section $\sigma_{\rm NRQCD}(\Lambda)$. 

\vskip 5mm \noindent
{\bf \underline {Contributions related to the 
imaginary part of the Coulomb Green function}}\\[2mm] 
\noindent
We first analyze the phase space matching contributions originating from the
imaginary part of the Coulomb Green function $\mbox{Im}[G^c]$ determined in 
Eqs.~(\ref{tildeCVA0}), (\ref{tildeCVA01}), (\ref{tildeCVA1}) and
(\ref{leadingswavecurrent}). 
Since the imaginary part of the Coulomb Green function constitutes the leading
order contribution of the factorization formula~(\ref{crosssection}), these
phase space matching terms represent the numerically dominant contributions.
We define the NRQCD cross section associated to the imaginary part of the
Coulomb Green function as
\begin{align}
\sigma_{\rm NRQCD}^{{\rm Im}[G^c]}(\Lambda,\nu) & = \,
N_c \, \big((C_{V,1}^{\rm Born})^2 + (C_{A,1}^{\rm Born})^2 \big) \,
\Gamma_t^2 \!\!\!\int\limits_{\Delta(\Lambda)
}\!\!\frac{d^4p}{(2\pi)^4}\,  
\frac{(2m_t)^4}{(t_1^2+m_t^2\Gamma_t^2)(t_2^2+m_t^2\Gamma_t^2)}
\,\Big |f_{v,m_t,\nu}(|\bmp|)\Big |^2,
\label{sigmaImG}
\end{align}
where $t_1$ and $t_2$ are the invariant mass variables of Eq.~(\ref{t12def}) and
$f_{v,m_t,\nu}$ is the Coulomb vertex factor. The Coulomb vertex factor can
be written in the form
\begin{align}
f_{v,m_t,\nu}(|\bmp|) & \, = \, \left[\frac{|\bmp|^2}{m_t} - (E+i\Gamma_t)\right]\,\tilde
G^0_{v,m_t,\nu}(0,|\bmp|)\,.
\label{vertexfunction}
\end{align}
Here $\tilde G^0_{v,m_t,\nu}(0,|\bmp|)$ is the partially Fourier transformed Coulomb
Green function with the first argument in position space at $\bmx=0$ and the
second in momentum space with the momentum $\bmp$. Because of $\bmx=0$ there is
no dependence on the direction of $\bmp$. At LL order, i.e. accounting only for
iterations of the LL Coulomb potential of Eq.~(\ref{Lpot}),
$\tilde G^0_{v,m_t,\nu}(0,|\bmp|)$ is known analytically:
\begin{align}
\tilde G^0_{v,m_t,\nu}(0,|\bmp|) &\, =  -\frac{i\, m_t}{4 k |\bmp|}\frac{1}{1-\lambda}  
 \bigg[\,_2F_1\left(2,1;2-\lambda;\frac{1}{2}\left(1+\frac{i\,
     |\bmp|}{k}\right)\right)\nonumber\\[2mm]
&{}\hspace{3cm} -\,
   _2F_1\left(2,1;2-\lambda;\frac{1}{2} \left(1-\frac{i\,
         |\bmp|}{k}\right)\right)\bigg]
\,,
\label{G0pLL}
\end{align} 
where $k\equiv \sqrt{-m_t(E+i\Gamma_t)}$ and 
$\lambda\equiv C_F\alpha_s(m_t\nu)\, m_t/2k$, and 
$_2F_1(a,b;c;z)$ is the hypergeometric function.
At NLL and NNLL order we use numerical results for the form factor
$f_{v,m_t,\nu}$ using the computational techniques we also employ for the
determination of the Coulomb Green function at the respective orders, see
Refs.~\cite{Strassler:1990nw,Jezabek:1992np}. These numerical methods have been
applied and thoroughly tested before for predictions of the top three-momentum
distribution, see e.g. Refs.~\cite{Jezabek:1992np,Hoang:1999zc}. Without
phase space restrictions, i.e.\ for $\Lambda=\infty$, Eq.~(\ref{sigmaImG})
reduces to the form
\begin{align}
\sigma_{\rm NRQCD}^{{\rm Im}[G^c]}(\infty) & = \,
2N_c \, \big((C_{V,1}^{\rm Born})^2 + (C_{A,1}^{\rm Born})^2 \big) \,
{\rm Im}\Big[
G^c(a,v,m_t,\nu)
\Big]\,.
\label{sigmaImGinfty}
\end{align}

\begin{figure}[t]
  \begin{center}
  \includegraphics[width=0.49\textwidth]{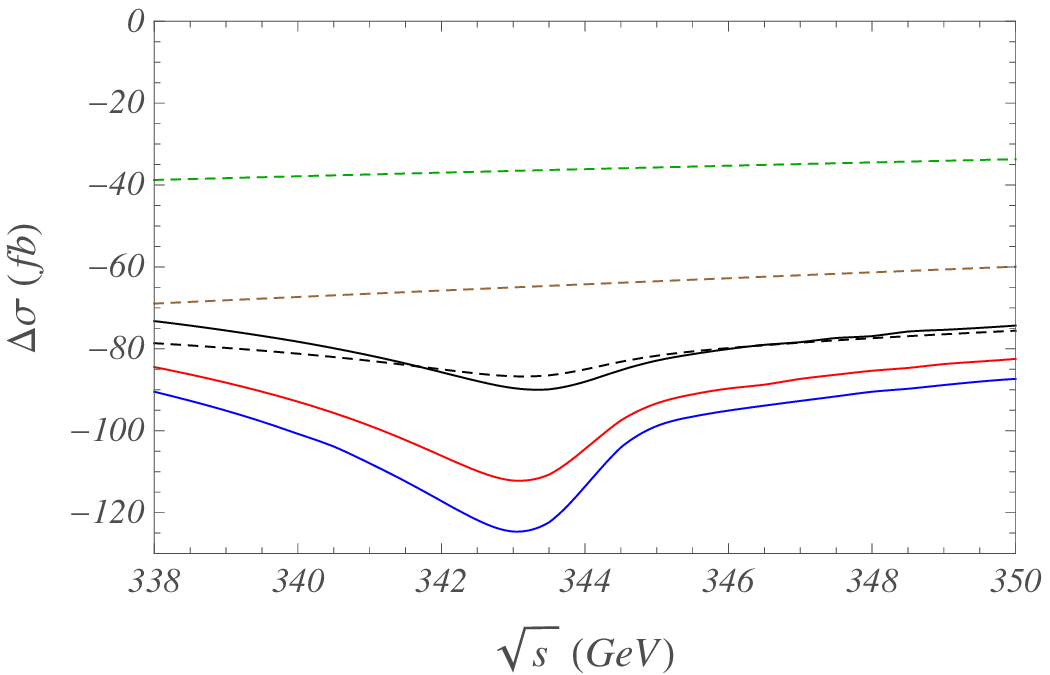}\;\;
  \includegraphics[width=0.49\textwidth]{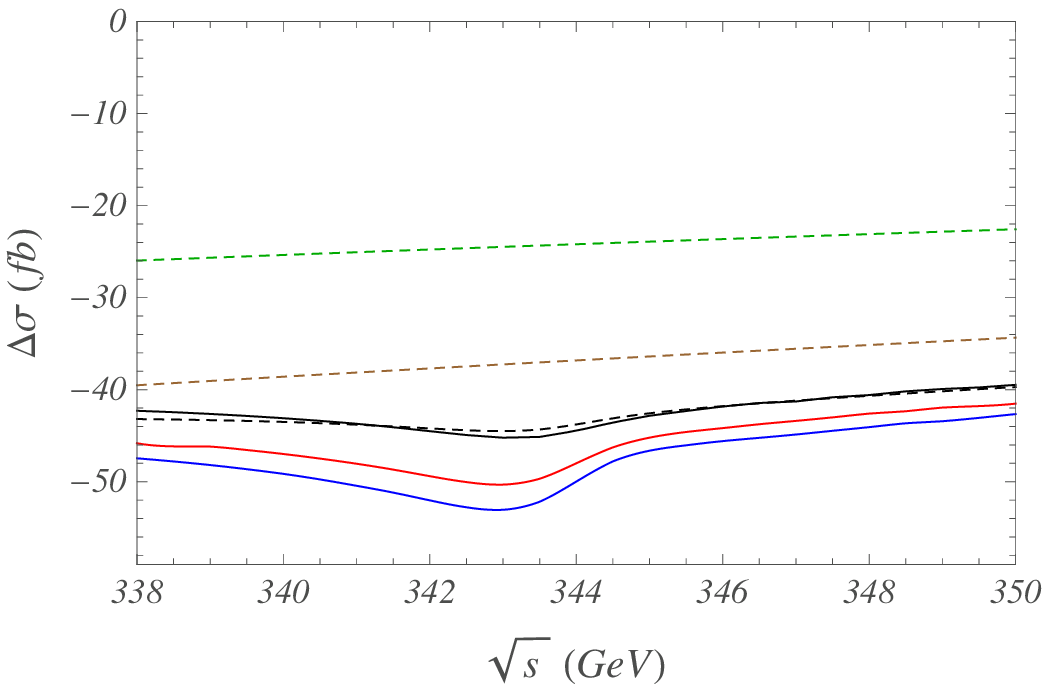}
  \caption{Contributions to the inclusive cross section from the
NLL, NNLL and N$^{3}$LL order phase space matching corrections 
in $\Delta\sigma_{\rm psm}^{{\rm Im}[G^c]}$
 (green, brown
and black dashed lines, respectively) as a function of the total center of mass 
energy for $\Delta M_t=15$ (left) and $\Delta M_t=35$~GeV (right).
The solid lines correspond to the phase space matching contributions 
$\Delta\sigma^{{\rm Im}[G^c]}$
obtained from the imaginary part of the exact cut Coulomb Green function at 
LL, NLL and NNLL (black,
red and blue, respectively).}
  \label{fig:ImG}
  \end{center}
\end{figure}

In Fig.~\ref{fig:ImG} we display 
$\Delta\sigma^{{\rm Im}[G^c]}=\sigma_{\rm NRQCD}^{{\rm Im}[G^c]}(\Lambda)-
\sigma_{\rm NRQCD}^{{\rm Im}[G^c]}(\infty)$ for $\Delta M_t=15$~GeV (left panel)
and $\Delta M_t=35$~GeV (right panel) using the LL (black solid line), NLL (red
solid line) and NNLL (blue solid line) Coulomb Green functions,
and the 1S top mass scheme~\cite{Hoang:1999zc,Hoang:2001rr}
with $m_t=172$~GeV. The strong
coupling is evaluated at the hard scale $\nu=1$, $\alpha_s(m_t)=0.1077$ and the other input parameters are given in Eq.~(\ref{parameters}), except for 
the electromagnetic coupling constant, which is also taken at the hard scale  $\alpha_{\rm qed}(m_t)=1/125.9$. For simplicity we
neglect here and in the following examinations of Sec.~\ref{sec:expansiontest}
the hard QCD corrections to the $t\bar t$ current matching coefficient, i.e.\ we
set $c_1(\nu=1)=1$. The lines for $\Delta\sigma^{{\rm Im}[G^c]}$ represent the
effects of the invariant mass cuts including iterations of the Coulomb potential
to all orders. From our examinations in Sec.~\ref{sec:potentials}
we know that $\Delta\sigma^{{\rm Im}[G^c]}$ exhibits hard and soft contributions
that can only be separated by the phase space matching procedure. Thus
$\Delta\sigma^{{\rm Im}[G^c]}$ unavoidably contains large logarithmic terms for
any choice of the renormalization scale. The size of the higher order
corrections discussed in the following should therefore
be interpreted with some care as they might not reflect the quality of the
$\alpha_s$-expansion when all logarithms are properly summed up. Despite this
fact we see that the effects of 
including the ${\cal O}(\alpha_s)$ and ${\cal O}(\alpha_s^2)$ corrections to the
Coulomb potential in $\Delta\sigma^{{\rm Im}[G^c]}$ show good convergence. For 
$\Delta M_t=15$~GeV ($35$~GeV) the  ${\cal O}(\alpha_s^2)$
corrections to the Coulomb potential (difference between the red and blue solid
lines) lead to a shift between $-10$~fb 
($-4$~fb) and $-5$~fb ($-2$~fb). For $\Delta M_t=15$~GeV this exceeds our
theoretical precision aim (green dashed lines in the lower panel of
Fig.~\ref{fig:sigttbar}) for $\sqrt{s}-2m_t\lsim -2$~GeV. The shift caused by the 
${\cal O}(\alpha_s^2)$ corrections to the Coulomb potential, however, quickly
drops below the precision aim for $\Delta M_t>15$~GeV for all values of $E$, and
we conclude that phase 
space matching corrections related to insertions of the ${\cal O}(\alpha_s^2)$
corrections to the Coulomb potential, which contribute at N${}^4$LL order and
higher, do not have to be considered. 

We now examine $\Delta\sigma^{{\rm Im}[G^c]}$ computed from the phase space matching procedure carried out in
the previous sections. Up to the N${}^3$LL order it has the form
\begin{align}
\Delta\sigma_{\rm psm}^{{\rm Im}[G^c]} \, = &\,
\im \bigg[ \tilde C^{0}_V(\Lambda) + \tilde C^{0}_A(\Lambda) \bigg]_{\rm NLL} \,+\,
\im\bigg[ \tilde C^{1}_V(\Lambda) + \tilde C^{1}_A(\Lambda) \bigg]_{\rm NNLL} 
\nonumber\\[2mm]& \,+\,
\im\bigg[  \tilde C^{2}_V(\Lambda) + \tilde C^{2}_A(\Lambda) \, + \,
 \frac{E}{m_t}\Big(\tilde C^{(1),0}_V(\Lambda) + \tilde C^{(1),0}_A(\Lambda)\Big) 
\nonumber\\[2mm]& \quad\quad\quad +\,
4\,N_c\,\Big( ( C_{V,1}^{\rm Born} )^2
+ ( C_{A,1}^{\rm Born} )^2 \Big)
\,i\,\delta\tilde c_1(\Lambda)\,G^c(a,v,m_t,\nu)  \bigg]_{{\rm
   N}^3{\rm LL}}
\,,
\label{sigmaImGexp}
\end{align}
where we have indicated the NLL, NNLL and N${}^3$LL order corrections.
The results for $\tilde C^{0}_{V/A}$, $\tilde C^{1}_{V/A}$, 
$\tilde C^{(1),0}_{V/A}$ and $\delta\tilde c_1$ have been given in 
Eqs.~(\ref{tildeCVA0}), (\ref{tildeCVA1}), (\ref{tildeCVA01}), and 
(\ref{leadingswavecurrent}), respectively. The N${}^3$LL corrections depend on
the ($\overline{\rm MS}$ renormalized) real part of the Coulomb Green function  
which at this order can be replaced by its LL expression $G^0$, see
Eq.~(\ref{deltaGCoul}). Note that for actual NRQCD predictions the Green
function $G^c(a,v,m_t,\nu)$
in Eq.~(\ref{sigmaImGexp}) has to be evaluated with $\nu\sim \alpha_s$ to
properly sum large logarithms. As mentioned above, we set $\nu=1$ for the
following examinations.  
In Fig.~\ref{fig:ImG} the NLL and the NNLL approximations for 
$\Delta\sigma_{\rm psm}^{{\rm Im}[G^c]}$ are displayed as the green and brown
dashed lines, respectively. For $\Delta M_t=35$~GeV (right panel) we find that
the NNLL corrections are about half the size of the NLL contributions. For
$\Delta M_t=15$~GeV (left panel), where we expect a worse $\alpha_s$-expansion
due to the $\Lambda$-dependence of $\tilde C^{0}_{V/A}\sim \Gamma_t/\Lambda$ and
$\tilde C^{1}_{V/A}\sim a\, m_t\Gamma_t/\Lambda^2$, the NNLL corrections are
only about $15$\% smaller than the NLL contributions. The difference of 
$\Delta\sigma_{\rm psm}^{{\rm Im}[G^c]}$ at NNLL order and
$\Delta\sigma^{{\rm Im}[G^c]}$ accounting for the LL Coulomb potential (black
solid lines) is between $3$ and $7$~fb for $\Delta M_t=35$~GeV and between $7$
and $25$~fb for $\Delta M_t=15$~GeV. For  $\Delta M_t=15$~GeV this exceeds 
our theoretical precision aim, visualized by the green dashed line in the lower
panel of Fig.~\ref{fig:sigttbar}. The difference is even larger 
with respect to $\Delta\sigma^{{\rm Im}[G^c]}$ accounting for the NLL Coulomb
potential (red solid lines). It is therefore required to also account for the
full set of N${}^3$LL phase space matching contributions displayed in
Eq.~(\ref{sigmaImGexp}). Unfortunately, at this time the full expressions for 
$\tilde C^{2}_{V/A}$ are unknown. They get contributions from two insertions of
the leading Coulomb potential (contained in the LL Coulomb Green function) and
from one insertion of the ${\cal O}(\alpha_s)$
correction to the Coulomb Green function (contained in the NLL Coulomb Green
function). We have computed the contribution from 
two insertions of the Coulomb potential using the methods described in
Sec.~\ref{sec:potentials}. The result reads
\begin{align}
\tilde C^{2,V_cV_c}_{V/A}(\Lambda) = 2\,i\,N_c\,(C_{V/A,1}^{\rm Born})^2\,
\frac{m_t^2}{4\pi}\bigg[ a^2\,\frac{4\sqrt{2}}{3\pi}
\,\bigg(
\ln\Big(\frac{\mu^2}{\Lambda^2}\Big) -\frac{7}{3} - \frac{\pi^2}{4}+ \frac{2}{3}\ln 2
\bigg) \,\frac{m_t^2\Gamma_t}{\Lambda^3} \bigg]\,.
\label{tildeCVA2VcVc}
\end{align}
The logarithmic term is related to a NNLL order contribution to the imaginary 
anomalous dimension of $\tilde C_{V/A}(\Lambda,\nu)$. The fact that the
logarithm vanishes for $\mu=\Lambda\sim m_t$ reconfirms that the phase space 
matching contributions are hard effects.
Although we do not have the complete result for $\tilde C^{2}_{V/A}(\Lambda)$,
using instead the result of Eq.~(\ref{tildeCVA2VcVc}) allows us to compare
$\Delta\sigma_{\rm psm}^{{\rm Im}[G^c]}$ at N${}^3$LL order (black dashed lines)
with the numerical results for $\Delta\sigma^{{\rm Im}[G^c]}$ accounting for the
LL Coulomb Green function (black solid lines). We see that including the
N${}^3$LL phase space matching 
contributions leads to a considerably improved agreement with the Coulomb
resummed numerical results. For $\Delta M_t=15$~GeV and $35$~GeV the
difference is always smaller than $5$~fb, except when $E<-5$~GeV for $\Delta
M_t=15$~GeV. Since this is acceptable for our precision aim, we believe that the
full set of N${}^3$LL phase space matching corrections should be adequate for the
precision expected at a future linear collider. 
Since the full result for $\tilde C^{2}_{V/A}(\Lambda)$ is unknown, we use for
the time being as a substitute for the N${}^3$LL order terms in 
$\Delta\sigma_{\rm psm}^{{\rm Im}[G^c]}$ 
the numerical Coulomb-resummed expression of Eq.~(\ref{sigmaImG}), evaluated 
with the NLL Coulomb vertex factor at the hard scale for $\alpha_s$ minus the
NLL and NNLL order terms of Eq.~(\ref{sigmaImGexp}):
\begin{align}
(\Delta\sigma_{\rm psm}^{{\rm Im}[G^c]})_{{\rm N}^3{\rm LL}} \, = &\,
\sigma_{\rm NRQCD}^{{\rm Im}[G^c]}(\Lambda,1)
\, - \,\im \bigg[ \tilde C^{0}_V(\Lambda) + \tilde C^{0}_A(\Lambda) \bigg]_{\rm NLL} 
\, - \,
\im\bigg[ \tilde C^{1}_V(\Lambda) + \tilde C^{1}_A(\Lambda) \bigg]_{\rm NNLL}
\,.
\label{sigmaImGexpN3LL}
\end{align}

\vskip 5mm \noindent
{\bf \underline {Contributions related to the real part 
of the Coulomb Green function}}\\[2mm] 
\noindent
We now analyze the phase space matching contributions related to the real part
of the Coulomb Green function. These phase space matching contributions are
proportional to the type-1 imaginary Wilson coefficients $i C_{V/A,1}^{\rm int}$
in Eqs.~(\ref{CintV}) and (\ref{CintA}), which describe the interference of the
$e^+e^-\to t\bar t\to b\bar b W^+W^-$ diagram with diagrams for 
$e^+e^-\to b\bar b W^+W^-$ with only either  $t$ or $\bar t$ at intermediate
stages. As we have shown in Sec.~\ref{subsectionphasespacematchingNRQCD}, the
interference effects cause the largest phase space matching contributions among
the ${\cal O}(v^2)$ relativistic corrections in the factorization 
formula~(\ref{crosssection}), and we therefore examine them separately. In
analogy to the previous section we first define the Coulomb-resummed NRQCD cross
section with invariant mass restrictions arising from the real part of the
Coulomb Green function: 
\begin{align}
\sigma_{\rm NRQCD}^{{\rm Re}[G^c]}(\Lambda) & = \,
2N_c\,\big(C_{V,1}^{\rm Born}C_{V,1}^{\rm int}+C_{A,1}^{\rm
    Born}C_{A,1}^{\rm int}\big)\,
\Gamma_t^2\!\int\limits_{\Delta(\Lambda)
}\!\!\frac{d^4p}{(2\pi)^4}\,  
\frac{(2m_t)^4}{(t_1^2+m_t^2\Gamma_t^2)(t_2^2+m_t^2\Gamma_t^2)}\nonumber\\
&\qquad{} \times \left[-\frac{t_1+t_2}{2m_t\Gamma_t}\,
  \re\left[f_{v,m_t,\nu}(|\bmp|)\right] - 
  \im\left[f_{v,m_t,\nu}(|\bmp|)\right] \right]
\,.
\label{sigmaReG}
\end{align}
Since the interference contributions are related to ${\cal O}(v^2)$ operator
insertions in the factorization theorem~(\ref{crosssection}), we only consider
the vertex factor in the LL approximation 
as given in Eq.~(\ref{vertexfunction}). The expression in the brackets involves
the Coulomb-resummed generalization of the function $\Delta^{\rm int,0}$ given
in Eqs.~(\ref{Deltav2def}).
The corresponding phase space matching contributions at N${}^3$LL order can be
derived from Eq.~(\ref{crosssection}) and read
\begin{align}
\Delta\sigma_{\rm psm}^{{\rm Re}[G^c]} & \, = \,
\im \bigg[ \tilde C^{{\rm int},0}_V(\Lambda) + \tilde C^{{\rm int},0}_A(\Lambda) \bigg]_{\rm NLL} \,+\,
\im \bigg[ \tilde C^{{\rm int},1}_V(\Lambda) + \tilde C^{{\rm int},1}_A(\Lambda) \bigg]_{\rm NNLL} 
\nonumber\\[2mm]& \,+\,
\im \bigg[  \tilde C^{{\rm int},2}_V(\Lambda) + \tilde C^{{\rm int},2}_A(\Lambda) \,+ \,
 \frac{E}{m_t}\Big(\tilde C^{(1),{\rm int},0}_V(\Lambda) + 
\tilde C^{(1),{\rm int},0}_A(\Lambda) \Big)
\nonumber\\[2mm]& \quad\quad\quad+\,  
4\,N_c\,
\Big( C_{V,1}^{\rm Born}C_{V,1}^{\rm int} 
+  C_{A,1}^{\rm Born} C_{A,1}^{\rm int}  \Big)
\,i\,\delta\tilde c_1^{{\rm int},1}(\Lambda)\,G^c(a,v,m_t,\nu)  \bigg]_{{\rm
   N}^3{\rm LL}}
\,,
\label{sigmaReGexp}
\end{align}
where we have again indicated the NLL, NNLL and N${}^3$LL order corrections.
The results for the $\tilde C^{{\rm int},0}_{V/A}$, $\tilde C^{{\rm int},1}_{V/A}$, 
$\tilde C^{(1),{\rm int},0}_{V/A}$ and $\delta\tilde c_1^{{\rm int},1}$ have
been given in  Eqs.~(\ref{relativisticresults1}), 
(\ref{tildeCVA1v2}), 
(\ref{relativisticresults2}) and 
(\ref{deltacint1}), respectively. The results for $\tilde C^{{\rm int},2}_{V/A}$
are currently unknown. In the following analyses we neglect them, i.e. we set them
to zero. 
\begin{figure}[t]
  \begin{center}
  \includegraphics[width=0.49\textwidth]{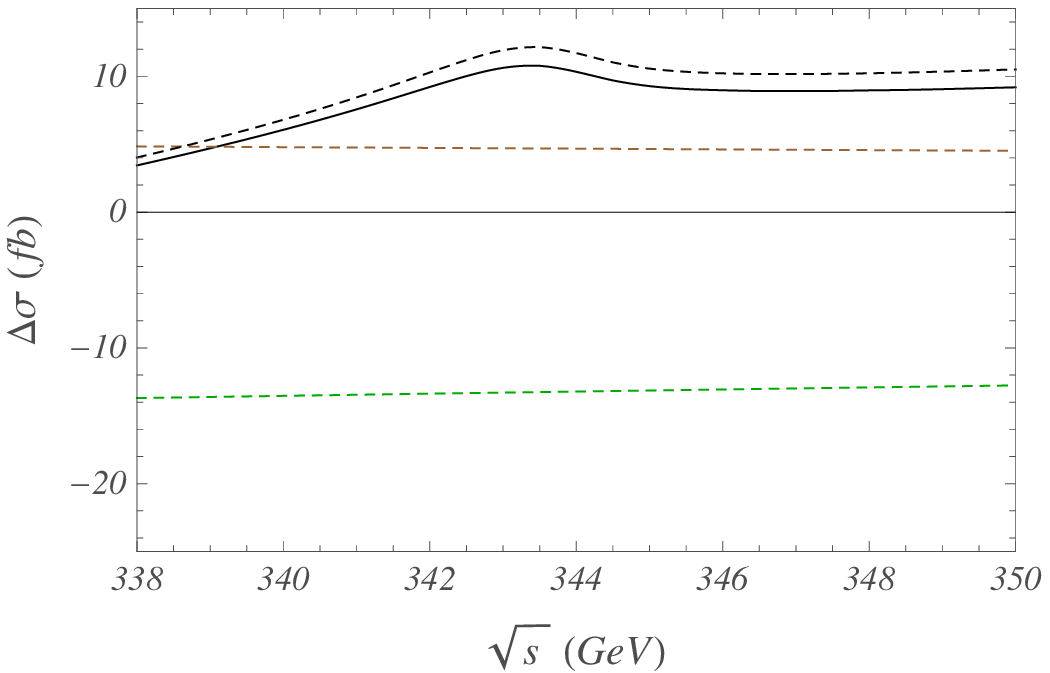}\;\;
  \includegraphics[width=0.49\textwidth]{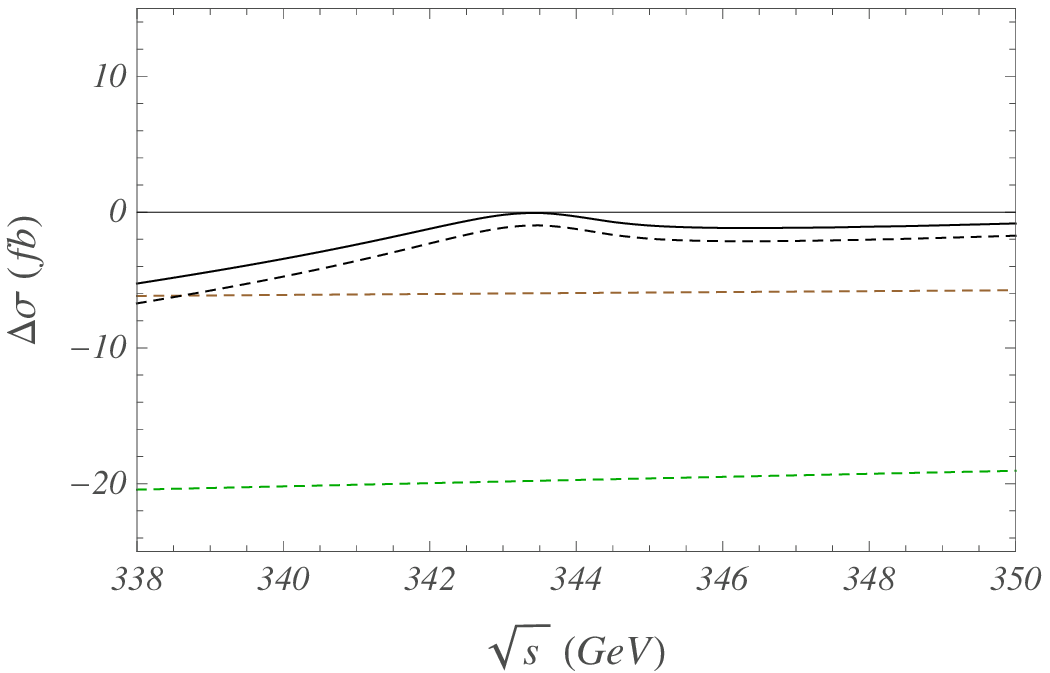}
  \caption{Contributions to the inclusive cross section from the
NLL, NNLL and N$^{3}$LL order phase space matching corrections 
in $\Delta\sigma_{\rm psm}^{{\rm Re}[G^c]}$
 (green, brown
and black dashed lines, respectively) as a function of the total center of mass 
energy for $\Delta M_t=15$ (left) and $\Delta M_t=35$~GeV (right).
The black solid lines correspond to the phase space matching contributions 
$\Delta\sigma^{{\rm Re}[G^c]}$
obtained from the real part of the exact cut Coulomb Green function at LL.
}
  \label{fig:ReG}
  \end{center}
\end{figure}

In Fig.~\ref{fig:ReG} the numerical LL Coulomb-resummed phase space effects
$\Delta\sigma^{{\rm Re}[G^c]}\equiv\sigma_{\rm NRQCD}^{{\rm Re}[G^c]}(\Lambda)-
\sigma_{\rm NRQCD}^{{\rm Re}[G^c]}(\infty)$ (solid black lines) and the
corresponding NLL (green dashed lines), NNLL (brown dashed lines) and N${}^3$LL
(black dashed lines) approximations from the phase space matching procedure are
shown. The left panel refers to $\Delta M_t=15$~GeV and the right panel to
$\Delta M_t=35$~GeV. The strong and QED couplings are again evaluated at the
hard scale $\nu=1$, i.e.\ $\alpha_s(m_t)=0.1077$ and $\alpha_{\rm
  qed}(m_t)=1/125.9$, and the  
other input parameters are given in Eq.~(\ref{parameters}). As in the previous
subsection we again neglect the hard QCD corrections to the $t\bar t$ current
matching coefficient, i.e.\ we 
set $c_1(\nu=1)=1$. The situation we find is quite similar to the one discussed
before for the imaginary part of the Coulomb Green function. The NLL order phase
space matching contributions amount to around $-13$ ($-20$)~fb, and the NNLL
order contributions to about $+18$ ($+15$)~fb for $\Delta M_t=15$~($35$)~GeV.
The NNLL corrections are quite sizeable and even exceed the NLL contributions
for $\Delta M_t=15$~GeV. It is conspicuous that the NLL and NNLL phase space
matching corrections have opposite signs and cancel each other to a large
extent. The sum of the NLL and NNLL phase space matching corrections differ from
$\Delta\sigma^{{\rm Re}[G^c]}$ by at most $5$~fb.
The N${}^3$LL order phase space matching corrections,
on the other hand, are much smaller than the NLL and NNLL order ones. Although
we have neglected the $\tilde C^{{\rm int},2}_{V/A}$, which arise from diagrams
with two insertions of the Coulomb potential, the difference of the phase space
matching contributions up to N${}^3$LL order and the exact LL Coulomb-resummed
result amounts to less than $2$~fb for all $\Delta M_t$ between $15$ and
$35$~GeV. We conclude that the  $\tilde C^{{\rm int},2}_{V/A}$ are numerically
small and that the N${}^3$LL phase space matching contributions are
more than adequate for our theoretical precision aim.

\vskip 5mm \noindent
{\bf \underline {Contributions related to the other ${\cal O}(v^2)$ relativistic
  corrections}}\\[2mm] 
\noindent
As the third class of phase space matching contributions we now examine the
corrections that arise from insertions of ${\cal O}(v^2)$ suppressed operators
other than the interference corrections just discussed above. Up to N${}^3$LL
order the contributions of these matching corrections to the inclusive NRQCD
cross section have the form ($i=\{{\rm kin}, {\rm dil}, v^2, 
{\mbox{$P$-wave}}\}$, 
$j=\{r, s, {\rm kin}, {\rm dil}, v^2, {\mbox{$P$-wave}}\}$, 
$k=\{r, s, {\rm kin}, {\rm dil}, v^2\}$)
\begin{align}
\Delta\sigma_{\rm psm}^{{\cal O}(v^2)} & \, = \,
\im\bigg[ \sum_i\,\Big(\tilde C^{i,0}_V(\Lambda) + \tilde C^{i,0}_A(\Lambda) \Big)\bigg]_{\rm NLL} \,+\,
\im\bigg[ \sum_j\,\Big( \tilde C^{j,1}_V(\Lambda) + \tilde C^{j,1}_A(\Lambda)  
\Big)\bigg]_{\rm NNLL} 
\nonumber\\[2mm]& \,+\,
\im \bigg[  \sum_j\,\Big( \tilde C^{j,2}_V(\Lambda) + \tilde C^{j,2}_A(\Lambda)  \Big)
\,+ \,
 \frac{E}{m_t}\, \sum_i\,\Big(\tilde C^{(1),i,0}_V(\Lambda) + 
\tilde C^{(1),i,0}_A(\Lambda) \Big)
\nonumber\\[2mm]& \quad\quad\quad+\,
4\,N_c\,
\Big( (C_{V,1}^{\rm Born})^2 + (C_{A,1}^{\rm Born})^2 \Big)
\sum_k\,\,i\,\delta\tilde c_1^{k,1}(\Lambda)\,G^c(a,v,m_t,\nu)  \bigg]_{{\rm
   N}^3{\rm LL}}
\,,
\label{sigmav2exp}
\end{align}
where we have again indicated the NLL, NNLL and N${}^3$LL order corrections.
The results for the $\tilde C^{i,0}_{V/A}$, $\tilde C^{j,1}_{V/A}$, 
$\tilde C^{(1),i,0}_{V/A}$ and $\delta\tilde c_1^{k,1}$ have
been given in  Eqs.~(\ref{relativisticresults1}), 
(\ref{tildeCVA1v2}), 
(\ref{relativisticresults2}) and 
(\ref{deltacint1}), respectively. As for the interference coefficients the
results for the $\tilde C^{j,2}_{V/A}$ 
are currently unknown and are neglected in the following analyses.

\begin{figure}[t]
  \begin{center}
  \includegraphics[width=0.49\textwidth]{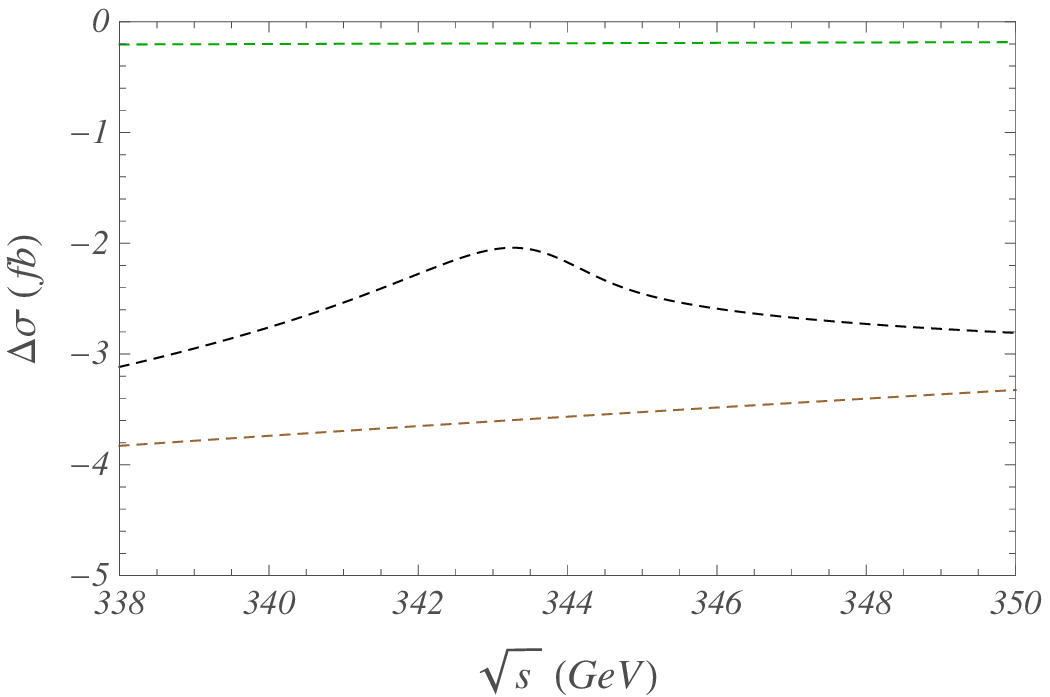}\;\;
  \includegraphics[width=0.49\textwidth]{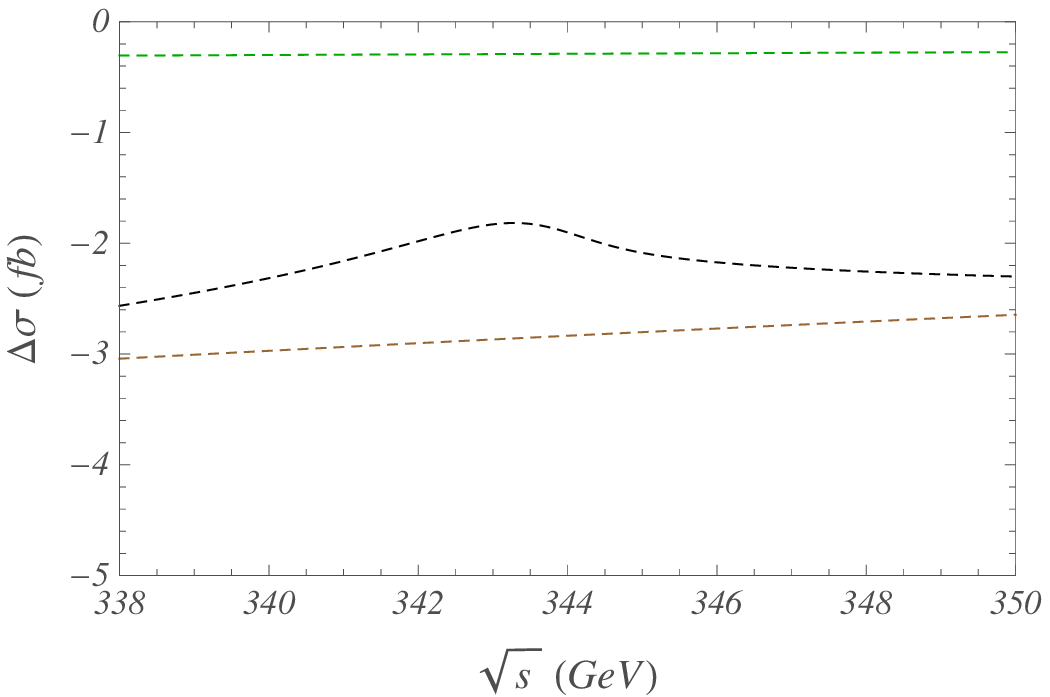}
  \caption{Contributions to the inclusive cross section
from the sum of the phase space matching arising from insertions 
of ${\cal O}(v^2)$ relativistic corrections other than the
interference contributions 
at NLL (green), NNLL (brown) and N${}^3$LL (black) order, as a function of the total center of mass 
energy for $\Delta M_t=15$ (left) and $\Delta M_t=35$~GeV (right).
}
  \label{fig:sumv2}
  \end{center}
\end{figure}

In Fig.~\ref{fig:sumv2} $\Delta\sigma_{\rm psm}^{{\cal O}(v^2)}$ is displayed at
NLL (green dashed lines), NNLL (brown dashed line) and N${}^3$LL (black dashed
lines). It is striking that the NLL contributions are at the sub-fb level and an
order of magnitude smaller than the NNLL terms. The small size of the NLL terms
is, however, due to an almost complete cancellation in the sum of the individual
$\tilde C^{i,0}_{V/A}$ coefficients (see
Fig.~\ref{fig:relatcor}). The size of the individual coefficients is at the
level of $1-3$~fb and comparable in size to the individual NNLL coefficients
$\tilde C^{j,1}_{V/A}$ (see Fig.~\ref{fig:relatcoras}). The NNLL coefficients, on
the other hand, do not cancel to the same extent due to the additional contribution from the potential ${\cal V}_r^{(s)}$, see
Eq~(\ref{tildeCVA1v2}). Thus the situation concerning the $\alpha_s$-expansion
for the third class of the phase space matching contributions is similar
to the other classes of phase space matching contributions discussed above.
This is also confirmed by the size of the N${}^3$LL
corrections which amount to at most $1$ to $2$~fb. Although we do not have a
numerical Coulomb-resummed expression for a more thorough comparison, the
results indicate that the $\alpha_s$-expansion is also well under control for
the third class of phase space matching contributions. Due to the overall small
size of these contributions we conclude again that keeping the phase space
matching contributions up to N${}^3$LL order is more than adequate to reach our
theoretical precision goal.

\section{Numerical Analysis}
\label{sectionanalysis}
In Sec.~\ref{subsectionfulltheory} we have demonstrated for the case
$\alpha_s=0$ that the phase space matching contributions are essential in order
to compensate for the fact that the previous NRQCD predictions for top threshold
production overestimate the full theory cross section by a substantial 
amount. The problem of the previous NRQCD predictions is related to the fact
that upon shifting the energy by $E\to E+i\Gamma_t$, in order to account for the
top quark finite lifetime, the NRQCD $t\bar t$ phase space becomes unrestricted
and extends to unphysical regions as a consequence of the nonrelativistic
expansion. The phase space matching procedure removes these unphysical phase
space contributions and implements the information on possible experimental cuts
into the NRQCD predictions. These phase space
matching contributions are represented by imaginary contributions to the Wilson
coefficients of NRQCD, see Sec.~\ref{sectionnotation}.

In this section we analyze the complete set of N${}^3$LL phase space 
matching contributions determined in the previous sections for predictions of
the inclusive NRQCD top pair threshold cross section with cuts on the invariant 
masses of the top and the antitop quark defined in Eq.~(\ref{eq:lambdac}). We
also compare the size of the phase space matching contributions to the other
and previously known types of electroweak effects relevant for the top pair
threshold cross section: the NNLL QED corrections, the hard electroweak
corrections~\cite{Hoang:2006pd} and the (type-1) finite lifetime corrections,
which are not related to phase space restrictions~\cite{Hoang:2004tg}. Together
with the phase space matching contributions, which we call type-2 finite
lifetime corrections, these three classes of contributions constitute all
effects of the electroweak interactions on the threshold cross 
section.\footnote{
We do not discuss here the effects of the $e^+e^-$ luminosity spectrum since it
is determined for the most part from experimental measurements and simulations.
}

To start we collect all phase space matching contributions to the inclusive
NRQCD cross section up to N${}^3$LL order. In the previous sections we have for
simplicity neglected the hard QCD and QED matching corrections contained in the
Wilson coefficient $c_1(\nu=1)$ of the leading $(e^+e^-)(t\bar t)$ top pair
production operator. Accounting for these matching corrections the complete set
of N${}^3$LL phase space matching contributions can be derived from the
factorization formula in Eq.~(\ref{crosssection}) and takes the form
\begin{align}
\Delta &\sigma_{\rm PSM}(\Lambda,\nu=1) \,= \,
\Delta\sigma^{\rm NLL}(\Lambda,1) \, + \,
\bigg[\,\Delta\sigma^{\rm NNLL}(\Lambda,1) + 
2\,h_1^{(1)}\,\Delta\sigma^{\rm NLL}(\Lambda,1)
\,\bigg]_{\rm NNLL} 
\nonumber \\[2mm]& \, + \,
\bigg[\,\Delta\sigma^{{\rm N}^3{\rm LL}}(\Lambda,1) + 
2\,h_1^{(1)}\,\Delta\sigma^{\rm NNLL}(\Lambda,1) + 
\Big(\,2\,h_1^{(2)}+(h_1^{(1)})^2\,\Big)\,\Delta\sigma^{\rm NLL}(\Lambda,1)
\,\bigg]_{{\rm N}^3{\rm LL}}
\,,
\label{DeltasigmaPSM}
\end{align}
where we have specifically indicated by brackets the NNLL and N${}^3$LL order
contributions. Explicit expressions for the NLL and NNLL hard QCD/QED matching
conditions $h_1^{(1)}$ and  $h_1^{(2)}$ are given in Eq.~(\ref{c1hard}).
The terms $\Delta\sigma^{{\rm N}^k{\rm LL}}(\Lambda,1)$
are the N${}^k$LL phase space matching contributions to the inclusive cross
section with the $h_{1}^{(1,2)}$ set to zero. The result for 
$\Delta\sigma^{\rm NLL}$ reads 
($i={\rm int}, {\rm dil}, {\rm kin}, v^2, \mbox{P-wave}$)
\begin{align}
\Delta\sigma^{\rm NLL}(\Lambda,1) \,= \,
\im\bigg[\tilde C^0_V(\Lambda) + \tilde C^0_A(\Lambda) +
\sum_i\,\Big(\tilde C^{i,0}_V(\Lambda) + \tilde C^{i,0}_A(\Lambda) \Big)\bigg]
\,,
\label{DeltasigmaNLL}
\end{align}
where expressions for $\tilde C^{0}_{V/A}$ and the 
$\tilde C^{i,0}_{V/A}$  have been given in 
Eqs.~(\ref{tildeCVA0}) and (\ref{relativisticresults1}), respectively.
The term $\Delta\sigma^{\rm NNLL}$ has the form
($j=r,s,{\rm int}, {\rm dil}, {\rm kin}, v^2, \mbox{$P$-wave})$
\begin{align}
\Delta\sigma^{\rm NNLL}(\Lambda,1) & \,= \,
\im\bigg[\tilde C^1_V(\Lambda) + \tilde C^1_A(\Lambda) +
\sum_j\,\Big(\tilde C^{j,1}_V(\Lambda) + \tilde C^{j,1}_A(\Lambda) \Big)
\nonumber \\[2mm]& \quad\quad\quad -\,
h_1^{(1)}\,\Big(\tilde C^{v^2,0}_V(\Lambda) + \tilde C^{v^2,0}_A(\Lambda) \Big)
\bigg]
\,,
\label{DeltasigmaNNLL}
\end{align}
where expressions for $\tilde C^{1}_{V/A}$ and the 
$\tilde C^{j,1}_{V/A}$  have been given in 
Eqs.~(\ref{tildeCVA1}) and (\ref{tildeCVA1v2}), respectively.
Finally, $\Delta\sigma^{{\rm N}^3{\rm LL}}$ reads
($i=\{{\rm int}, {\rm dil}, {\rm kin}, v^2, \mbox{P-wave}\},
j=\{r, s, {\rm int}, {\rm dil}, {\rm kin}, v^2, \mbox{P-wave}\},
k=\{r, s, {\rm dil}, {\rm kin}, v^2 \})$
\begin{align}
\Delta\sigma^{{\rm N}^3{\rm LL}}(\Lambda,1) & \, = \,
(\Delta\sigma_{\rm psm}^{{\rm Im}[G^c]})_{{\rm N}^3{\rm LL}} 
\, + \, 
\im \bigg[
 \frac{E}{m_t}\, \sum_i\,\Big(\tilde C^{(1),i,0}_V(\Lambda) + 
\tilde C^{(1),i,0}_A(\Lambda) \Big)\bigg]
\nonumber\\[2mm]
& \,+\,
\im \bigg[  \sum_j\,\Big( \tilde C^{j,2}_V(\Lambda) + \tilde C^{j,2}_A(\Lambda)  \Big) \bigg]
\nonumber\\[2mm]
& \,+\, \im \bigg[
4\,N_c\,
\Big( (C_{V,1}^{\rm Born})^2 + (C_{A,1}^{\rm Born})^2 \Big)
\sum_k\,\,i\,\delta\tilde c_1^{k,1}(\Lambda)\,G^c(a,v,m_t,\nu)  \bigg]
\nonumber\\[2mm]
& \,+\, \im \bigg[
4\,N_c\,
\Big(  C_{V,1}^{\rm Born}C_{V,1}^{\rm int} + C_{A,1}^{\rm Born}C_{A,1}^{\rm int} \Big) \,i\,\delta\tilde c_1^{\rm int,1}(\Lambda)\,G^c(a,v,m_t,\nu)  \bigg]
\nonumber \\[2mm]& \, + \, \im \bigg[
\Big( (h_1^{(1)})^2-h_1^{(2)}\Big)\,\Big(\tilde C^{v^2,0}_V(\Lambda) + \tilde C^{v^2,0}_A(\Lambda) \Big)
\,-\, h_1^{(1)}\,
\Big(\tilde C^{v^2,1}_V(\Lambda) + \tilde C^{v^2,1}_A(\Lambda) \Big) \bigg]
\,,
\label{DeltasigmaN3LL}
\end{align}
where $(\Delta\sigma_{\rm psm}^{{\rm Im}[G^c]})_{{\rm N}^3{\rm LL}}$ is a 
numerical expression defined
in Eq.~(\ref{sigmaImGexpN3LL}), and the results for 
$\tilde C^{(1),k,0}_{V/A}$ and $i\delta\tilde c^{k,1}$ have been given in Eqs.~(\ref{relativisticresults2}) and 
(\ref{deltacint1}), respectively.
The results for the $\tilde C^{j,2}_{V/A}$ are currently unknown. For the
analysis we carry out in the following we set them to zero. Together with an
analytic determination of  
$(\Delta\sigma_{\rm psm}^{{\rm Im}[G^c]})_{{\rm N}^3{\rm LL}}$ we plan to
compute them in a separate publication. 

In the following analysis we use $m_t=172$~GeV  for the top mass in the 1S mass
scheme~\cite{Hoang:1999zc,Hoang:2001rr}, 
and all matching coefficients are evaluated at the scale $m_t$
($\nu=1$). For the QCD coupling we use $\alpha_s(m_t)=0.1077$
and for the QED ($\overline{\mbox{MS}}$) coupling 
$\alpha_{\rm qed}(m_t)=1/125.9$. All soft matrix element contributions are
evaluated for the QCD and QED couplings at the velocity renormalization parameter
$\nu=0.2$ which corresponds to $\mu_{\rm soft}=34.4$~GeV for the soft and to
$\mu_{\rm usoft}=6.88$~GeV for the ultrasoft scales. For the evaluation
of the hard one-loop electroweak corrections we choose $m_{\rm Higgs}=130$~GeV.
All other parameters are given in Eqs.~(\ref{parameters}). 

\begin{figure}[t]
  \begin{center}
  \includegraphics[width=0.49\textwidth]{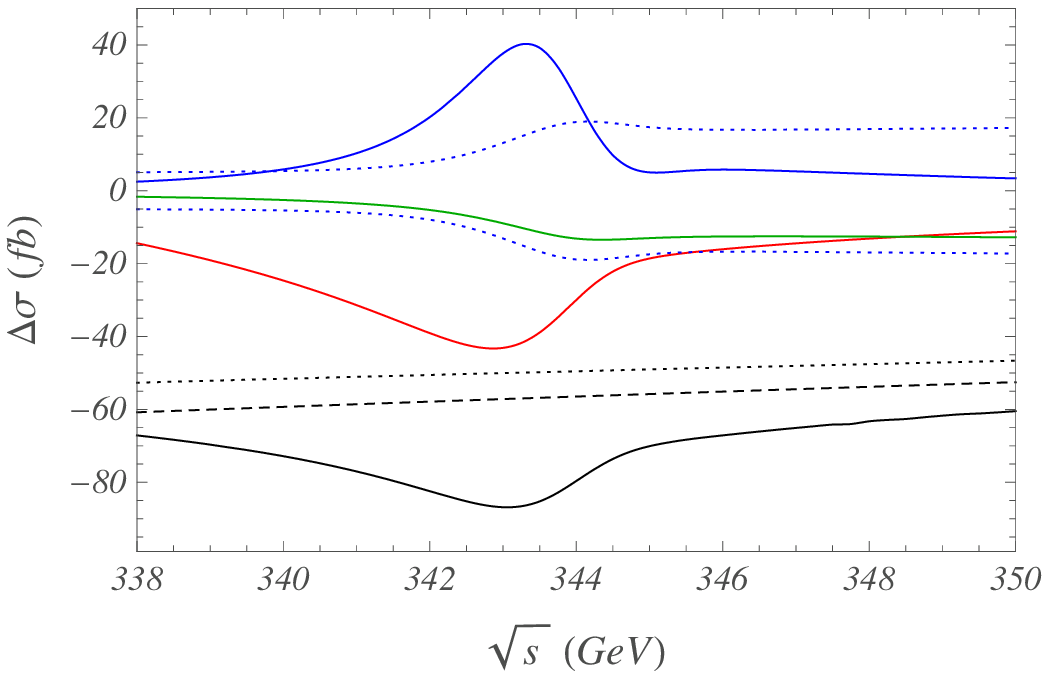}\;\;
  \includegraphics[width=0.49\textwidth]{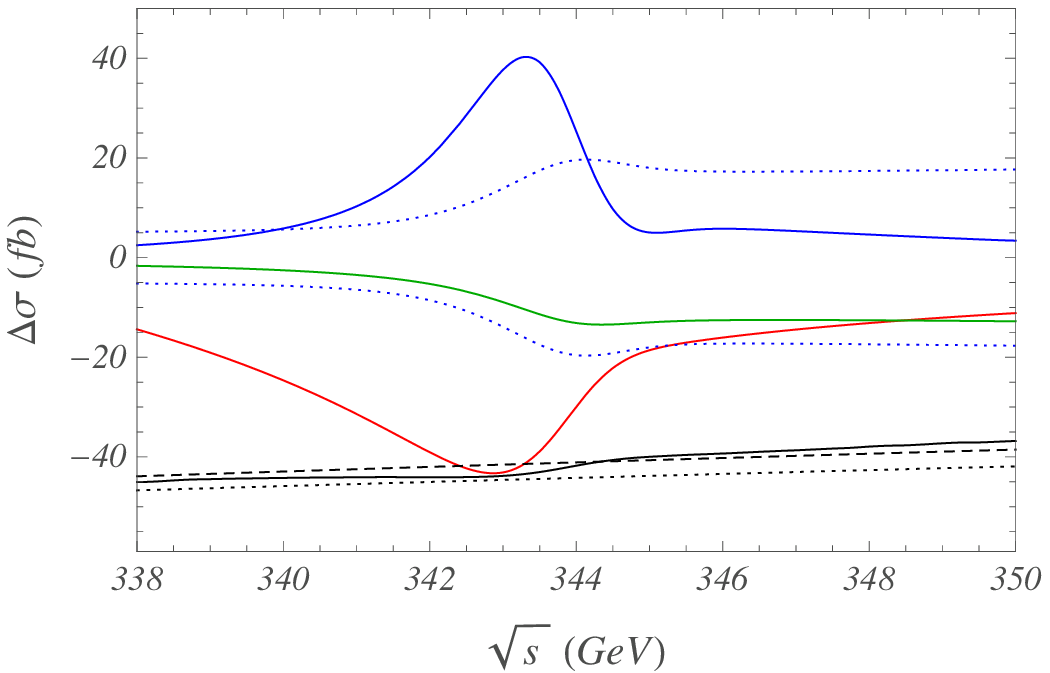}
  \caption{Sizes of the different contributions to the inclusive cross section 
arising from electroweak interactions as a function of the total center of
mass energy for $\Delta M_t=15$~GeV (left) and $\Delta M_t=35$~GeV (right):
(green line) NNLL hard one-loop electroweak effects
from Ref.~\cite{Hoang:2006pd}, (red line) NNLL finite lifetime corrections 
from Ref.~\cite{Hoang:2004tg}, (blue line) NNLL QED effects, and phase
space matching corrections at NLL, NNLL and N${}^3$LL (dotted, dashed and
solid black lines, respectively). The blue dotted lines correspond to the expected
experimental uncertainties at the LC.
}
  \label{fig:SigmaCorrec}
  \end{center}
\end{figure}

In Fig.~\ref{fig:SigmaCorrec} we show the phase space matching corrections to
the inclusive $t\bar t$ threshold cross section at NLL (black dotted lines),
NNLL (black dashed lines) and N${}^3$LL order (black solid lines), the NNLL QED
corrections (blue lines), the NNLL hard one-loop electroweak corrections 
(green lines) and the type-1 finite lifetime
corrections (red lines) as a function
of the c.m.\ energy $\sqrt{s}$. The left panel shows the results for an invariant
mass $\Delta M_t=15$~GeV and the right panel for  $\Delta M_t=35$~GeV. The QED
effects arise from the electromagnetic correction to the QCD Coulomb potential
(see text after Eq.~(\ref{Vccoeff})) and the one-loop QED matching correction to the
Wilson coefficient $c_1$ of the $t\bar t$ current, see
Eq.~(\ref{wilsoncoeff}). The hard one-loop electroweak corrections are encoded
in the coefficients $C_{V/A,1}^{\rm 1 loop}$ also shown in
Eq.~(\ref{wilsoncoeff}). The result for $C_{V/A,1}^{\rm 1 loop}$ has been
obtained in Ref.~\cite{Hoang:2006pd}.
The type-1 finite lifetime corrections represent all
finite lifetime corrections which are not related to phase space
constraints. They consist of the
corrections generated by the imaginary interference matching coefficient $i
C_{V/A,1}^{\rm int}$, see Eq.~(\ref{wilsoncoeff}), the time dilation corrections to
the Green function shown in Eq.~(\ref{greendilfirst}) and the contributions from
the renormalization group summation of phase space logarithms contained in the
coefficients $\tilde C_{V/A}$ of the $(e^+e^-)(e^+e^-)$ forward scattering operators 
given in Eq.~(\ref{tildeCav}). 
The matching and time dilation corrections are known at NNLL order
and the summation of phase space logarithms at NLL order~\cite{Hoang:2004tg}. 
The QED, hard electroweak and type-1 finite lifetime corrections do not depend
on phase space restrictions and are therefore identical in both panels. 
In Fig.~\ref{fig:SigmaCorrec} the blue dotted lines represent a rough (and
likely optimistic) estimation of the expected experimental uncertainties at a
future linear collider consisting of an energy-independent error of $5$~fb and a
$2$\% relative uncertainty with respect to the full prediction, both being added
quadratically.

\begin{figure}[!t]
  \begin{center}
  \includegraphics[width=1\textwidth]{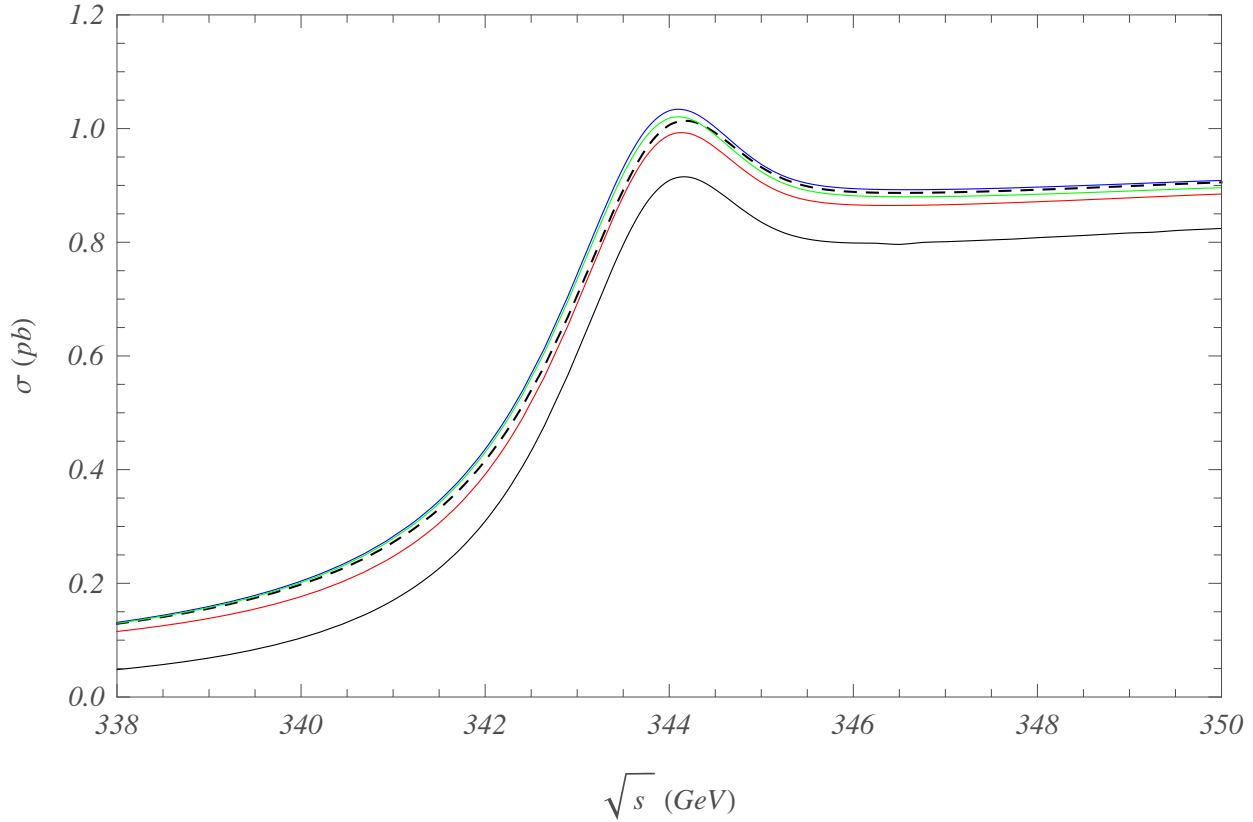}
  \caption{Total inclusive top pair production cross section from NRQCD: starting
from the pure QCD NNLL prediction (black dashed line), we add step-by-step
the QED corrections (blue line), the hard electroweak corrections (green line),
the type-1 finite lifetime corrections (red line) and the N${}^3$LL phase space
corrections (black solid line) for $\Delta M_t=15$~GeV.
}
  \label{fig:FinalQRIP15GeVCutnu02}
  \end{center}
\end{figure}
\begin{figure}
  \begin{center}
  \includegraphics[width=1\textwidth]{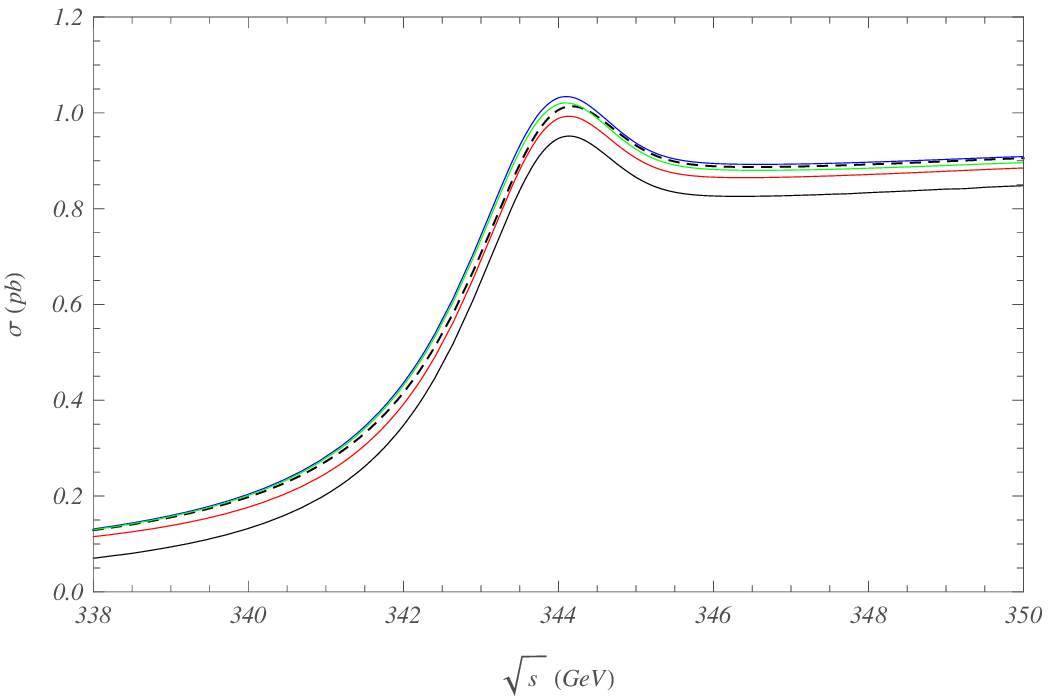}
  \caption{Total inclusive top pair production cross section from NRQCD: starting
from the pure QCD NNLL prediction (black dashed line), we add step-by-step
the QED corrections (blue line), the hard electroweak corrections (green line),
the type-1 finite lifetime corrections (red line) and the N${}^3$LL phase space
corrections (black solid line) for $\Delta M_t=35$~GeV.
}
  \label{fig:FinalQRIP35GeVCutnu02}
  \end{center}
\end{figure}

We see that the QED (blue lines) and the type-1 finite lifetime corrections (red
lines) are sizeable (at the level of $40$~fb) only in the peak region just below
$\sqrt{s}=2 m_t$. Above and below the peak region the QED corrections are quite
small and do not exceed  $5$~fb. Above and below the peak the type-1 finite
lifetime corrections amount to $-15$ to $-10$~fb. Due to their different signs
the QED corrections and the type-1 finite lifetime corrections cancel each other
to a large extent in the peak region. The hard electroweak corrections (green
lines) represent a multiplicative factor of -1.2\% to the total
cross section and are therefore very small below the peak and at the level of
$12$-$13$~fb above the peak region.\footnote{
The small size of the hard electroweak corrections displayed in
Fig.~\ref{fig:SigmaCorrec} is obtained for the QED coupling defined at the scale
of the top mass $m_t$.} 
We see that the phase space matching contributions represent the largest of the
four classes of electroweak effects. In contrast to the other classes of
electroweak effects they do not decrease strongly for energies below the peak
region. For $\Delta M_t=15$~GeV the N${}^3$LL phase space matching contributions
amount between $-85$ and $-65$~fb and for $\Delta M_t=35$~GeV they are
between $-45$ and $-35$~fb. The overall size of the phase space matching
corrections decreases for larger values of the top invariant mass cut $\Delta
M_t$. We emphasize, however, that the results obtained in this work are 
valid only for moderate values of $\Delta M_t$ in the region between 
$15$ and $35$~GeV. For invariant mass cuts below $15$~GeV the phase space
constraints are not related anymore to hard effects and for invariant mass cuts
substantially above $35$~GeV matching contributions that need to be computed
from full theory diagrams have to be included. 
The relatively flat behavior of the phase space matching contributions is
related to the fact that the dominant phase space matching contributions are
energy-independent. The small linear dependence on $\sqrt{s}$ is 
related to the $\sqrt{s}$ dependence of the virtual $\gamma$ and $Z$ propagators
of the basic $e^+e^-\to t\bar t$ process and the peak-like structure comes
from an imaginary phase space matching contribution to the $(e^+e^-)(t\bar t)$
top pair production operator which enters the N${}^3$LL inclusive cross section
in terms of a time-ordered product. This time-ordered product leads to a
non-analytic dependence on 
the energy, see Eqs.~(\ref{crosssection}) and (\ref{DeltasigmaN3LL}). In
Fig.~\ref{fig:SigmaCorrec} we have also displayed the phase space matching
contribution to the inclusive cross section at NLL (black dotted lines), NNLL
(black dashed lines) and N${}^3$LL order (black solid lines) in order to show
the convergence of the phase space matching procedure. The results show that the
expansion related to the phase space matching procedure is particularly good for
larger values of $\Delta M_t$ and still well under
control for $\Delta M_t=15$~GeV. We note that the rather small size of the NNLL
corrections (difference of black dotted and dashed lines) for $\Delta
M_t=15$~GeV arises from a 
cancellation between different independent NNLL corrections, see
Sec.~\ref{sec:expansiontest}.

In Figs.~\ref{fig:FinalQRIP15GeVCutnu02} and \ref{fig:FinalQRIP35GeVCutnu02} the
size of the four different types of electroweak corrections is shown for
predictions of the total inclusive cross section. Starting from the pure QCD
NNLL cross section (black dashed lines), which accounts only for the electroweak
effects from the basic $e^+e^-\to\gamma,Z\to t\bar t$ process and the LL finite
lifetime effects through the energy replacement rule $E\to E+i\Gamma_t$), we add
step-by-step the QED corrections (blue lines), the  
hard electroweak corrections (green lines), the type-1 finite lifetime
corrections (red lines) and the N${}^3$LL phase space corrections (black solid
lines). In Fig.~\ref{fig:FinalQRIP15GeVCutnu02} we have $\Delta M_t=15$~GeV and
in Fig.~\ref{fig:FinalQRIP35GeVCutnu02} we have $\Delta M_t=35$~GeV. Only the
phase space matching corrections depend on $\Delta M_t$. We again see that the
phase space matching contributions exceed by far the other electroweak
corrections. Since we have already
discussed the size of the individual types of electroweak corrections in our
analysis of Fig.~\ref{fig:SigmaCorrec}, we concentrate here only on the
behavior of the predictions for energies below the peak region where
the cross section is small. Here the phase space matching corrections are
very large and amount to changes of more than $50$\% percent for $\Delta
M_t=15$. These large corrections are related to the unphysical phase space
contributions contained in the pure QCD prediction which are a consequence of
the nonrelativistic expansion and the energy replacement rule $E\to
E+i\Gamma_t$. We stress that these unphysical effects cannot 
be cured by adding more of the higher order QCD corrections because they
originate from modifications to the nonrelativistic $t\bar t$ phase space 
products caused by the top width. Thus in order to obtain realistic predictions
for the top threshold cross section it is essential to account for the phase
space matching corrections.

\section{Conclusion}
\label{sectionconclusion}

The effects of the finite top quark lifetime are an essential ingredient for
predictions of the top pair production rate close to threshold $\sqrt{s}\approx
2m_t$, where $m_t$ is the top quark mass. Because the top width $\Gamma_t$ has
approximately the same size as the typical top quark kinetic energies, finite
lifetime effects already enter the leading-order predictions and cannot be
treated as corrections. An important consequence is that the top pair cross
section is only well-defined with a given set of prescriptions how the observed
top decay final states are accounted for in the cross section. This entails that
the cross section also accounts for non-$t\bar t$ processes which lead to the
same observed final state, and that the cross section can depend on experimental
cuts on kinematic variables such as the reconstructed invariant masses. 
When matching to the NRQCD effective theory these finite lifetime effects can be
integrated out for observables that are inclusive on the top and antitop decays,
and lead to imaginary matching contributions to the Wilson coefficients of NRQCD
operators. These imaginary matching coefficients are much like the complex
indices of refraction in the Maxwell theory of light propagation in an
absorptive medium. The cross section including the finite lifetime effects can
then be obtained from the absorptive part of the $e^+e^-\to e^+e^-$ forward
scattering amplitude using the optical theorem. 

One can distinguish two types of imaginary NRQCD matching coefficients. The
type-1 contributions~\cite{Hoang:2004tg} account for the (Cutkosky) cuts through
the top decay final states in full theory diagrams. They lead e.g.\ to the well
known quark bilinear top width term and also account for the interference
effects mentioned above. Insertions of the associated operators also cause
UV divergences in the nonrelativistic $t\bar t$ phase space integrations since
the resulting unstable top propagator $i/(p_0-\bmp^2/2m_t+i\Gamma_t/2)$ lifts
the stable particle dispersion relation $p_0=\bmp^2/2m_t$ and allows for
arbitrarily large final state top invariant masses. These UV divergences require
the introduction of $(e^+e^-)(e^+e^-)$ forward scattering operators, which
acquire an imaginary anomalous dimension and sum large logarithms of the top
velocity in the 
$t\bar t$ final state phase space. The type-2 imaginary matching contributions
are the matching conditions of the Wilson coefficients of these
$(e^+e^-)(e^+e^-)$ forward scattering operators. They encode the information on
the experimental cuts used for the cross section measurement. Since the type-2
matching contributions are not related to hard virtual fluctuations, but to real
final state configurations with large top quark off-shellness, we call them 
{\it phase space matching contributions}. 

In this work we have determined and analyzed the type-2 phase space matching
conditions up to N${}^3$LL order in the nonrelativistic expansion for cuts on
the reconstructed top and antitop invariant masses $M_t$, $M_{\bar t}$ of the
form $|M_{t,\bar t}-m_{t}|\,\le\, \Delta M_t$, with $\Delta M_t$ between $15$
and $35$~GeV and neglecting the $W$ boson width. We have demonstrated that the
numerically dominant effect of the phase space matching is to remove the
unphysical NRQCD phase space contributions that do not pass 
the cut. This is because the nonrelativistic unstable top propagator 
$i/(p_0-\bmp^2/2m_t+i\Gamma_t/2)$ overestimates by far top and
antitop fluctuations with large off-shellness. The remaining numerical
contributions to the phase space matching conditions coming from calculations of
relativistic full theory diagrams were found to be well below $5$~fb for the
cross section, which is negligible for the experimental precision one can expect
at a future linear collider. From the field theoretic point of view, the
procedure of carrying out the phase space matching agrees with the common
matching and renormalization methods for stable particle theories. 
Thus at higher orders in the nonrelativistic expansion it is required to account
for the phase space 
matching contributions of subdiagrams to remove non-analytic matrix element
terms from the matching equations and to achieve that the matching coefficients 
are analytic in the external energy. In the phase space matching procedure at
N${}^3$LL order for top pair production at threshold one has to also include 
the phase space matching for $(e^+e^-)(t\bar t)$ top production
operators. 

Since the phase space matching procedure we have carried out involves the
computation of NRQCD phase space integrations with a hard cutoff related to
$\Delta M_t$, our results contain power-counting breaking contributions. This
means that phase space matching contributions coming from insertions of higher
order operators can give contributions that are formally lower order. We have
shown that such power-counting breaking contributions do not spoil the
nonrelativistic expansion and that power-counting breaking can be ignored from the
practical point of view. As far as the $\alpha_s$ expansion is concerned, we have
found that the N${}^3$LL (${\cal O}(\alpha_s^2)$) corrections to the phase
space matching contributions need to be determined to meet the experimental
precision expected at a future linear collider. At this time these  N${}^3$LL
corrections are not yet fully known analytically, and their determination shall
be addressed in subsequent work. 

Our final numerical results have been given in
Sec.~\ref{sectionanalysis}. Leaving aside the effects from the $e^+e^-$
luminosity spectrum, which are known to distort the cross section shape and
normalization in a quite substantial way, we find that the phase space matching
contributions to the cross section  exceed by far the other types of electroweak
corrections, which are known from previous work. The phase space matching
contributions are between $-85$ and $-35$~fb for invariant mass cuts $\Delta
M_t$ between $15$ and $35$~GeV and are essential for realistic theoretical
predictions. In the peak and the continuum region ($\sqrt{s}\gsim 2m_t$) they
amount to $6$ to $10$\%. They are particularly important in the region below the
peak ($\sqrt{s}\lsim 2m_t$) where the cross section decreases and the
unphysical off-shell contributions of the NRQCD $t\bar t$ phase space become
dominant. Here the phase space matching contributions can amount to more than
$50$\%, and they ensure that the cross section has the correct physical
behavior. 

Phase space matching is also important for predictions of the top pair threshold
cross section if no kinematic cuts are imposed, since the NRQCD
phase space contributions involving off-shell top and antitop quarks still lead
to large unphysical contributions. Here the phase space matching contributions
from relativistic full theory diagrams are numerically important and cannot be
neglected (Sec.~\ref{subsectionfulltheory}). In this work we have determined
these full theory contributions for $\alpha_s=0$, i.e.\ at NLL order. We finally
note that the phase space matching procedure can also be carried out using
exclusively  full theory computations with kinematic cuts to determine the
imaginary type-2 matching 
coefficients. This approach does not involve any power-counting breaking
contributions and allows to determine the phase space matching contributions
more easily for $\Delta M_t>35$~GeV. At this time these full theory computations
are only known for $\alpha_s=0$, which allows to carry out this phase space
matching approach at NLL order. To go beyond the NLL level the results for the
${\cal O}(\alpha_s)$ and ${\cal O}(\alpha_s^2)$ corrections to the $e^+e^-$
cross section are required 
for the final states that arise in top pair production. Such results are not
available at this time.

\begin{acknowledgments} 
This work was supported in part  by the 
DFG Sonder\-forschungsbereich/Transregio~9 
``Computergest\"utzte Theoretische Teilchenphysik'' 
and the EU network contract MRTN-CT-2006-035482 (FLAVIAnet).
P.R. thanks M.~Beneke for useful discussions.
Feynman diagrams have been drawn with the packages 
{\sc Axodraw}~\cite{Vermaseren:1994je} and 
{\sc Jaxo\-draw}~\cite{Binosi:2003yf}.
\end{acknowledgments}

\appendix

\section{QCD interference effects}
\label{app:QCDinterference}

In this appendix we compute the ${\cal O}(\alpha_s)$ ultrasoft corrections to
the imaginary phase space matching coefficients $\tilde C_{V/A}$ of the 
$(e^+e^-)(e^+e^-)$ forward scattering operators $\tilde{\cal O}_{V/A}$
for the invariant mass prescription explained in Sec.~\ref{sec:QCDusoft}. The
corresponding diagrams in Coulomb gauge are shown in Fig.~\ref{fig6}. For the
interference diagram \ref{fig6}d we need to define the top and antitop
invariant masses in the presence of an additional ultrasoft gluon in the final
state. As a toy prescription that can be easily implemented analytically we
assume that we can resolve the gluon down to an infrared scale $\lambda$. For
gluon energies larger than $\lambda$ we define the top and antitop 4-momentum as
the sum of 4-momenta of their decay products, $bW^+$ and $\bar{b}W^-$, 
respectively. The top and antitop invariant masses are then defined exactly as
in Eq.~(\ref{eq:invariantmassdef}).
The result for the ultrasoft phase space matching corrections obtained from this
prescription should also be generic for the typical size of corrections for
other, more realistic invariant mass prescriptions. 
We can write the ${\cal O}(\alpha_s)$ contributions to
the imaginary phase space matching coefficients $\tilde C_{V/A}$  
from each of the diagrams in Fig.~\ref{fig6}a--d as ($i=a,b,c,d$)
\begin{align}
\tilde C^{\rm us}_{V/A,(i)} \, = \, 
i N_c (C^{\rm Born}_{V/A,1})^2\,c_{(i)}^{\rm us}
\,.
\end{align}
The resulting contributions to the inclusive cross section read
\begin{align}
\Delta\sigma_{(i)}^{\mathrm{us}} = N_c \left[ (C_{V,1}^{\rm Born})^2 +  
(C_{A,1}^{\rm Born})^2 \right] c^{\mathrm{us}}_{(i)} 
\,.
\end{align}

The computation of the ultrasoft corrections for energies close to threshold is 
performed using NRQCD Feynman rules for the top/antitop propagators and  
ultrasoft gluon couplings.
The gluon momentum $k$ is neglected when appropriate according to the scaling
$k\sim m_t v^2$. In Coulomb gauge the time-like gluon propagator has the form
$i/\bmk^2$ and the transverse propagator is 
$i(\delta^{ij}-{\rm k}^i{\rm k}^j/\bmk^2)/(k^2+i\epsilon)$.
We cut the diagrams as indicated by the red dashed lines in Fig.~\ref{fig6}
using the well-known 
Cutkosky rules for the transverse gluon propagator and
Eq.~(\ref{propcuttingrule}) for the top and antitop propagators. We note that
our computation is quite similar to the one presented some time ago in
Ref.~\cite{Melnikov:1993np}, where, however, no phase space cuts were considered.
The main conclusion in the work of Ref.~\cite{Melnikov:1993np} was that for the total cross
section (i.e. without phase space restrictions) the contributions from the
ultrasoft diagrams in Fig.~\ref{fig6} cancel. This serves as an important cross
check of the computations we carry out here.

The contribution from the time-like gluon exchange 
between the $b\bar{b}$ pair (Fig.~\ref{fig6}a plus the conjugated diagram)
reads:\footnote{
We disagree with the corresponding result given in Ref.~\cite{Melnikov:1993np} with respect
to the sign of the term $|{\mathbf B}(k^0,\bmk)|^2$. However, this does not
alter the conclusion that the contribution cancels in the absence of phase space
restrictions.
} 
\begin{eqnarray}
c^{\mathrm{us}}_{(a)}&=& \frac{4}{3} i\,C_F m_t^4 \int \frac{d^4p}{(2\pi)^4}\int 
\frac{d^4k}{(2\pi)^4}
\frac{\big( 3|A(k^0,\bmk)|^2-|{\mathbf B}(k^0,\bmk)|^2 \big) }{\bmk^2}\nn\\
&&\times
\frac{1}{(t_1-i m_t \Gamma_t)(t_2-i m_t \Gamma_t)(t_1-2m_t k^0+i m_t 
\Gamma_t)(t_2+2m_t k^0+i m_t \Gamma_t)} + {\rm c.c.}
,\quad
\label{eq:Cusa}
\end{eqnarray}
where the invariant mass variables $t_1,\,t_2$, are defined in
Eq.~(\ref{t12def}). The same variables are employed for the
computation of the
diagrams in Figs.~\ref{fig6}b and c.  
In Eq.~(\ref{eq:Cusa}) we have already performed the integrations 
over the $bW^+$ and $\bar{b}W^-$ phase space variables, which yields the 
functions 
\begin{eqnarray}
A(k^0,\bmk) &=& -\frac{g_s\Gamma_t}{\bmag k}
\ln \left(\frac{k^0-\bmag k-i\epsilon}{k^0+\bmag k-i\epsilon} \right)\,, 
\nn\\[3mm]
{\mathbf B}(k^0,\bmk) &=& 2\,\frac{g_s\Gamma_t}{\bmag k}\,\frac{2x-1}{2x+1}
\left[ 1 + \frac{k^0}{2\bmag k} 
\ln \left(\frac{k^0-\bmag k-i\epsilon}{k^0+\bmag k-i\epsilon} \right) \right]
\frac{\bmk}{\bmag k}\,,
\end{eqnarray}
where $x=M_W^2/m_t^2$.
It is straightforward to check that Eq.~(\ref{eq:Cusa}) 
vanishes if no bounds are imposed on the integration over the top energy:
carrying out the $p^0$ integration by residues one obtains a purely
imaginary number, which cancels out when adding the conjugate diagram.  
This confirms that the diagram vanishes if there are no cuts on the phase space
integration. 
For the invariant mass cuts of Eq.~(\ref{t1t2limits}) we proceed by performing 
the integration over the 3-momentum and the energy of the virtual gluon.
After the trivial integration over the $\bmp$ angles we obtain
a representation of the time-like gluon exchange diagram of the form
\begin{eqnarray}
c^{\mathrm{us}}_{(i)}&=& \frac{m_t^3 \Gamma_t^2}{2\pi^3}
\int_{\tilde \Delta(\Lambda)} dt_1 dt_2 \,
\frac{ \sqrt{m_t E-(t_1+t_2)/2} } {(t_1^2+(m_t\Gamma_t)^2)\,
(t_2^2+(m_t\Gamma_t)^2)}\,\Delta_{(i)}(t_1,t_2) 
\,.
\label{eq:t1t2dist}
\end{eqnarray}
We use this generic form for all the QCD interference diagrams, $i=a,b,c,d$. 
Note that expression~(\ref{eq:t1t2dist}) is compatible with
Eq.~(\ref{eq:sigmaNRQCDas0generic}), i.e. the $\Delta(t_1,t_2)$ functions in
both expressions have the same normalization. For the time-like 
gluon exchange from diagram \ref{fig6}a
we obtain
\begin{eqnarray}
\Delta_{(a)}(t_1,t_2) &=& 2C_F\alpha_s \bigg( 1-\frac{1}{9}\bigg( 
\frac{2x-1}{2x+1} \bigg)^2 \bigg)
\frac{1}{(t_1+t_2)^2+4(m_t\Gamma_t)^2}\nn\\[3mm]
&&\times \bigg\{ \Big( 2(m_t\Gamma_t)^2+t_2(t_1+t_2) \Big) \arctan 
\frac{t_1}{m_t\Gamma_t}
+\frac{m_t\Gamma_t}{4}(t_1-t_2)\ln \left( \frac{t_2^2+(\mG)^2}{t_1^2+(\mG)^2} 
\right) \nn\\[3mm]
&& \quad\; +\; \{t_1 \leftrightarrow t_2\} \bigg\}\,.
\end{eqnarray}
The time-like gluon exchange 
between $t\bar{b}$ and $\bar{t}b$, Fig.~\ref{fig6}b, is computed analogously:
\begin{eqnarray}
c^{\mathrm{us}}_{(b)}&=& 16 \,i\, C_F g_s m_t^5 \int \frac{d^4p}{(2\pi)^4}\int 
\frac{d^4k}{(2\pi)^4}
\frac{A(k^0,\bmk)}{\bmk^2}\frac{1}{(t_1-i m_t \Gamma_t)(t_2^2 + 
(\mG)^2)}\nn\\[3mm]
&&\times
\frac{1}{(t_1-2 m_t k^0+i m_t \Gamma_t)
(t_2 + 2m_t k^0+i m_t \Gamma_t)} 
+ \bigg\{ \begin{array}{c} t_1\leftrightarrow t_2 \\ k^0 \to -k^0 
\end{array} \bigg\}+{\rm c.c.}\,.
\label{eq:Cusb}
\end{eqnarray}
The terms shown explicitly in Eq.~(\ref{eq:Cusb}) represent the
contribution from  the $t\bar{b}$  gluon exchange (first diagram in
Fig.~\ref{fig6}b). The contribution from the $\bar{t}b$ gluon exchange is
obtained with the replacements $t_1\leftrightarrow t_2$, and $k^0 \to -k^0$,
as indicated in Eq.~(\ref{eq:Cusb}). 
For this contribution we find it more convenient to perform first the 
$k^0$-integral.
Again, it is easy to check that if the $p^0$-integration is done by residues, 
one obtains a result which is purely imaginary, so this contribution vanishes for
the total cross section without phase space restrictions.
The expression can be cast into the form of Eq.~(\ref{eq:t1t2dist}) with
\begin{eqnarray}
\Delta_{(b)}(t_1,t_2) &=& -\bigg( 1-\frac{1}{9}\bigg( \frac{2x-1}{2x+1} \bigg)^2 
\bigg)^{-1}
\Delta_{(a)}(t_1,t_2)\,.
\end{eqnarray}

Let us now turn to the space-like gluon exchange between the final state $b\bar
b$ pair in Fig.~\ref{fig6}c. The result reads
\begin{eqnarray}
c^{\mathrm{us}}_{(c)}&=& - \frac{8}{3}i\, C_F m_t^4 \int \frac{d^4p}{(2\pi)^4}\int 
\frac{d^4k}{(2\pi)^4}
\frac{|C(k^0,\bmk)|^2}{k^2+i\epsilon}\frac{1}{(t_1-i m_t \Gamma_t)(t_2-i\mG)}\nn\\[3mm]
&&\times
\frac{1}{(t_1-2 m_t k^0+i m_t \Gamma_t)
(t_2 + 2m_t k^0+i m_t \Gamma_t)} 
+ {\rm c.c.}\,,
\end{eqnarray}
where
\begin{eqnarray}
C(k^0,\bmk) &=& -\frac{g_s\Gamma_t}{\bmag k}\,\frac{2x-1}{2x+1}
\left[ \frac{k^0}{\bmag k} + \frac{{(k^0)}^2-\bmk^2}{2\bmk^2} 
\ln \left(\frac{k^0-\bmag k-i\epsilon}{k^0+\bmag k-i\epsilon} \right) 
\right]\,.
\end{eqnarray}
The expression can be cast into the form of Eq.~(\ref{eq:t1t2dist}) with
\begin{eqnarray}
\Delta_{(c)}(t_1,t_2) &=& -\frac{C_F\alpha_s}{12\pi} \bigg( \frac{2x-1}{2x+1} 
\bigg)^2
\bigg\{ \Big( \ln \left( \frac{t_1^2+(\mG)^2}{(2m_t)^4} \right)
-\frac{4\pi}{3}\arctan \frac{t_1}{m_t\Gamma_t} + \{t_1 \leftrightarrow t_2\} 
\Big)
\nn\\[3mm]
&&+ \frac{(t_1-t_2)(4\mG + \frac{4\pi}{3} (t_1+t_2) ) }{(t_1+t_2)^2+4(m_t\Gamma_t)^2} 
\Big( \arctan \frac{t_1}{m_t\Gamma_t} - \arctan \frac{t_2}{m_t\Gamma_t}\Big) 
\nn\\[3mm]
&&+ \frac{(t_1-t_2)( t_1+t_2 - \frac{4\pi}{3} \mG)}{(t_1+t_2)^2+4(m_t\Gamma_t)^2}
\ln \left( \frac{t_2^2+(\mG)^2}{t_1^2+(\mG)^2} \right)
-2\ln \frac{\lambda^2}{(2m_t)^2}  \bigg\}\,.
\end{eqnarray}
The $|\bmk|$ integration in $c^{\rm us}_{(c)}$ yields an infrared divergence, 
which we have regularized with the cutoff $\lambda$ mentioned at the beginning of
this appendix. This IR divergence is cancelled by a 
corresponding IR divergence in the real gluon emission diagram in
Fig.~\ref{fig6}d, as shown below. 
Attaching a space-like ultrasoft gluon to the top or antitop lines such as in
diagram \ref{fig6}e yields an additional $v$ factor, 
so the corresponding contributions are suppressed in the nonrelativistic 
expansion with respect to those in Figs.~\ref{fig6}a-d. 
This way we only need to consider real gluon emission from the bottom quark 
lines. Due to the additional gluon in the final state the relation between the
variables $t_{1,2}$ and the nonrelativistic loop momenta $(p^0,\bmp^2)$ has to
be modified. For the momentum routing displayed in Fig.~\ref{fig6}d,
$t_1$ and $t_2$ are defined as
\begin{align}
t_{1} & = \,2m_t\Big( \frac{E}{2} + p^0 - \frac{\bmp^2}{2m_t} - k^0 \Big)
\,,
\nonumber\\[2mm]
t_{2} & = \,2m_t\Big( \frac{E}{2} - p^0 - \frac{\bmp^2}{2m_t} \Big)
\,.
\end{align}
The result from diagram~\ref{fig6}d then reads
\begin{eqnarray}
c^{\mathrm{us}}_{(d)}&=& \frac{4}{3}\, C_F  m_t^4 \int \frac{d^4p}{(2\pi)^4}\int 
\frac{d^3\bmk}{(2\pi)^4}
\frac{C(\bmag k,\bmk)^2}{\bmag k}\frac{1}{(t_1-i m_t \Gamma_t)(t_2 
+i\mG)}\nn\\[3mm]
&&\times
\frac{1}{(t_1+2 m_t \bmag k +i m_t \Gamma_t)
(t_2 + 2m_t \bmag k-i m_t \Gamma_t)} +{\rm c.c.}\,.
\label{eq:Cusd}
\end{eqnarray}
For this contribution the $\bmag k$-integration extends to values such that the 
phase space
factor $\bmag p= \sqrt{m_t E-m_t{\bmag k} -(t_1+t_2)/2}$ remains a real number.
As anticipated above, the $\bmag k$-integration is IR-divergent and we 
introduce the cutoff $\lambda$ as in the case of diagram~\ref{fig6}c.
The expression can be cast into the form of Eq.~(\ref{eq:t1t2dist}) with
\begin{align}
\Delta_{(d)}(t_1,t_2) &= \frac{C_F\alpha_s}{12\pi} \bigg( \frac{2x-1}{2x+1} 
\bigg)^2
\bigg\{ 4\ln\left( \frac{2m_t E -(t_1+t_2)}{m_t^2} \right)
-2\ln\frac{\lambda^2}{(2m_t)^2}  \nn\\[2mm]
 +\bigg( \bigg[ &\frac{2\,(t_2 -i\,\mG)}{t_1-t_2 +2i\mG}
\,\frac{h(t_2)}{\widetilde h(t_1+t_2)}\,
\ln \left( \frac{h(t_2)+ \widetilde h(t_1+t_2)}
{h(t_2)- \widetilde h(t_1+t_2) } \right) 
+ \{t_1 \leftrightarrow t_2\} \bigg] 
+{\rm c.c.} \bigg)  \bigg\} \,,\nn\\
\end{align}
where
\begin{align}
h(y) &= \sqrt{m_t E -(y-i\mG)/2}\,,
\nn\\[2mm]
\widetilde h(y) &= \sqrt{m_t E -y/2}\,.
\end{align}
We explicitly see the cancellation of the infrared divergent terms in the sum of
$c^{\rm us}_{(c)}$
and $c^{\rm us}_{(d)}$. Again it is easy to check that after 
integrating over the top energy $p^0$ without restrictions, both contributions
cancel completely. 

Expanding the ultrasoft phase space matching contributions  
$m_t E,m_t \Gamma_t\ll \Lambda^2$
up to terms of order $1/\Lambda^3$, we obtain
\begin{eqnarray}
c^{\rm us}_{(a)} &=& \frac{\sqrt{2}m_t^2}{3\pi^2}C_F\alpha_s \bigg( 
1-\frac{1}{9}\bigg( \frac{2x-1}{2x+1} \bigg)^2 \bigg)
\frac{m_t \Gamma_t^2}{\Lambda^3}
\left\{ 3\ln \frac{m_t\Gamma_t}{\Lambda^2} + \ln 2 -1-\sqrt{2}-\sinh^{-1}(1) 
\right\}
\,,\nn\\[3mm]
c^{\rm us}_{(b)} &=& - \frac{\sqrt{2}m_t^2}{3\pi^2}C_F\alpha_s 
\frac{m_t \Gamma_t^2}{\Lambda^3}
\left\{ 3\ln \frac{m_t\Gamma_t}{\Lambda^2} + \ln 2 -1- \sqrt{2} -\sinh^{-1}(1) 
\right\}
\,,\nn\\[3mm]
c^{\rm us}_{(c)} &=&  \frac{\sqrt{2}m_t^2}{3\pi^3}C_F\alpha_s \bigg( 
\frac{2x-1}{2x+1} \bigg)^2
\frac{\Gamma_t}{\Lambda} \bigg\{ \ln \frac{\Gamma_t}{\lambda} \left(1+\frac{m_t 
E}{3\Lambda^2} \right) 
+\frac{1}{3\pi}\frac{m_t\Gamma_t}{\Lambda^2}\Big( \sqrt{2}+\sinh^{-1}(1) \Big) 
\ln\frac{\lambda m_t}{2\Lambda^2} 
\nn\\[3mm]
&&\qquad \qquad \qquad \qquad \qquad \qquad \quad +\frac{m_t\Gamma_t}{\Lambda^2}\left(
\frac{\pi}{3}\ln\frac{m_t\Gamma_t}{\Lambda^2} - \frac{k_1}{9\pi} \right)
 \bigg\}\,,
\nn\\[3mm]
c^{\rm us}_{(d)} &=& -\frac{\sqrt{2}m_t^2}{3\pi^3}C_F\alpha_s \bigg( 
\frac{2x-1}{2x+1} \bigg)^2
\frac{\Gamma_t}{\Lambda} \bigg\{ \ln \frac{\Gamma_t}{\lambda} \left(1+\frac{m_t 
E}{3\Lambda^2} \right) 
+\frac{1}{3\pi}\frac{m_t\Gamma_t}{\Lambda^2}\Big( \sqrt{2}+\sinh^{-1}(1) \Big) 
\ln\frac{\lambda m_t}{2\Lambda^2} 
\nn\\[3mm]
&&\qquad \qquad \qquad \qquad \qquad \qquad \quad +\frac{m_t\Gamma_t}{\Lambda^2} 
\frac{k_2}{9\pi} \bigg\}\,
\end{eqnarray}
with
\begin{eqnarray}
k_1 &=& d_1 + 5\sqrt{2} - 6\sqrt{2}\ln 2 - \frac{3\ln^2 2}{2}
+\pi(3\ln 2 -\frac{1}{2} ) + \pi^2( \sqrt{2}+\frac{5}{12}-2\ln 2) \nn\\[3mm]
&& +(5-\pi^2-9\ln 2)\sinh^{-1}(1) + \frac{3}{2} \big[\sinh^{-1}(1) \big]^2
- 3 {\rm Li}_2\Big(\frac{1-\sqrt{2}}{2}\Big)\,,
\nn\\[3mm]
k_2 &=& d_2 + \frac{\pi}{2}(5-12\ln 2)
+ \frac{3\pi^2}{4}\,.
\end{eqnarray}
The constants $d_1$ and $d_2$ have been evaluated numerically and read 
\begin{align}
d_1=-4.961 \quad,\quad d_2=-17.75\,.
\end{align}
Adding up the results of all diagrams we are - as anticipated from the general
arguments discussed in Sec.~\ref{sec:QCDusoft} - left with a correction to the
cross section of order $m_t\Gamma_t/\Lambda^2$ with respect to the NLL phase space 
correction:
\begin{equation}
\sum_{i=a,b,c,d} c_{(i)}^{\mathrm{us}} = \frac{m_t^2}{4\pi}\frac{4\sqrt{2}}{\pi}\frac{\Gamma_t}{\Lambda}\Delta^{\rm 
us}\,,
\end{equation}
where
\begin{eqnarray}
\label{deltausfinal}
\Delta^{\rm us} &=&
  \frac{C_F\alpha_s }{27\pi^2} \bigg( \frac{2x-1}{2x+1} \bigg)^2
\frac{m_t\Gamma_t}{\Lambda^2} \left\{ \pi^2\Big( 1+\sqrt{2}-\ln 2+\sinh^{-1}(1) 
\Big) - k_1 - k_2 \right\}\,.
\end{eqnarray}
Apart from the cancellation of the $\ln\lambda$ infrared divergences already
pointed out above, we also find that all logarithms of $\Gamma_t$ cancel in the
sum of all diagrams. This is expected because the phase space matching
contributions represent hard effects. A proper evaluation of the ultrasoft phase
space matching corrections therefore also requires to employ the strong coupling
in Eq.~(\ref{deltausfinal}) at the hard scale. For 
$m_t=172$~GeV, $M_W=80.425$~GeV and $C_F\alpha_s(m_t)=0.1436$ we find 
that $\Delta^{\rm us}= 0.004 \, m_t\Gamma_t/\Lambda^2$. We thus find that the
ultrasoft corrections based on our invariant mass prescription have an
additional strong numerical suppression factor. Although this result might not
be generalized to other invariant mass definitions, it nevertheless supports the
conclusion that ultrasoft effects are in general irrelevant at the level of
precision we aim for in this work.

\bibliography{phasespace}

\end{document}